\documentclass[aps, prb, superscriptaddress, twocolumn, a4paper]{revtex4-2}
\usepackage{hyperref}
\usepackage{graphicx}
\usepackage{amsmath}
\usepackage{amssymb}
\usepackage{amsfonts}   
\usepackage{multirow}
\usepackage[b]{esvect} 
\usepackage{braket}
\usepackage{color}
\usepackage{bm}
\usepackage{lipsum}
\renewcommand{\Re}{\operatorname{Re}}
\renewcommand{\Im}{\operatorname{Im}}

\newcommand{\bea}{\begin{eqnarray}}
\newcommand{\eea}{\end{eqnarray}}
\newcommand{\up}{\uparrow}
\newcommand{\dn}{\downarrow}

\begin{document}
\title{Photoinduced intradomain dynamics and nonthermal switching of metastable states in the one-dimensional extended Peierls-Hubbard model}

\author{Junichi Okamoto}
\affiliation{Institute of Physics, University of Freiburg, Hermann-Herder-Str. 3, 79104 Freiburg, Germany}
\affiliation{EUCOR Centre for Quantum Science and Quantum Computing, University of Freiburg, Hermann-Herder-Str. 3, 79104 Freiburg, Germany}

\author{Sajad Mirmohammadi}
\affiliation{Institute of Physics, University of Freiburg, Hermann-Herder-Str. 3, 79104 Freiburg, Germany}

\date{\today}

\begin{abstract}
We investigate the microscopic dynamics at the initial stage of photoinduced phase transitions in tetrathiafulvalene-$p$-chloranil by exact diagonalization. We first show that the one-dimensional extended Peierls-Hubbard model exhibits a neutral phase with small ionicity and negligible dimerization and an ionic phase with moderate ionicity and dimerization. Aside from these phases, we find a doubly ionized phase with strong dimerization that we call the ``dipole" phase. These ground-state phases are characterized by various order parameters and the Zak phase, which is relevant to electronic polarization. We further explore the microscopic dynamics of the three phases triggered by short monocycle optical pulses. The electronic order parameters and lattice displacement suggest that the neutral--ionic, ionic--neutral, and dipole--ionic transitions are induced. Furthermore, clear spectroscopic changes are observed in the time-dependent spectral density and pump-probe conductivity. A detailed analysis of the spectroscopy demonstrates the generation of coherent charge-transfer strings via multiphoton absorption and the crucial roles of the excited states and the metastable ground state at the new lattice position for the ultrafast dynamics.  
\end{abstract}

\maketitle
\section{introduction}\label{sec:introduction}
Intense optical excitation is a novel tool to investigate the intriguing regimes of a free-energy landscape that is hidden under the equilibrium condition \cite{tokura2006}. The recent progress of experimental techniques such as ultrafast lasers, nanoscale fabrication, and heterostructure engineering is rapidly expanding the realm of ``nonequilibrium material science” \cite{basov2017}. The detailed knowledge of the states far away from equilibrium is used to steer a system into a metastable state with a unique attribute, which serves as a basis for future technological application. A few examples of nonequilibrium material engineering are: optical modulation of Berry phase in graphene \cite{mciver2020}, light-induced superconductivity in high-$T_c$ superconductors \cite{fausti2011, hu2014, mitrano2016}, and ultrafast structural transitions in transition metal dichalcogenides \cite{stojchevska2014}. While the light sources in these examples can be considered classical, strong coupling between a material and quantum light is also a fascinating field of study. For instance, novel exciton-polariton states are created in van der Waals materials \cite{basov2016} or organic semiconductors \cite{orgiu2015, nagarajan2020}.

Photoinduced phase transitions in tetrathiafulvalene-$p$-chloranil (TTF-CA)---an organic charge-transfer complex---are one of the earliest demonstrations of optical manipulation of solid-state systems \cite{nasu2004, yonemitsu2008, dressel2017}. Near the equilibrium transition temperature, bidirectional transitions can be triggered by optical stimulation between the neutral phase with a centrosymmetric crystal structure and the ionic phase with a noncentrosymmetric structure \cite{koshihara1990, koshihara1999, collet2003}. Since the ionic phase of TTF-CA has large electronic polarization \cite{giovannetti2009, kobayashi2012}, such a conversion means optical switching of ferroelectricity, which may be used for new types of capacitors or sensors. The material in the ionic phase also shows a bulk photovoltaic effect (shift current generation) due to the noncentrosymmetric nature \cite{nakamura2017}.

Starting from the early studies using nanosecond lasers~\cite{koshihara1990, suzuki1999, tanimura2001}, the recent focus is shifted towards more rapid excitation of the system using femtosecond lasers. For example, ultrashort laser pulses are used to induce coherent electron-lattice dynamics~\cite{iwai2002, morimoto2017a, kinoshita2020} or to manipulate the polarization of photoinduced states~\cite{yonemitsu2004b, iwai2006a, miyamoto2013a}. Due to the strong correlation in TTF-CA and its slow energy dissipation to the environment, nonlinear dynamics that strongly depends on the initial states and pump pulses appears. Such ultrafast dynamics has been analyzed by various experiments using pump-probe spectroscopy~\cite{okamoto2004, tanimura2004, uemura2010, matsubara2011}. Thus, aside from the experimental progress, it is desirable to theoretically understand the microscopic dynamics at the initial stage of the photoinduced phase transitions and their relation to the time-resolved spectroscopy for the optical control of material properties. 

For this aim, in this paper, we investigate the photoinduced microscopic dynamics of TTF-CA by numerically studying the one-dimensional extended Peierls-Hubbard model. We focus on the initial nucleation dynamics within a small homogeneous domain (intra-domain) and ignore the subsequent slow domain-wall motion \cite{huai2000, luty2002, yonemitsu2004, yonemitsu2004a, soos2007, peterseim2015, kishida2009}. In other words, we consider the excitation of a charge-transfer (CT) string---the precursor for the (equilibrium) phase transitions \cite{lemee-cailleau1997, koshihara1999, suzuki1999, guerin2010}---of a fixed size without boundaries~\cite{cavatorta2015}. Such a situation may be relevant to the ultrafast dynamics within a few picoseconds. We demonstrate that femtosecond monocycle pulses can generate coherent CT strings leading to very efficient conversions between different phases, i.e., nonthermal switching of the electronic and lattice orders~\cite{mohr-vorobeva2011, porer2014, schuler2018}.

Using exact diagonalization, we first show that the model hosts a neutral phase with small ionicity (TTF$^{0}$CA$^{0}$) and negligible dimerization and an ionic phase with moderate ionicity (TTF$^{+1}$CA$^{-1}$) and moderate dimerization [see Figs.~\ref{fig:model}(b) and (c)]. While these two phases are experimentally relevant to TTF-CA, we find that a doubly ionized phase (TTF$^{+2}$CA$^{-2}$) with strong dimerization is also stable; we call this phase the ``dipole" phase [see Fig.~\ref{fig:model}(d)]. Aside from the various local order parameters, we characterize these phases by the Zak phase, which is related to electronic polarization \cite{zak1989, resta2002}. While the Zak phase in the neutral phase is zero, as predicted before, the other two phases are topologically distinct and their electronic polarization is nonzero. 

We then study the dynamics induced by ultrashort monocycle pulses from these ground-state phases. While the lattice motion is subject to damping, we find that the electronic part reaches quasi-steady states. The completeness of photoinduced phase transitions depends nonlinearly on the pump pulse's duration and strength. Within the parameters that we use, neutral--ionic, ionic--neutral, and dipole--ionic transitions are achieved. The temporal changes of order parameters present correlated dynamics between the electronic and lattice degrees of freedom. Furthermore, we calculate the time-resolved spectral density and the pump-probe conductivity to obtain a microscopic picture of the photoexcited states. After the photoexcitation, these quantities exhibit rapid spectroscopic changes that may be experimentally detected. We argue that the intermediate Franck-Condon states are multiphoton excited states that already have the properties of the final target phase. We also clarify the non-negligible role of the adiabatic ground states.

The remainder of our paper is organized as follows. Section \ref{sec:formalism} introduces the model that we consider and the employed method. In Sec.~\ref{sec:GS}, we discuss the ground-state properties of the model revealed by numerical diagonalization and by a simple ansatz. Linear optical conductivity is calculated for the ground states as well. Section~\ref{sec:dynamics} is devoted to the photoinduced dynamics of the model induced by short monocycle pulses. The steady states, the real-time dynamics, the time-dependent spectral density, and the pump-probe conductivity are discussed. In Sec.~\ref{sec:discussion}, we summarize the obtained picture of photoinduced microscopic dynamics and compare it to the previous studies. Section \ref{sec:conclusion} is the conclusion.

\section{Formalism}\label{sec:formalism}
\subsection{Model}\label{sec:model}

\begin{figure}[!tb]
\begin{center}
\includegraphics[width= \columnwidth]{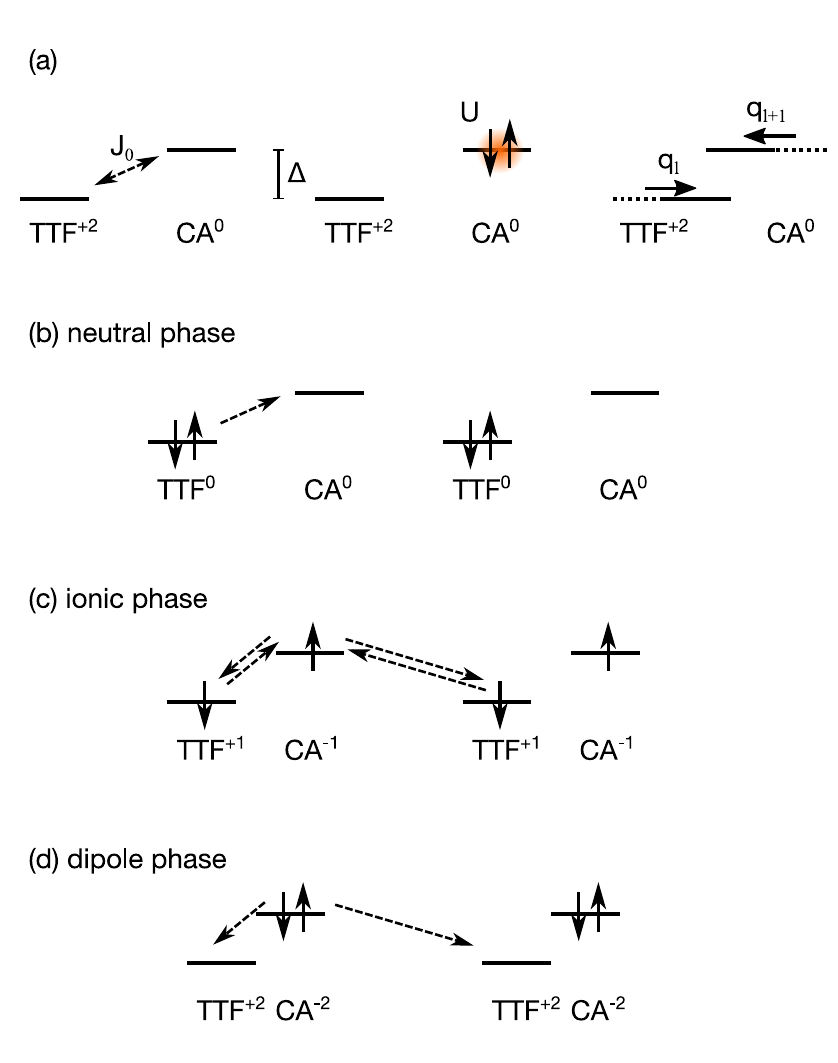}  
\caption{(a) Schematics of the model and (b)-(d) possible ground-state phases. The dashed arrows represent the dominant optical excitation processes.}
\label{fig:model}
\end{center}
\end{figure}
TTF-CA is a quasi-one-dimensional organic material that consists of chains of alternatively stacked TTF molecules and CA molecules [Fig.~\ref{fig:model}(a)]. TTF molecules are electron donors and CA electron acceptors. Above the transition temperature ($81$ K in equilibrium under ambient pressure), the material is in the neutral phase, with a small charge transfer, $\rho \approx 0.3$, from the donors to the acceptors without lattice modulation. Below the transition temperature, the material transforms into the ionic phase, where more electrons move to the acceptors, $\rho \approx 0.6$, and lattice dimerization occurs~\cite{torrance1981, torrance1981a}. Upon increasing pressure, an ionic phase without dimerization appears~\cite{lemee-cailleau1997}, while we will not consider such a phase in this work. 

Various models have been proposed to understand the underlying physics of the equilibrium and nonequilibrium phase transitions in charge-transfer materials \cite{soos1974}. A typical example is the ionic Hubbard model, which contains only the electronic degrees of freedom \cite{bruinsma1983, avignon1986}. On top of such a model, it is common to introduce lattice degrees of freedom to better characterize the interplay between the electrons and the phonons.

In this work, we consider the one-dimensional extended Peierls-Hubbard model, whose specific form reproduces the basic experimental properties of TTF-CA. The inter-chain interaction is ignored for simplicity. Such a model was proposed and solved by the Hartree-Fock method in Refs.~\cite{iizuka-sakano1996, huai2000}. Its photoinduced dynamics was studied by the time-dependent Hartree-Fock method in Refs.~\cite{miyashita2003, yonemitsu2004, yonemitsu2004a, yonemitsu2004b}. These studies focus on thermal and spatial fluctuations, while quantum fluctuations are not fully taken into account. Here, we solve the model by exact diagonalization, where quantum fluctuations arising from the competing phases are properly included. The method is also suitable to calculate various (time-dependent) spectral functions, as shown below.

There are other models that also describe the neutral--ionic phase transitions in organic mixed-stack compounds. A commonly used one includes the electron-phonon coupling on the electron's hopping energy as in the Su-Schrieffer-Heeger model~\cite{su1980, horovitz1987, luty1987, gomi2017, ohmura2019}. Compared to the model in this work, the broken centrosymmetry on the single-particle level presumably produces more complex behavior in electronic polarization~\cite{resta1995, resta1999} and, hence, in infrared spectra~\cite{kumar2010}. Here, considering that the kinetic energy is small compared to the interaction energy in the material, we will not include such a coupling. Effects of Holstein phonons, which couple to the local electron density, have been also studied in Refs.~\cite{painelli1988, caprara2000, soos2007, yonemitsu2011, cavatorta2015, davino2017}.

First, we take the ionic Hubbard model as the electronic part [see Fig.~\ref{fig:model}(a)] 
\begin{multline}
H_{\rm el}  = -\sum_{l, \sigma} J_0 \left[ e^{-i \frac{e}{\hbar} a A(t)}c^{\dagger}_{l+1, \sigma} c_{l \sigma} + \text{H.c.} \right] \\
+ U \sum_l n_{l \up}n_{l\dn} + \sum_l \frac{\Delta}{2} (-1)^l n_l,
\label{eq:H_el}
\end{multline}
where $J_0$ is the hopping amplitude, and $c_{l \sigma}$ and $c^{\dagger}_{l \sigma}$ annihilation and creation operators at site $l$ of spin $\sigma$. Here we take odd sites as the highest occupied molecular orbital of the donors (TTF) and even sites as the lowest unoccupied molecular orbital of the acceptors (CA). $\Delta$  is the staggered potential representing the difference of site energies (i.e., the redox energy), and $U$ the on-site Hubbard interaction. The density operators are given by $n_{l \sigma} = c^{\dagger}_{l \sigma} c_{l \sigma}$ and $n_{l} = n_{l \up} + n_{l \dn}$. We assume no net spin polarization and focus on half-filling, where the number of particles $N = \sum_l n_l$ is equal to the number of sites $L$. This model has been investigated to understand organic crystals or transition metal oxides \cite{nagaosa1986, egami1993, ishihara1994, maeshima2005a}. 

\begin{table}[!tb]
\caption{\label{tab:parameters} Parameters of the model in the unit of eV taken from Ref.~\cite{huai2000}.}
\begin{ruledtabular}
\begin{tabular}{cccccccc}
$J_0$ & $\Delta$ & $U$ & $s_1$ & $s_2$ & $V_0$ & $\beta_1$ & $\beta_2$ \\
\hline 
0.17 & 2.716 & 1.528 & 4.86 & 3400 & 0.604 & 1 & 8.54
\end{tabular}
\end{ruledtabular}
\end{table}
 
Optical fields are introduced by a spatially uniform time-dependent vector potential $A(t)$, entering the model via Peierls' substitution as Eq.~\eqref{eq:H_el}. This corresponds to the electric field parallel to the stacking axis ($a$ axis). In the remainder of the paper, we use $e = \hbar = c = 1$ and include the lattice constant $a$ in the vector potential. The electric field is given by $E(t) = -\partial_t A(t)$. Motivated by recent experiments using short monocycle pulses~\cite{morimoto2017}, we consider the following monocycle pump pulses for $t \in [t_{\rm pump}, t_{\rm pump} + T_{\rm pump}]$ [outside this range, $A(t) = 0$],
\begin{equation}
A_{\rm pump} (t, t_{\rm pump}) =  A_0 \left\{ \cos\left[ \omega_{\rm pump} (t-t_{\rm pump}) \right] - 1 \right\} ,
\label{eq:A_pump}
\end{equation}
where $A_0$ is the pump strength, $\omega_{\rm pump}$ the pump frequency, $t_{\rm pump}$ the pump starting time, and $T_{\rm pump} = 2\pi/\omega_{\rm pump}$ the width of the pump. As a probe pulse (if included in the simulations) \cite{shao2016}, we use a step function at time $t_{\rm probe}$,
\begin{equation}
A_{\rm probe} (t, t_{\rm probe}) = -E_0 \Theta(t - t_{\rm probe}),
\label{eq:probe}
\end{equation}
which gives an electric field of a delta function ${E_{\rm probe}(t, t_{\rm probe}) = E_0 \delta(t - t_{\rm probe}) }$. Both pump and probe fields are polarized parallel to the stacking axis. The current operator is given by
\begin{equation}
j = - \sum_{l, \sigma} \left[ i J_0 e^{-i A(t)} c^{\dagger}_{l+1, \sigma} c_{l \sigma} + {\rm H.c.} \right].
\label{eq:current}
\end{equation}

The phonon part is given by
\begin{equation}
H_\text{ph} = \sum_l \left[ \frac{P_l^2}{2 M_l} + \frac{S_1}{2} (Q_l - Q_{l+1})^2 + \frac{S_2}{4} (Q_l - Q_{l+1})^4 \right],
\end{equation}
where $M_l$ is the mass of the molecule at site $l$, $Q_l$ and $P_l$ the corresponding displacement and momentum. For the sake of simplicity, we use dimensionless displacements $q_l \equiv Q_l/a$ measured by the equilibrium lattice constant $a \approx 3.6$ \AA; other quantities are also renormalized as $m_l \equiv M_l a^2$, $p_l \equiv P_l a$, $s_1 \equiv S_1 a^2$, and $s_2 \equiv S_2 a^4$,
\begin{equation}
H_\text{ph} = \sum_l \left[ \frac{p_l^2}{2 m_l} + \frac{s_1}{2} (q_l - q_{l+1})^2 + \frac{s_2}{4} (q_l - q_{l+1})^4 \right].
\label{eq:H_ph}
\end{equation}
$s_1$ and $s_2$ represent harmonic and anharmonic potentials.

The displacement of molecules modifies the nearest-neighbor Coulomb interaction, which gives the electron-phonon coupling as 
\begin{equation}
H_\text{el-ph} = \sum_l V(q_l, q_{l+1}) \rho_l \rho_{l+1}.
\label{eq:H_el-ph}
\end{equation}
Here $\rho_l$ is the total charge density at site $l$,
\begin{equation}
\rho_l  = 
\begin{cases}
2- n_l & \text{at the donor sites (odd $l$)} \\
- n_l & \text{at the accepter sites (even $l$)}
\end{cases}.
\end{equation}
The coupling constant depends on the lattice displacement as
\begin{equation}
V(q_l, q_{l+1}) = V_0 + \beta_1 (q_l - q_{l+1}) + \beta_2 (q_l - q_{l+1})^2. 
\label{eq:Vq}
\end{equation}

The parameters that we use are summarized in Table \ref{tab:parameters}. The bare optical phonon frequency is given by $\omega_\text{opt} = \sqrt{2s_1(m_1 + m_2)/(m_1 m_2)} \approx 5.43$ meV, which corresponds to the period of $\approx 0.78$ ps. Once the electron-phonon coupling is included, the frequency of phonon oscillation estimated by a real-time simulations is reduced to $\approx 0.5$ ps, which is closer to the observed period $\approx 0.6$ ps \cite{iwai2006a}.

\subsection{Order parameters}
To characterize the phases, we consider the order parameters of the charge-density wave (CDW) and the spin-density wave (SDW),
\begin{equation}
\begin{split}
O_\text{CDW} = \frac{1}{L^2} \sum_{k,l} \braket{n_k n_{l} (-1)^{k-l}}, \\
O_\text{SDW} = \frac{1}{L^2} \sum_{k,l} \braket{S^z_k S^z_{l} (-1)^{k-l}},
\end{split}
\end{equation}
where $S^z_l = n_{l\up} - n_{l\dn}$ is the spin density operator at site $l$. They represent the spatial charge and spin density modulation with wave vector $\pi/a$. We also measure the degree of charge transfer (or ionicity) by
\begin{equation}
\rho = \frac{1}{L} \sum_{l} \Braket{\rho_l}(-1)^{l+1}.
\end{equation}
Similarly, dimerization of the lattice is quantified by 
\begin{equation}
O_\text{Peierls} = \frac{1}{L^2} \sum_{k,l} \braket{q_k q_l (-1)^{k-l}}.
\end{equation}

\subsection{Method}\label{sec:method}
We solve the model, $H = H_\text{el} + H_\text{ph} + H_\text{el-ph}$, by exact diagonalization with the Lanczos method \cite{lanczos1950}. The periodic boundary condition is employed. The system is time-evolved after initializing the wave function to be a ground state. The phonons are treated semiclassically, and their equation of motion follows the Ehrenfest dynamics,
\begin{equation}
\begin{split}
\frac{d q_l}{dt} &= \frac{p_l}{m_l}, \\
\frac{d p_l}{dt} &= - \Braket{\frac{\delta H(t)}{\delta q_l}}-\gamma \frac{d q_l}{dt},
\end{split}
\end{equation}
where $\langle \cdots \rangle$ is the expectation value in terms of the wave function at time $t$, the force from the electrons takes the form of the Hellmann-Feynman force, and a phenomenological damping constant $\gamma$ is also included. The positions of ions are time-dependent parameters in the electronic Hamiltonian, and the wave function of the electrons follows the time-dependent Schr\"odinger equation,
\begin{equation}
i \hbar \frac{d}{d t} \ket{\psi(t)} = \left[ H_\text{el} + H_\text{el-ph} \right] \ket{\psi(t)}. 
\end{equation}
We use the fourth-order Runge-Kutta method for the numerical integration of these equations. The time step is fixed to $dt = 0.002$; the relatively small time step here ensures that the norm of the wave function is preserved for the entire simulation time.

\section{Ground-state properties}\label{sec:GS}
\subsection{Potential energy surfaces and order parameters}\label{sec:PES}

\begin{figure}[!tb]
\begin{center}
\includegraphics[width= \columnwidth]{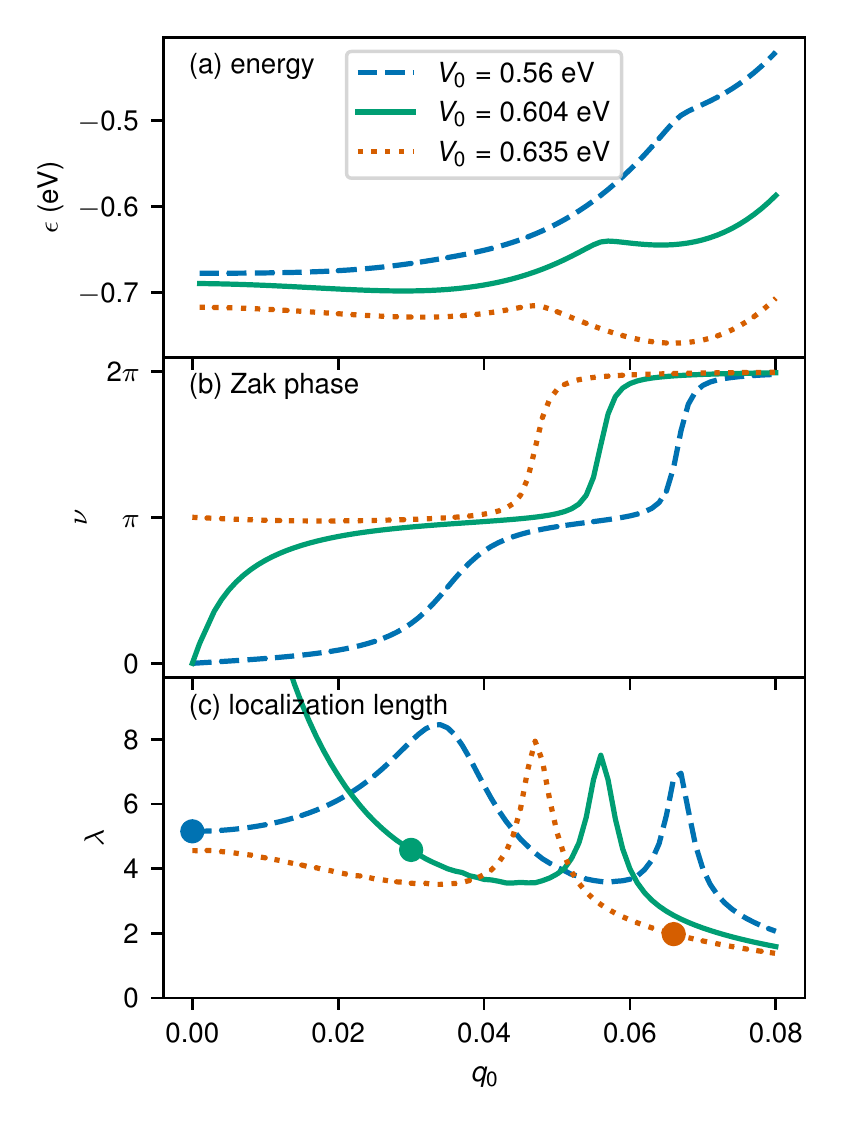}  
\caption{(a) Adiabatic potential energy surfaces, (b) the Zak phase $\nu$, and (c) the localization length $\lambda$ in terms of dimensionless dimerization $q_0$ at various $V_0$'s calculated by exact diagonalization of $L=12$ sites. The dots mark the points corresponding to the global minimum of each PES.}
\label{fig:GS_ED}
\end{center}
\end{figure}

Before we look at the photoinduced dynamics, we first elaborate on the ground-state properties of the model. While the two essential phases, i.e., the neutral (N) and ionic (I) phases, were found for the model previously, we show that the model possesses another possible state that we call the dipole (D) phase characterized by strong dimerization and large charge transfer. 

To obtain the adiabatic potential energy surfaces (PESs), we calculate the ground states of the model by exact diagonalization using the Lanczos method for a fixed amount of dimerization $q_0$, i.e., 
\begin{equation}
q_l = q_0  (-1)^l.
\label{eq:q_ansatz}
\end{equation}
The dimerization order parameter is given by $O_\text{Peierls} = q_0^2$ in this case. Figure~\ref{fig:GS_ED}(a) shows the PESs for three different values of $V_0$. Due to the insulating nature of these ground states, the finite-size effects are small. At $V_0= 0.56$ eV, the global minimum is located at $q_0 = 0$, corresponding to the neutral phase. For $q_0 \gtrsim 0.065$, the ground state has strong dimerization, which is the dipole phase, as we will discuss further below. The ionic phase with moderate dimerization, $q_0 \approx 0.03$, is not stable in this regime. At $V_0 = 0.604$ eV (i.e., the experimentally relevant case), the ionic phase obtains the global minimum at $q_0 \approx 0.03$, while the local minimum of the dipole phase is nearly degenerate to the global minimum. Finally, at $V_0 = 0.635$ eV, the dipole phase becomes the most stable phase. Therefore, as $V_0$ increases, the most stable phase changes from the neutral phase to the ionic one and then to the dipole one.

\begin{figure}[!tb]
\begin{center}
\includegraphics[width= \columnwidth]{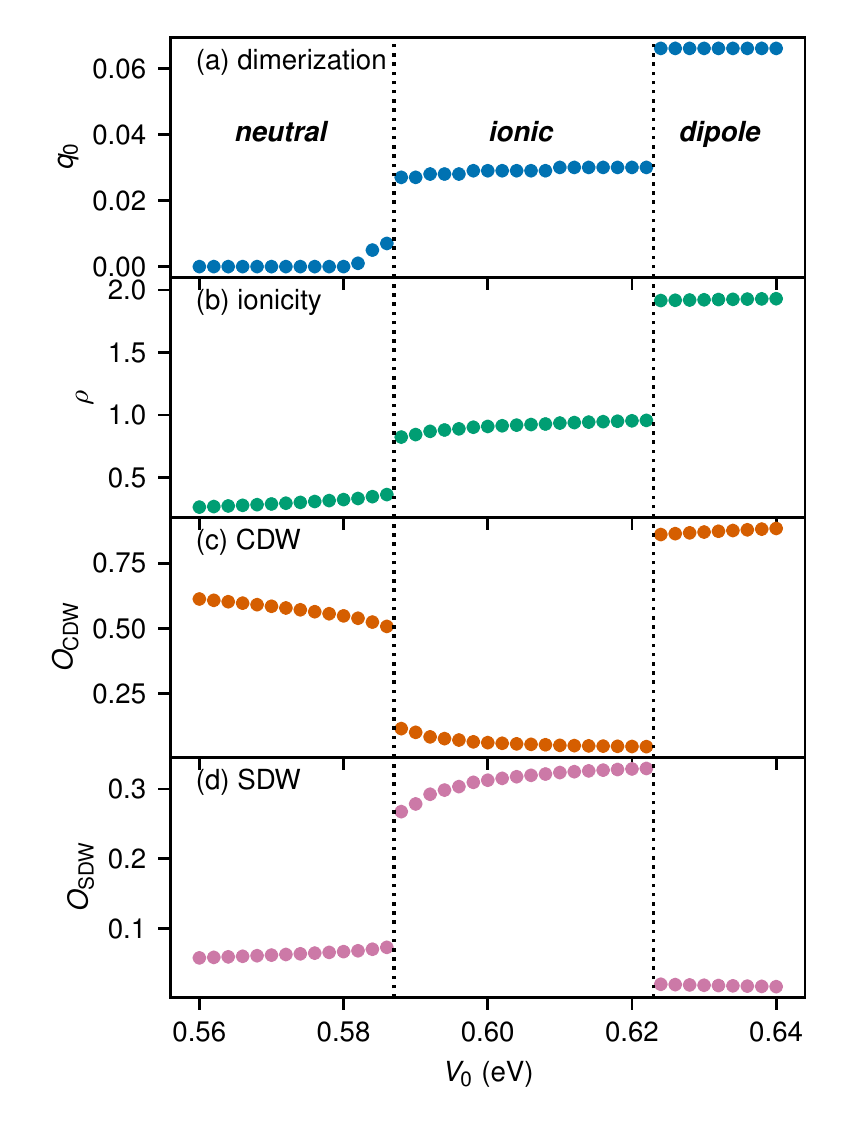}  
\caption{Order parameters of the ground state at the global minimum of each PES for various $V_0$: (a) dimerization $q_0$, (b) ionicity $\rho$, (c) CDW order, and (d) SDW order. For $V_0 < 0.587$ eV, the system is in the neutral phase. For $0.587$ eV $\leq V_0 < 0.623$ eV, the system is in the ionic phase. For $V_0 > 0.623$ eV, the dipole phase is the most stable. The dotted vertical lines indicate the phase boundaries.}
\label{fig:GS_order_parameter}
\end{center}
\end{figure}

We characterize the three possible phases by lattice dimerization, ionicity, and the electronic order parameters. Figure~\ref{fig:GS_order_parameter} plots these order parameters of the ground state (corresponding to the global minimum of the PES) as functions of the bare electron-phonon coupling $V_0$. For $V_0 < 0.587$ eV, dimerization is nearly absent, $q_0 \approx 0$, and the electrons are localized at the donor sites, $\rho \approx 0.3$, which gives a finite CDW order. On the other hand, the SDW order is absent. Thus, this phase corresponds to the neutral phase of TTF-CA. For $0.587$ eV $< V_0 <$ $0.623$ eV, moderate dimerization is induced, $q_0 \approx 0.03$, and the electron occupation is nearly equal on donor and acceptor sites leading to moderate ionicity, $\rho \approx 1.0$. While the CDW order is absent, the SDW is developed; the system is in the Mott insulating state with antiferromagnetic correlation. Therefore, this state corresponds to the ionic phase of TTF-CA. Finally, for $V_0 > 0.623$ eV, strong dimerization, $q_0 \approx 0.066$, occurs and the CDW order becomes finite again. However, the electrons are mostly localized on the acceptor sites, leading to large ionicity, $\rho \approx 2.0$. Such a dimer naively possesses a strong dipole moment, and we thus call this phase the dipole phase.

The order parameters change abruptly between these phases, which indicates the first-order nature of the phase transitions \cite{mitani1984}. In the ionic Hubbard model, $H_{\rm el}$, the transition between the neutral phase and the ionic phase occurs as $U$ increases, and the existence of a bond order between the two is suggested \cite{manmana2004}. However, we do not observe such a bond order in this work, probably due to the small system size.

The forms of the adiabatic potential energy surfaces are readily understood by considering the case of the vanishing hopping amplitude $J_0 =0$. Since the realistic value of $J_0 \approx 0.17$ eV is much smaller than the other energies ($U$, $V_0$, and $\Delta$), this limit gives a simple but reliable picture for the ground states. In the absence of hopping, any Fock state (in the real-space occupation basis) is an eigenstate of the Hamiltonian. Here we assume spin-dependent staggered occupation
\begin{equation}
\Braket{n_{l\sigma}} = \frac{1}{2} + (-1)^l \delta n_{\sigma}.
\end{equation}
This leads to $O_\text{CDW} = \left(\delta n_{\up} + \delta n_{\dn}\right)^2$ and $O_\text{SDW} = \left(\delta n_{\up} - \delta n_{\dn}\right)^2$.

The resultant energy density $\epsilon = \braket{H}/L$ is found to be
\begin{multline}
\epsilon(\delta n_{\up}, \delta n_{\dn}, q_0) = U \left(\frac{1}{4} + \delta n_{\up} \delta n_{\dn} \right) + \frac{\Delta}{2} \left(\delta n_{\up} + \delta n_{\dn}\right) \\
- \tilde{V}(q_0) \left(1 + \delta n_{\up} + \delta n_{\dn}\right)^2 + 2s_1 q_0^2 + 4 s_2 q_0^4,
\label{eq:epsilon}
\end{multline}
where $\tilde{V}(q_0) = V_0 + 4 \beta_2 q_0^2$ represents the effective nearest-neighbor Coulomb interaction modified by lattice dimerization. We see that the onsite interaction $U$ tends to suppress double occupation. The staggered potential $\Delta$ prefers large occupation on the donor (odd) sites, i.e., $\delta n_{\up, \dn} < 0$. On the other hand, the negative sign in front of $\tilde{V} (q_0)$ indicates that this interaction favors larger occupation on the acceptor (even) sites, i.e., $\delta n_{\up, \dn} > 0$. The competition between these three terms leads to the nearly degenerate phases, and quantum phase transitions among them. 

\begin{figure}[!tb]
\begin{center}
\includegraphics[width= \columnwidth]{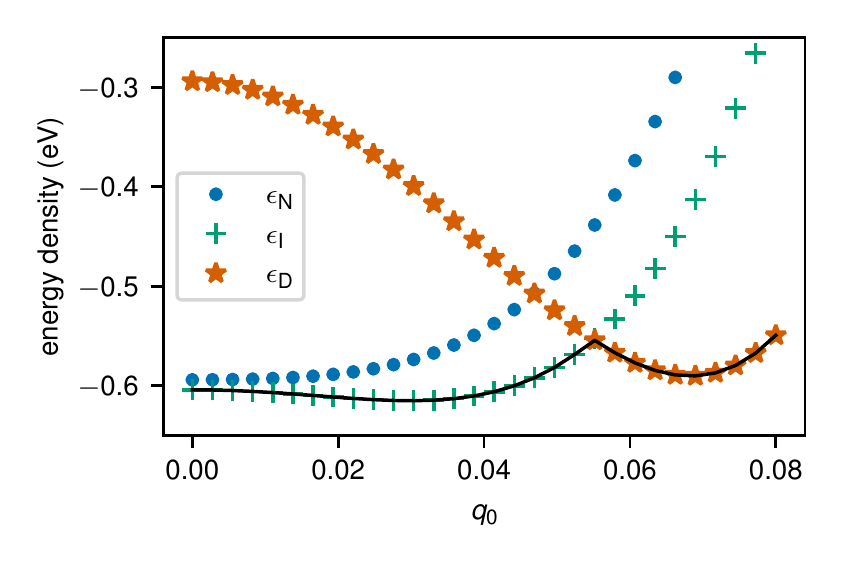}  
\caption{The energy densities in Eq.~\eqref{eq:energy_ansatz} obtained in the limit of $J_0 = 0$ at $V_0 = 0.604$ eV. The lowest energy among the three (black line) qualitatively reproduces the PES in Fig.~\ref{fig:GS_ED}(a).}
\label{fig:PES_ansatz}
\end{center}
\end{figure}

To find the ground state, we numerically minimize Eq.~\eqref{eq:epsilon} in terms of $\delta n_{\sigma} \in [-0.5, 0.5]$ at each $q_0$. The obtained diabatic PESs are plotted in Fig.~\ref{fig:PES_ansatz}, which qualitatively reproduces the results with $J_0 >0$. However, contrarily to the finite hopping case, the PESs with $J_0 = 0$ show sharp kinks between different phases. They are the energy-level crossings of the three energy eigenstates given by
\begin{enumerate}
\item $\delta n_{\up} = \delta n_{\dn} = -0.5$ \ (neutral states),
\item $\delta n_{\up} = -\delta n_{\dn} = \pm 0.5$ \ (ionic states) ,
\item $\delta n_{\up} = \delta n_{\dn} = 0.5$  \ (dipole states).
\end{enumerate}
The corresponding energies are 
\begin{equation}
\begin{split}
\epsilon_\text{N} &= \frac{U}{2} -  \frac{\Delta}{2} + 2 s_1 q_0^2 + 4 s_2 q_0^4, \\ 
\epsilon_\text{I} &= -\tilde{V}(q_0) + 2 s_1 q_0^2 + 4 s_2 q_0^4, \\ 
\epsilon_\text{D} &= \frac{U}{2} + \frac{\Delta}{2} - 4 \tilde{V}(q_0) + 2 s_1 q_0^2 + 4 s_2 q_0^4.
\end{split}
\label{eq:energy_ansatz}
\end{equation}
From these expressions, we see that the effective interaction $\tilde{V} (q_0)$ becomes more dominant as $V_0$ or $q_0$ increases and prefers the dipole state.

Finally, we note the direct relevance of the dipole phase to TTF-CA is yet unclear. For example, the additional next-nearest-neighbor interaction effectively enhances the onsite Hubbard interaction $U$ and the staggered potential $\Delta$. Therefore, such a term favors the neutral phase and hinders the dipole phase. Nevertheless, the precise degree of the suppression of the dipole phase depends on the actual parameters of materials, and it can still remain as a local energy minimum. Therefore, it may appear in other systems or as a metastable state during a photoinduced phase transition.

\subsection{Zak phase and localization length}
Aside from various local order parameters, electronic polarization is another important quantity to characterize the ground states. The modern formulation of macroscopic polarization developed since the early 1990s sheds new light on the concept of electronic polarization \cite{king-smith1993, resta1992}. Now, there is experimental and theoretical evidence that the polarization of the ionic phase in TTF-CA is mainly of the electronic origin, not the ionic origin, even though it is accompanied by lattice dimerization.

Here, we follow the approach by Resta and Sorella~\cite{resta1999}, which is shown to be also useful for strongly correlated systems~\cite{soos2004}. The quantity of interest is a complex number $z_N$ defined as
\begin{equation}
z_N \equiv \Braket{\Psi |e^{-i\frac{2 \pi}{L} \hat{X}} | \Psi},
\end{equation}
where the expectation value is taken for a ground state $\ket{\Psi}$. $\hat{X} = \sum_l l n_l$ is the position operator in the open boundary condition, and the exponential operator is a proper  position operator for the periodic boundary condition. The phase of $z_N$ is related to the Zak phase \cite{zak1989, ortiz1994, resta1995, resta1998, resta1999, souza2000, resta2002}
\begin{equation}
\nu = i\int dk \braket{\Phi_k^0 | \partial_k \Phi^0_k} \simeq \Im \ln z_N,
\end{equation}
where $\ket{\Phi^0_k}$ is the ground state of a Hamiltonian with an additional phase on the hopping amplitude $J_0 \rightarrow J_0 e^{i k/L}$. An intriguing topological property given by the Zak phase in the ionic Hubbard model was revealed by Ref.~\cite{resta1995}, and experimentally realized by using ultracold atoms~\cite{atala2013}. The macroscopic polarization is given by the Zak phase as~\cite{resta2002}
\begin{equation}
P = \frac{e \nu}{2\pi}.
\label{eq:polarization}
\end{equation}
The amplitude of $z_N$ is related to the localization length $\lambda$ of the ground state wave function as (at half-filling),
\begin{equation}
\lambda = \sqrt{-N \ln |z_N|^2}/(2\pi).
\end{equation}
We calculate $z_N$ using the periodic ansatz from Ref.~\cite{resta1998} to circumvent the severe finite-size effects. We take the number of the supercells $M = 100$. 

Figure~\ref{fig:GS_ED}(b) shows the Zak phase $\nu$ corresponding to the three PESs in Fig.~\ref{fig:GS_ED}(a). At $V_0 = 0.560$ eV, we see that the neutral phase at small $q_0$ has $\nu = 0$ and thus is topologically trivial. However, the ionic phase at large $q_0$ obtains an additional phase $\nu = \pi$ and thus is topologically nontrivial. This result is consistent with the previous result for the pure ionic Hubbard model \cite{resta1995}. By increasing $q_0$ further, we reach the dipole phase, which has $\nu = 2\pi$, i.e., another topologically distinct state. Therefore, the Zak phase clearly distinguishes the three quantum phases. In particular, the transition between the neutral and the ionic phases is more evident in the Zak phase compared with the PESs in Fig.~\ref{fig:GS_ED}(a). Furthermore, the relation in Eq.~\eqref{eq:polarization} suggests that the macroscopic polarization is nonzero for the ionic and dipole phases (bigger for the latter), while the neutral phase has no polarization; this agrees with the paraelectric/ferroelectric nature of the neutral/ionic phase in TTF-CA. At $V_0= 0.604$ eV, the neutral  and ionic states are nearly degenerate, and $\nu$ increases slowly from $0$ to $\pi$ at small $q_0 \lesssim 0.02$. However, as $q_0$ further increases, the Zak phase becomes $2\pi$ (the dipole phase). Finally, at $V_0 = 0.635$ eV, we only observe the ionic phase ($\nu = \pi$) and the dipole phase ($\nu = 2\pi$). 

Figure~\ref{fig:GS_ED}(c) shows the localization length $\lambda$ calculated from $z_N$. As discussed in Sec.~\ref{sec:PES}, the transitions among different phases are caused by level crossings or gap closing. Such gap closing signals metallic behavior \cite{resta1998}, and correspondingly the localization length diverges. For the three values of $V_0$'s, the calculated $\lambda$'s are indeed enhanced at the phase boundaries. The data points corresponding to the global minimum of the PES for each $V_0$ are plotted as the dots. We find that the localization length is the largest in the neutral state at $V_0 = 0.56$ eV, and the smallest in the dipole phase at $V_0 = 0.635$ eV. This order agrees with the fact that the neutral (dipole) phase has the smallest (largest) optical charge gap, as shown in Sec.\ref{sec:sigma_GS}.

\subsection{Linear optical conductivity}\label{sec:sigma_GS}

\begin{figure}[!tb]
\begin{center}
\includegraphics[width= \columnwidth]{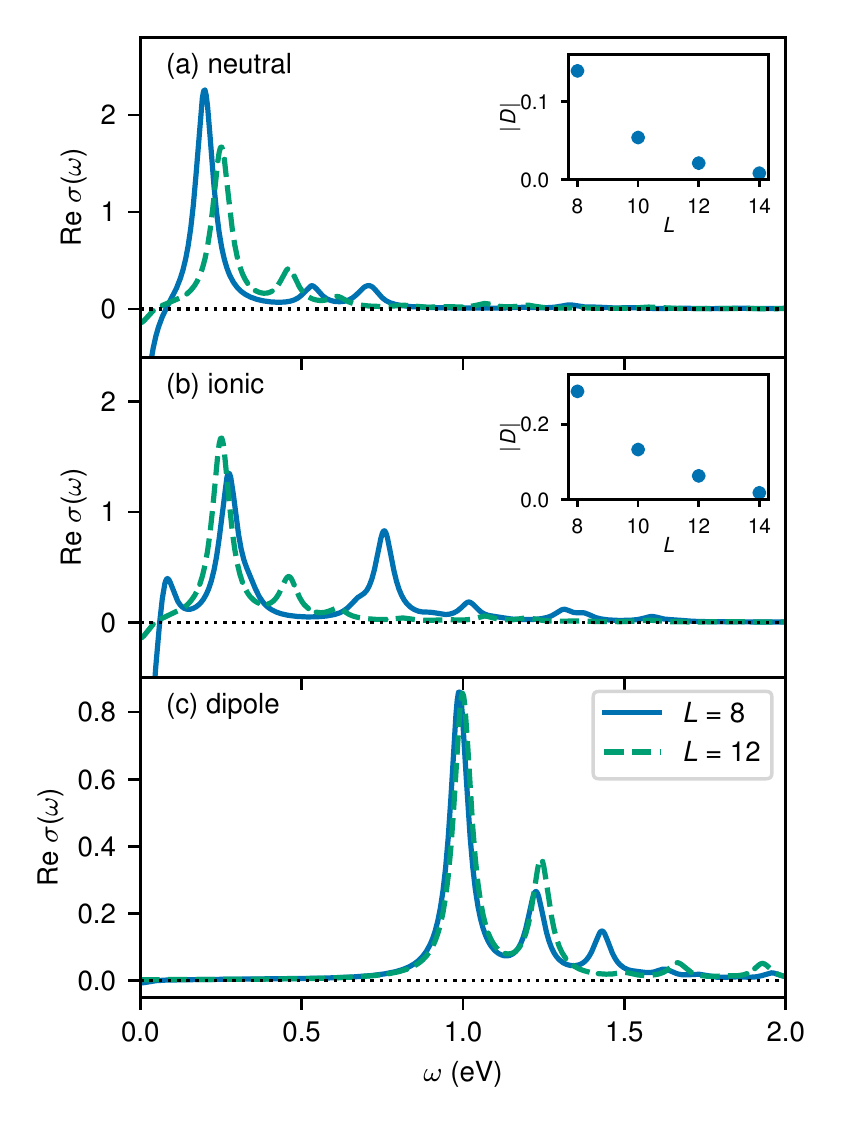}  
\caption{Real part of linear optical conductivity $\Re [\sigma(\omega)]$ for (a) a neutral state ($V_0 = 0.56$ eV), (b) an ionic state ($V_0 = 0.604$ eV), and (c) a dipole state ($V_0 = 0.635$ eV). Solid and dashed lines are for $L=8$ and $12$. The insets in (a) and (b) show the system-size dependence of the Drude weight.}
\label{fig:sigma_GS}
\end{center}
\end{figure}

In the next section, we calculate the pump-probe conductivity of the system after photoexcitation. To understand the spectral changes in this quantity, here, we present the linear optical conductivity of the three ground-state phases. The real part of the optical conductivity $\sigma(\omega)$ for an energy eigenstate $\ket{\rm i}$ with eigenenergy $E_{\rm i}$ can be written as \cite{dagotto1994, maeshima2005a, takahashi2008}
\begin{equation}
\begin{split}
\Re \left[ \sigma(\omega ) \right] &= D\delta(\omega) +  \sigma^{\rm reg}(\omega),\\
D &= - \frac{\pi}{L} \braket{K}_{\rm i}  - \frac{2\pi}{L}  \sum_{n\neq {\rm i}} \frac{|\braket{{\rm i}|j_0| n}|^2}{E_n - E_{\rm i} },\\
\sigma^{\rm reg}(\omega) &= \frac{\pi}{L} \sum_{n\neq {\rm i}} \frac{|\braket{{\rm i}| j_0 | n}|^2}{E_n - E_{\rm i} } \\
&\times \left[ \delta(E_n - E_{\rm i} - \omega) + \delta(E_n - E_{\rm i} + \omega) \right], \\
\end{split}
\label{eq:sigma}
\end{equation}
where $\braket{K}_{\rm i}$ is the expectation value of the kinetic energy in terms of $\ket{\rm i}$, $\ket{n}$ the $n$th eigenstate with eigenenergy $E_n$, and $j_0$ the current operator without the vector potential in Eq.~\eqref{eq:current}. $D$ is the Drude weight and $\sigma^{\rm reg} (\omega)$ is related to the absorption of a photon at $\omega$. When $\ket{\rm i}$ is equal to the ground state $\ket{0}$, the second term in $\sigma^{\rm reg}(\omega)$ vanishes for $\omega > 0$, and the expression is reduced to the conventional one. We will use this general expression to understand the pump-probe optical conductivity in Sec.~\ref{sec:sigma_pp}. The excited states are calculated based on the Franck-Condon principle; the lattice displacement is fixed to the initial-state configuration. In Fig.~\ref{fig:sigma_GS}, we plot the optical conductivity of the three ground-state phases for $L=8$ and $12$.

In the neutral phase, there is a dominant peak around $\omega \approx 0.2$ eV, and this corresponds to the transfer of an electron from a donor to a neighboring acceptor [see Fig.~\ref{fig:model}(b)]. In the case of $J_0 = 0$ eV, the excitation energy is given by
\begin{equation}
\omega^{\text{N}}_1 = \Delta - U - V_{12}^{\rm N} \approx 0.63 \text{ eV},
\end{equation}
which is close to the experimental value $0.6$ eV \cite{jacobsen1983}. Here $V_{12}^{\rm N}$ is the electron-phonon coupling in Eq.~\eqref{eq:Vq} between sites $1$ and $2$ with the equilibrium value of dimerization $q_0^{\rm N}$ for the neutral phase. The peak frequency of such excitation shifts to lower values as $J_0$ increases from $0$ eV. The negative values at $\omega = 0$ are the finite-size effect \cite{fye1991, maeshima2005a}; as shown in the inset, the Drude weight becomes exponentially small as the system size increases (for $L=10$ and $14$, we use the anti-periodic boundary condition). Therefore, the system is insulating.

The ionic phase has a more complex structure as depicted in Fig.~\ref{fig:sigma_GS}(b). While the system is also an insulating state (the inset shows the vanishing Drude weight as $L$ increases), we observe a few dominant peaks. They originate from the excitation paths shown in Fig.~\ref{fig:model}(c). For an ionic ground state, an electron can hop between a donor and an acceptor in both ways. Furthermore, the electron transfer can occur between a dimerized donor-acceptor pair or between a donor and an acceptor in neighboring dimers. In total, there are four possible excitation paths. The excitation energies for $J_0 = 0$ eV are 
\begin{equation}
\begin{split}
\omega^{\text{I}}_1 &= -\Delta + U + 2V_{12}^{\rm I} + V_{23}^{\rm I} \approx 0.65 \text{ eV}, \\
\omega^{\text{I}}_2 &= -\Delta + U + V_{12}^{\rm I} + 2V_{23}^{\rm I} \approx 0.77 \text{ eV}, \\
\omega^{\text{I}}_3 &= \Delta + U - 2V_{12}^{\rm I} - 3V_{23}^{\rm I} \approx 1.14 \text{ eV}, \\
\omega^{\text{I}}_4 &= \Delta + U - 3V_{12}^{\rm I} - 2V_{23}^{\rm I} \approx 1.02 \text{ eV}.
\end{split}
\end{equation}
Once $J_0$ takes the experimentally relevant value, {$0.17$ eV}, the peak locations are modified, while the four-peak structure is still visible. The change from the one-peak structure in the neutral phase to a two-peak structure in the ionic phase has been experimentally observed~\cite{tokura1982, jacobsen1983} and also theoretically predicted~\cite{huai2000}~using electric fields along the stacking axis. Since $\omega^{\text{I}}_{1,2}$ and $\omega^{\text{I}}_{3,4}$ are close, these may give the two-peak structure in the experiment. We note that, for $L=8$, weak absorption at very small energy $\omega \approx 0.1$ eV is possible, while it disappears for larger $L$'s.

Finally, for the dipole phase, absorption at low frequencies is suppressed, and the dominant peaks shift to larger values $\omega \gtrsim 1.0$ eV. Again, in the limit of $J_0 = 0$ eV, the possible excitation paths are moving an electron from an acceptor to a donor either close or away [Fig.~\ref{fig:model}(d)]:
\begin{equation}
\begin{split}
\omega^{\text{D}}_1 &= -\Delta - U + 4V_{12}^{\rm D} + 3V_{23}^{\rm D} \approx 1.11\text{ eV}, \\
\omega^{\text{D}}_2 &= -\Delta - U + 3V_{12}^{\rm D} + 4V_{23}^{\rm D} \approx 1.38\text{ eV}.
\end{split}
\end{equation}
Due to the strongly insulating nature of the phase, including a small value of $J_0 = 0.17$ eV does not shift the peak locations from the above values.

\section{Photoinduced dynamics}\label{sec:dynamics}
\subsection{Steady states}

\begin{figure*}[!htb]
\begin{center}
\includegraphics[width= \textwidth]{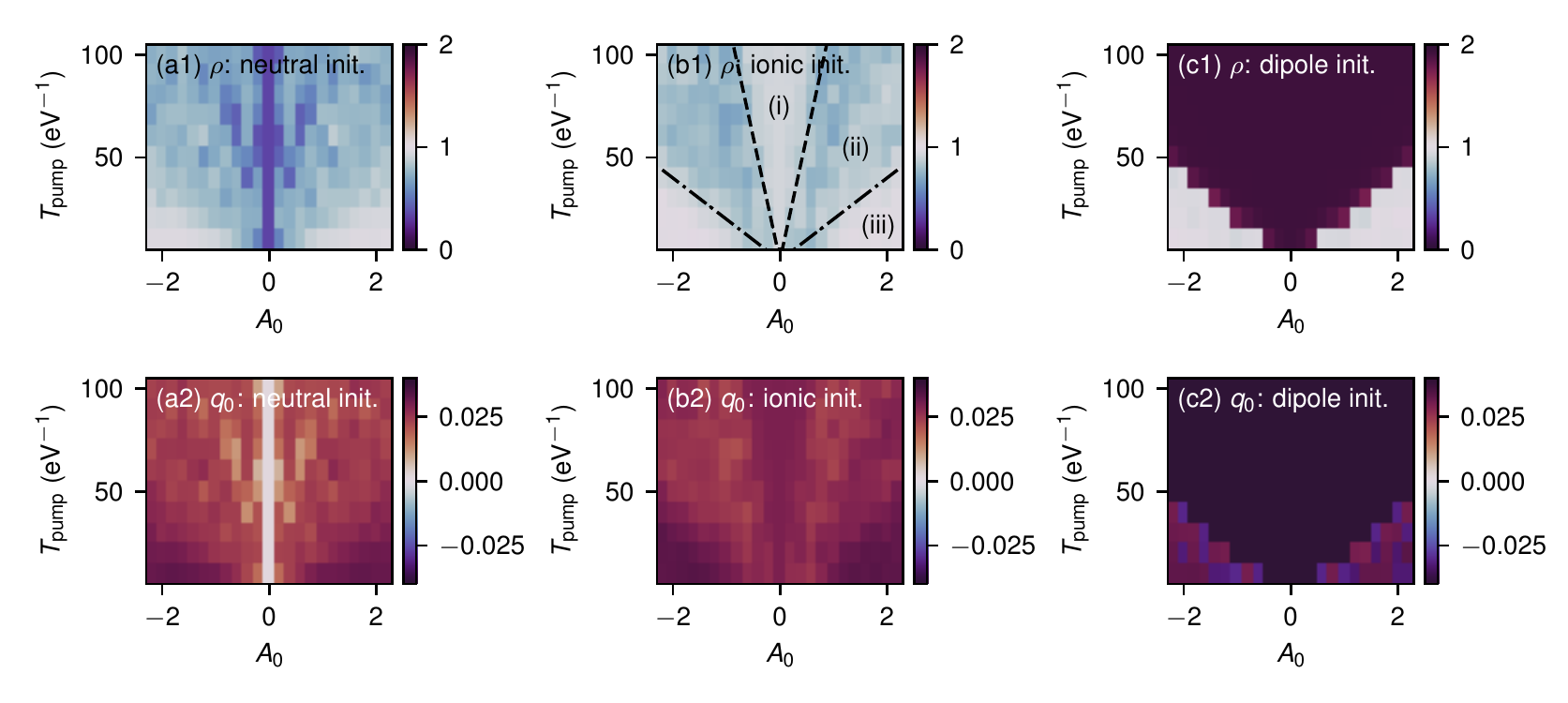}  
\caption{The time-averaged ionicity $\rho$ (upper panels) and dimerization $q_0$ for systems initialized as (a) a neutral state ($V_0 = 0.5825$ eV), (b) an ionic state ($V_0 = 0.604$ eV), and (c) a dipole state ($V_0 = 0.6275$ eV). Three distinct regimes in panel (b1) correspond to (i) the initial state, (ii) the photoinduced metastable state, and (iii) the infinite-temperature state. Lines are fitted by Eq.~\eqref{eq:heating_line}.}
\label{fig:2d_main}
\end{center}
\end{figure*}

We now turn to the dynamics of the model induced by monocycle optical pulses. We focus on the case with $L=8$ and the damping constant $\gamma = 600$. The initial states are the neutral, ionic, and dipole ground states at $V_0=0.5825$, $0.604$, and $0.6275$ eV, respectively. The initial lattice dimerization is located at the global minimum of each PES, i.e., $q_0 = 0.00108$, $0.02916$, and $0.0657$, respectively. The velocities of the ions are set to zero. The pump is induced around $t = 500$ eV$^{-1}$ and the simulation time is $t_\text{f} = 6000$ eV$^{-1}$. Due to the periodic boundary condition and spatially uniform optical excitation, the system remains in a translationally symmetric state; all the TTF sites are equivalent, and so are the CA sites. Therefore, the lattice displacement can be written in the form of Eq.~\eqref{eq:q_ansatz} after removing the degree of freedom corresponding to the translation of the whole system. 

First, Fig.~\ref{fig:2d_main} presents the average ionicity $\rho$ and dimerization $q_0$ after photoexcitation of different pump duration $T_{\rm pump}$ and amplitude $A_0$. The time average is taken for $t \in [t_{\rm f}/2, t_{\rm f}]$. As we will see below, the electronic observables almost immediately reach quasi-steady values after the optical excitation. Similarly, while the optical phonon mode oscillates, it approaches a long quasi-steady value due to the weak damping. Thus, the time-averaged values well represent the quasi-steady state after the pump pulse. This observation is also reinforced by the fact that the electronic observable $\rho$ (upper panels) and dimerization $q_0$ (lower panels) have the same dependence on $A_0$ and $T_{\rm pump}$ for each $V_0$. 

We see that the pump-pulse dependence of the quasi-steady states is similar regardless of the initializations. For a pulse with a small amplitude or long duration [case (i)], the system remains in the initial state. As the pulse becomes stronger or shorter [case (ii)], the system shows a photoinduced transition to a metastable state. As the pump pulse becomes too strong [case (iii)], the system is heated up to be the infinite-temperature state~\cite{okamoto2021}; there is no electronic order, and the double-occupation density $\sum_l \Braket{n_{l\up} n_{l\dn}}/L$ approaches to $1/4$. These three regimes (i)--(iii) are labeled in Fig.~\ref{fig:2d_main}(b1). The steady states depend sensibly on the pump parameters for the N--I transition, giving an intricate structure within regimes (i) and (ii).

The boundaries between the three regimes can be understood by the energy of the pump pulse, which is about $A_0^2 \omega_{\rm pump}^{2}/2$. If we assume that the transitions to metastable states occur when the amount of energy exceeds some critical energy scale $\epsilon_{\rm c}$, the phase boundaries are determined by
\begin{equation}
T_{\rm pump} \sim 2\pi |A_0|/\sqrt{2 \epsilon_{\rm c}}. 
\label{eq:heating_line}
\end{equation}
In Fig.~\ref{fig:2d_main}, we observe that the boundaries of the three regimes indeed follow a linear relation [highlighted by the lines in panel (b1)]. Such threshold behavior has been observed experimentally \cite{koshihara1999, suzuki1999, iwai2002}. We can associate the threshold energy scale $\epsilon_{\rm c}$ to the energy to create a sizable CT string while paying the lattice potential energy---a manifestation of the cooperative (or competitive) nature of strongly correlated systems \cite{miyashita2003, yonemitsu2004}. Finally, regardless of $V_0$, the transitions to the infinite-temperature state [case (iii)] occur around $T_{\rm pump} \sim 20 |A_0|$, or $\epsilon_{\rm c} \approx 0.05$.

Now let us focus on the intermediate regime [case (ii)]. At $V_0 = 0.5825$ eV, ionicity increases, $\rho \sim 0.5$, and dimerization becomes nonzero, $q_0 \sim 0.025$, which indicates a transformation of the system into an ionic state. At $V_0 = 0.604$ eV, ionicity changes from $\sim 1$ to a slightly smaller value $\sim 0.7$ and dimerization also diminishes. Thus, the photoinduced metastable state is partially neutral. Finally, at $V_0 = 0.6275$ eV, metastable ionic states appear in a very narrow range just before the infinite-temperature regime. This is because the energy barrier to the ionic state is larger than the one at the ionic--neutral phase boundary [Fig.~\ref{fig:GS_ED}(a)]. These metastable states in case (ii) can be considered the precursor CT strings that further grow to achieve bulk transitions~\cite{lemee-cailleau1997, koshihara1999, suzuki1999, guerin2010}.

In principle, two pulses with opposite signs, $\pm A_0$, applied to an ionic state or a dipole state can give different dynamics since these two phases break the inversion symmetry. However, the actual dynamics does not depend much on the sign of $A_0$. This is because the monocycle optical pulse is not directional in average, i.e., ${\int E_{\rm pump}(t) dt = 0}$. Instead, we have confirmed that directional pulses induce distinct dynamics for different signs of $A_0$.

\subsection{Real-time dynamics}
\begin{figure*}[!tb]
\begin{center}
\includegraphics[width= \textwidth]{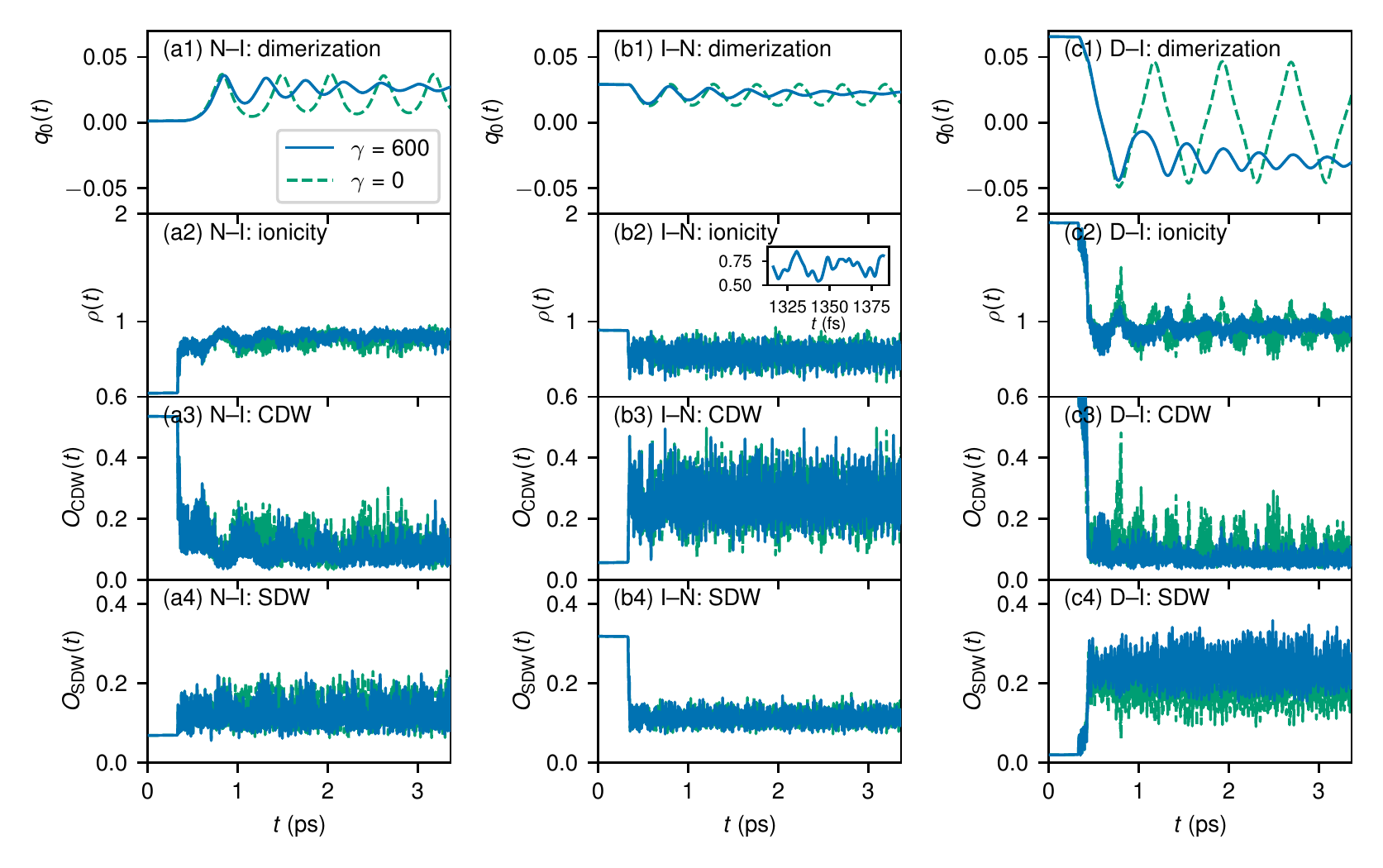}  
\caption{The real-time dynamics of order parameters (from top to bottom): (1) dimerization $q_0$, (2) ionicity $\rho$, (3) CDW order $O_{\rm CDW}$, and (4) SDW order $O_{\rm SDW}$. Three columns are for (a) N--I transition, (b) I--N transition, and (c) D--I transition. The inset in panel (a2) shows a femtosecond dynamics of $\rho(t)$.}
\label{fig:obs_dynamics}
\end{center}
\end{figure*}

\subsubsection{Order parameters}
To further characterize the photoinduced dynamics, we look at the real-time dynamics of several cases where clear photoinduced phase transitions are observed. Figure~\ref{fig:obs_dynamics} shows the real-time dynamics of several observables for (a) an N--I transition, (b) an I--N transition, and (c) a D--I transition. The parameters for the three cases are $(V_0, T_{\rm pump}, A_0) = (0.5825, 60, 1.2)$, $(0.604, 60, 1.0)$, and $(0.6275, 10, 0.6)$, respectively. For the comparison with experiments, we convert the time units to ps via 1 eV$^{-1} \approx 0.658$ fs.

In all three cases, the electronic order parameters, $\rho$ and $O_{\rm CDW/SDW}$, rapidly change after the optical excitation in a few femtoseconds and retain quasi-steady values afterward. In contrast, the lattice displacement $q_0$ shows slow oscillations in addition to gradual decay in a few picoseconds due to the finite damping. While the oscillation periods are similar to that of the bare optical phonon mode ($\approx 0.6$ ps), they sensibly depend on the modified ionicity, which renormalizes the optical phonon frequency through the electron-phonon coupling in Eq.~\eqref{eq:H_el-ph}. Experimentally, pump-dependent coherent oscillations were demonstrated in Ref.~\cite{iwai2006a}. Detailed studies of the coherent oscillations including Holstein phonons can be found in Refs.~\cite{yonemitsu2011, cavatorta2015}.

At $V_0 = 0.5825$ eV in Fig.~\ref{fig:obs_dynamics}(a), the transformation to the ionic phase is nearly complete in terms of dimerization, ionicity, and the CDW order. However, the SDW order parameter after the photoexcitation, $\approx 0.14$ is still below that of the equilibrium ionic phase, $\approx 0.3$. The D--I transition at $V_0 = 0.6275$ eV also leads to a nearly complete ionic state and the metastable SDW order, in this case, is close to that of the equilibrium ionic phase [Fig.~\ref{fig:obs_dynamics}(c4)]. As shown below, it is generally difficult to find an excited state that breaks the spin SU(2) symmetry, while the ground states can evolve adiabatically to an ionic state. Therefore, the difference between the two cases is due to the different ratios of the weights on the excited and ground states at the new lattice positions.

In contrast to the transition to an ionic state, the I--N transition is a partial transformation to a neutral state. Dimerization $q_0$ and ionicity $\rho$ are slightly reduced from their initial values but still away from the equilibrium values of the neutral phase. The CDW/SDW order is enhanced/suppressed only moderately. We consider that the difficulty to achieve a complete transformation to a neutral state stems from the instability at $q_0 = 0$ point [see Fig.~\ref{fig:GS_ED}(a)]. Since the PES has a local maximum at $q_0 = 0$, lowering $q_0$ is energetically hindered; only a slight reduction of $q_0$ occurs. On the other hand, in principle, we can imagine transforming an ionic state to the dipole phase, which has a local minimum in the PES. However, we do not observe such a transformation for the used parameters, probably due to the relatively high-energy barrier between the ionic and dipole phases.

\begin{figure}[!tb]
\begin{center}
\includegraphics[width= \columnwidth]{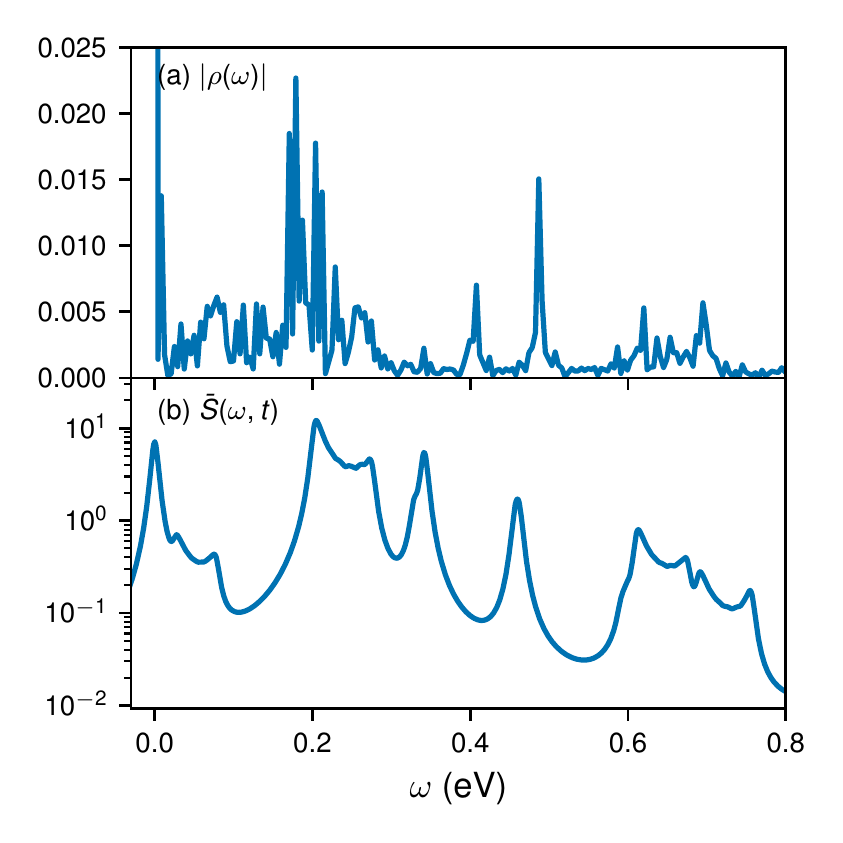}  
\caption{(a) Fourier spectrum of $\rho(t)$ for the I--N transition. (b) Averaged spectral density after the pump in the logarithmic scale. The delta function in Eq.~\eqref{eq:Sw} is approximated by the Lorentzian function with a small broadening factor $\eta = 0.05$.}
\label{fig:Fourier}
\end{center}
\end{figure}

The electronic observables show two types of oscillations: slow oscillations with picosecond frequencies and fast ones with femtosecond frequencies [see the inset of Fig.~\ref{fig:obs_dynamics}(b2)]. The slow oscillations come from the lattice displacement as indicated by the synchronized dynamics of $q_0(t)$ and $\rho(t)$. The most evident example is the D--I transition [Fig.~\ref{fig:obs_dynamics}(c)]. The origin of the fast oscillations can be understood by looking at the electronic spectral density
\begin{equation}
S(\omega; t) \equiv \sum_{n=0} \left| \Braket{\psi(t)| n(t)}\right|^2 \delta(\omega - E_n(t) + E_0 (t)), 
\label{eq:Sw}
\end{equation}
where $\ket{n(t)}$ is the $n$th energy eigenstate of an instantaneous Hamiltonian with $q_0(t)$ with eigenenergy $E_n(t)$. Figure~\ref{fig:Fourier} shows the Fourier spectrum of $\rho(t)$ and the time-averaged spectral density $\bar{S}(\omega; t)$ after the pump pulse. We see that the peaks in $\rho(\omega)$ agree with those of the spectral density. As we will see in more detail below, the slow lattice dynamics does not change the weights of the spectral density, $\left| \Braket{\psi(t)| n(t)}\right|^2$, but only shifts the peak positions slightly, $E_n(t)$; adiabaticity holds here. Therefore, we can consider the system to be in a quasi-stationary state, 
\begin{equation}
\begin{split}
\ket{\psi(t)} &= \sum_{n=0} \tilde{c}_n (t) \ket{n(t)}\\
&\simeq \sum_{n=0}c_n e^{- i E_n(t) t}  \ket{n(t)},
\end{split}
\label{eq:spectral_ansatz}
\end{equation}
where $\tilde{c}_n(t)$ is the time-dependent amplitude of the $n$th eigenstate, and $c_n$ is the corresponding approximate constant amplitude. The expectation value of an observable $\hat{O}$ can be approximated as \cite{okamoto2019}
\begin{equation}
\Braket{\hat{O}} \simeq \sum_{n, n'=0 } c_n c_{n'} e^{i [ E_n(t)-E_{n'}(t) ] t} \Braket{n(t)|\hat{O}|n'(t)}.
\end{equation}
The temporal fluctuations from the matrix element are presumably very slow, following the dynamics of the lattice. Instead, the phase factors induce interference between dominant eigenstates, giving rise to the fast oscillations in the electronic observables.

\subsubsection{Effect of damping}

As shown in Fig.~\ref{fig:obs_dynamics}(c1), the role of damping is not negligible for the lattice dynamics. In the absence of damping, lattice dimerization shows large persistent oscillations between $-0.05$ to $0.05$ when a dipole state is excited. This is due to the degeneracy of dimerized states in terms of the inversion symmetry ($q_0 \leftrightarrow -q_0$); the system oscillates between the different broken symmetry sectors. Such degeneracy can also induce intriguing phenomena such as soliton formation~\cite{sunami2018}. For a few parameter sets, we find similar behavior for the N--I transition where the photoinduced dimerization switches between the two possible ionic metastable states. However, we do not observe any switching of dimerization by monocycle pulses for the I-- N transition. In contrast, as is shown in Ref.~\cite{ohmura2019}, half-cycle pulses can induce switching behavior for both the I--N  and N--I transitions.

While the switching behavior seems an attractive option to control electronic polarization, as the lattice damping is included ($\gamma = 600$), it disappears after a few flips. For example, in Fig.~\ref{fig:obs_dynamics}(c1), dimerization first changes its signs, but then it quickly reaches a quasi-steady value $q_0 \approx -0.03$ (corresponding to the ionic phase). Therefore, we expect the switching to disappear completely in the overdamped regime, and the system remains in the initial broken-symmetry sector. Both degenerate states are reachable for weaker damping, while precise control to arbitrarily select a final state seems difficult.

We also note that the observables do not show a clear tendency of relaxation to the initial states within the timescale of our simulations; the system is trapped in a quasi-steady state. Since the model lacks thermal relaxation, the excited states cannot transit to lower energy eigenstates. Furthermore, thermal fluctuations are also absent in our simulations, which are required for the system to overcome an energy barrier between metastable states. Instead, the phenomenological damping in the phonon motion leads to the decay of the phonon oscillations in a few picoseconds, which agrees with the experimental observations \cite{iwai2006a}.

\subsubsection{Energy transfer}

\begin{figure*}[!tb]
    \begin{center}
    \includegraphics[width= \textwidth]{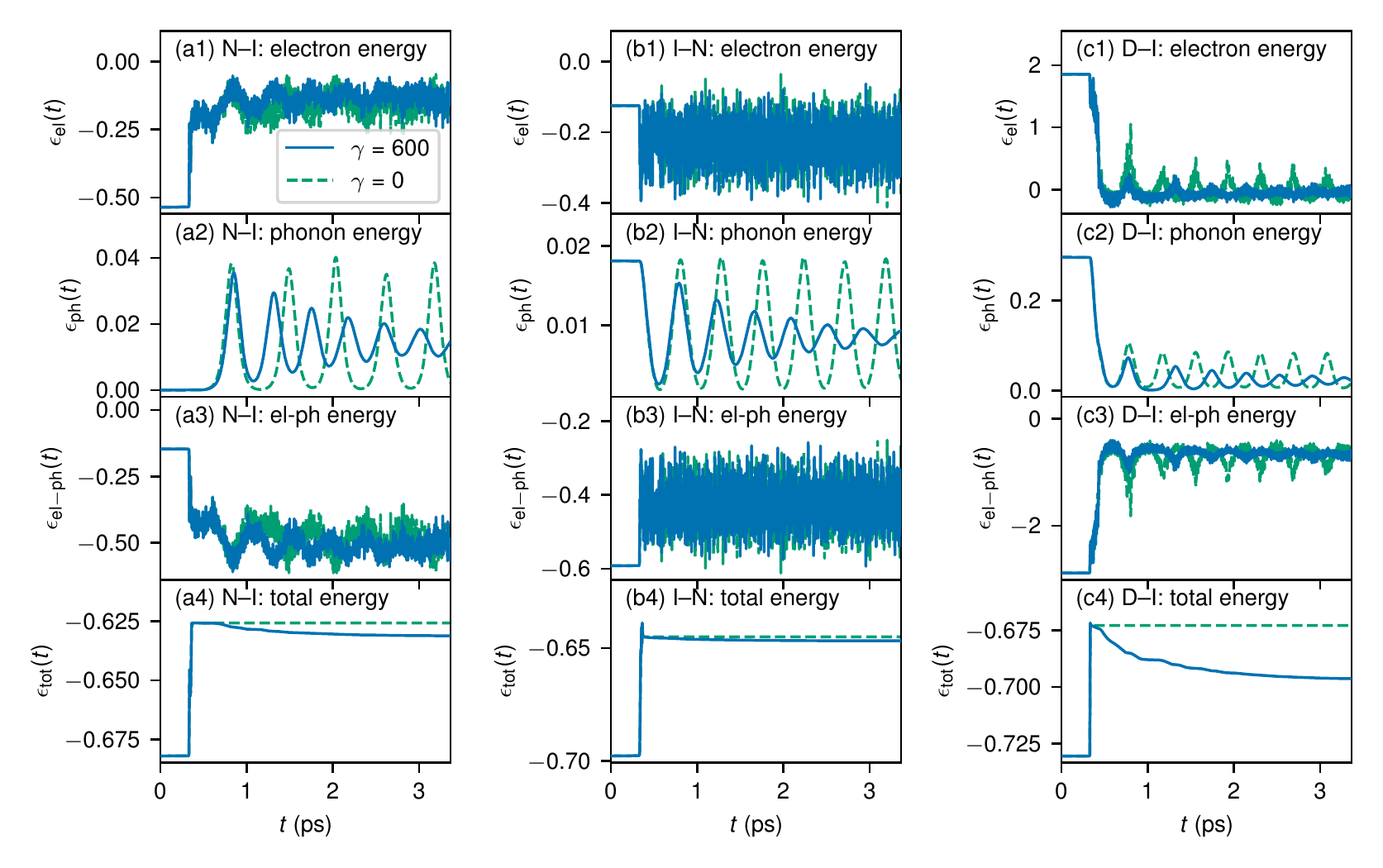}  
    \caption{Time evolution of energy densities in the units of eV (from top to bottom): (1) electronic energy $\epsilon_\text{el}$, (2) phononic energy $\epsilon_\text{ph}$, (3) electron-phonon interaction energy $\epsilon_\text{el-ph}$, and (4) total energy $\epsilon_\text{tot}$. Three columns are for (a) N--I transition, (b) I--N transition, and (c) D--I transition.}
    \label{fig:ens_dynamics}
    \end{center}
\end{figure*}

To better understand the energy flow after the photoexcitation, in Fig.~\ref{fig:ens_dynamics}, we plot the time evolution of the pure electronic energy density $\epsilon_{\rm el}$, the phonon energy density $\epsilon_{\rm ph}$, the electron-phonon interaction energy density $\epsilon_\text{el-ph}$, and the total energy density $\epsilon_{\rm tot}$. While the absorbed energy per site is about $0.05$ eV for all the cases, the heat distribution to each part of the system depends on the initial states. For the N--I transition, the electronic and phononic energies are increased while the electron-phonon interaction loses energy. On the other hand, the electron-phonon interaction gains energy for the I--N and D--I transitions, while the electronic and phononic energies are decreased. 

These qualitative features can be understood by the ansatz, Eq.~\eqref{eq:energy_ansatz}. The electronic energy densities of the three phases are approximately
\begin{equation}
\epsilon_{\rm el}^{\rm N} \simeq \frac{1}{2}(U - \Delta), \quad \epsilon_{\rm el}^{\rm I} \simeq 0, \quad \epsilon_{\rm el}^{\rm D} \simeq \frac{1}{2}(U + \Delta),
\end{equation}
which indicates $\epsilon_{\rm el}^{\rm N} < \epsilon_{\rm el}^{\rm I} < \epsilon_{\rm el}^{\rm D}$ for $0< U < \Delta$. This explains the reduction of $\epsilon_{\rm el}$ for the I--N and D--I transitions, and the increase for the N--I transition. Similarly, the electron-phonon interaction energies are 
\begin{equation}
\epsilon_\text{el-ph}^{\rm N} \simeq 0, \quad \epsilon_\text{el-ph}^{\rm I} \simeq -\tilde{V}(q_0^{\rm I}), \quad \epsilon_\text{el-ph}^{\rm D} \simeq -4\tilde{V} (q_0^{\rm D}),
\end{equation}
where $q_0^{\rm I,D}$ are the equilibrium dimerization for the ionic and dipole phases. These expressions indicate $\epsilon_\text{el-ph}^{\rm N} > \epsilon_\text{el-ph}^{\rm I} > \epsilon_\text{el-ph}^{\rm D}$, and thus the opposite trends to $\epsilon_{\rm el}$ are observed. Finally, the phononic energies are ${\epsilon_{\rm ph} = 2 s_1 q_0^{2} + 4 s_2 q_0^{4}}$ and the equilibrium $q_0$'s for the three phases satisfy $q_0^{\rm N} < q_0^{\rm I} < q_0^{\rm D}$. Thus the trend of $\epsilon_{\rm ph}$ is similar to that of $\epsilon_{\rm el}$. 

In accord with the adiabatic evolution of the electronic states, the energies of the electronic part and the electron-phonon interaction are nearly constant after the photoexcitation. On the other hand, the phononic energy shows large oscillations and slow decays in a few picoseconds; they include both the potential energy and the kinetic energy of ions. In particular, immediately after the photoexcitation, the kinetic energy is not ignorable. When the damping exists, the molecules lose their kinetic energy to the bath, and the positions of the molecules approach the quasi-steady values. For example, for the N--I transition, $\epsilon_{\rm ph}(t)$ reaches nearly $0.018$ eV, which is close to the equilibrium value of the ionic phase [see Fig.~\ref{fig:ens_dynamics}(a2)]. In contrast, for the I--N transition, the phonons have non-negligible residual energy (nearly half of the initial value) compared to the equilibrium neutral phase $\approx 0$ eV, indicating the incompleteness of the transition in this case.

While the total energy gradually decreases in the presence of damping, we find that the total energy does not go back to the initial values but rather converges to quasi-steady values. This observation again exemplifies the metastable nature of the obtained photoexcited states (we will discuss thermal relaxation to the initial states in Sec.~\ref{sec:discussion}). Furthermore, the total-energy loss is small compared to the absorbed energy (at most $40\%$ for the D--I transition). The reason is that the energy dissipation occurs only through the phonon damping in our model, and the phononic energy is an order of magnitude smaller than the electronic energies.

\subsection{Time-dependent spectral density}\label{sec:Sw}

\begin{figure}[!tb]
\begin{center}
\includegraphics[width= \columnwidth]{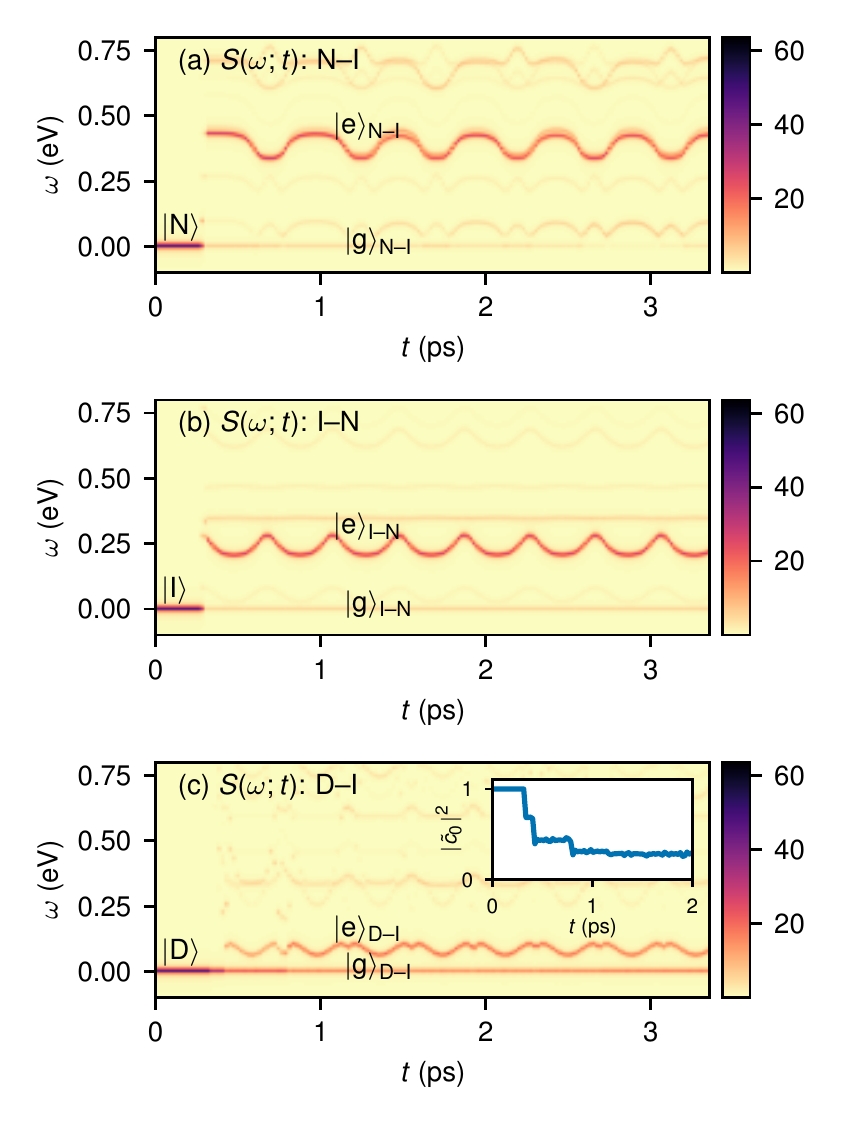}  
\caption{Time-dependent spectral densities for: (a) N--I transition, (b) I--N transition, and (c) D--I transition. $\gamma = 0$ is used. The initial ground states $\ket{\rm N, I, D}$, the new ground states $\ket{\rm{g}}$, and the most dominant excited states $\ket{\rm{e}}$ are labeled. The inset in panel (c) shows the weight of the ground state, i.e., $|\tilde{c}_{0}|^2$ in Eq.~\eqref{eq:spectral_ansatz} }
\label{fig:Sw}
\end{center}
\end{figure}

Let us now elaborate on the nature of photoexcited states by studying the time-dependent spectral density, Eq.~\eqref{eq:Sw}, more closely. In Fig.~\ref{fig:Sw}, we plot $S(\omega; t)$ for the three cases without damping corresponding to the dashed lines in Fig.~\ref{fig:obs_dynamics}. In all the presented cases, the weights transfer from the initial ground states to the excited states within femtosecond pump pulses. After that, the system evolves almost adiabatically, i.e., the weights on each adiabatic instantaneous eigenstate $|\Braket{\psi(t)|n(t)}|^2$ are preserved. The lattice displacement only modifies the adiabatic eigenvalues $E_n(t)$ slightly. Such adiabaticity holds even when the lattice displacement returns to the initial position. Only at the initial stage of the N--I or D--I transitions ($t \lesssim 1.0$ ps), we see slight non-adiabaticity and transfer of weights [see the inset of Fig.~\ref{fig:Sw}(c)]. 

Next, we discuss the distributions of the weights, $|\tilde{c}_n|^2$ in Eq.~\eqref{eq:spectral_ansatz}, in the time-evolved states after the the pump pulses. For the sake of simplicity, we focus on two classes of eigenstates. The first is the ground states at the new lattice positions, which we denote as $\ket{\rm{g}}_{\text{N--I}}$, $\ket{\rm{g}}_{\text{I--N}}$, and $\ket{\rm{g}}_{\text{D--I}}$ for the three cases. The second class is the most dominant excited states denoted as $\ket{\rm{e}}_{\text{N--I}}$, $\ket{\rm{e}}_{\text{I--N}}$, and $\ket{\rm{e}}_{\text{D--I}}$, which are located around $\omega \approx 0.4$ eV, $0.25$ eV, and $0.1$ eV in Fig.~\ref{fig:Sw}. For $t \gg t_\text{pump}$, the eigenstate decomposition in Eq.~\eqref{eq:spectral_ansatz} can be schematically rewritten as
\begin{equation}
    \ket{\psi(t)} \sim \tilde{c}_{\rm g} \ket{\rm{g}} + \tilde{c}_{\rm e} \ket{\rm{e}} + \sum_{n\neq \rm{g, e}} \tilde{c}_n (t) \ket{n(t)}  
\label{eq:g_e_approx}
\end{equation}
Table \ref{tab:weight} summarizes the weights of these two classes of eigenstates taken at $t=0.8$ ps. For the N--I transition and for the I--N transition [Figs.~\ref{fig:Sw}(a) and (b)], the weights on the new ground states are reduced to $|\tilde{c}_{\rm g}|^2 \approx 10\%$, and the dominant excited states have about half of the total weight. We will explain below why these similar distributions still give rise to a nearly complete photoinduced N-I transition, while the I-N transition is incomplete. We also note that the high concentration of the weights on a few states is characteristic of one-dimensional systems due to spin-charge separation~\cite{takahashi2008}. For the D--I transition [Fig.~\ref{fig:Sw}(c)], the weight of the new ground state and the dominant excited state are both around $30\%$. Because the N--I and D--I transitions are nearly complete, the different distributions of the weights are puzzling. 

\begin{table}[!tb]
    \caption{\label{tab:weight} The weights of the ground states at the new lattice positions and the most dominant excited states for the N--I, I--N, and D--I transitions taken at $t = 0.8$ ps.}
    \begin{ruledtabular}
    \begin{tabular}{cccc}
     & N--I & I--N & D--I\\
    \hline
     $|\tilde{c}_{\rm g}|^2$ & 0.05 & 0.10 & 0.33 \\
     $|\tilde{c}_{\rm e}|^2$ & 0.43 & 0.55 & 0.34 
    \end{tabular}
    \end{ruledtabular}
\end{table}


\begin{figure}[!tb]
    \begin{center}
    \includegraphics[width= \columnwidth]{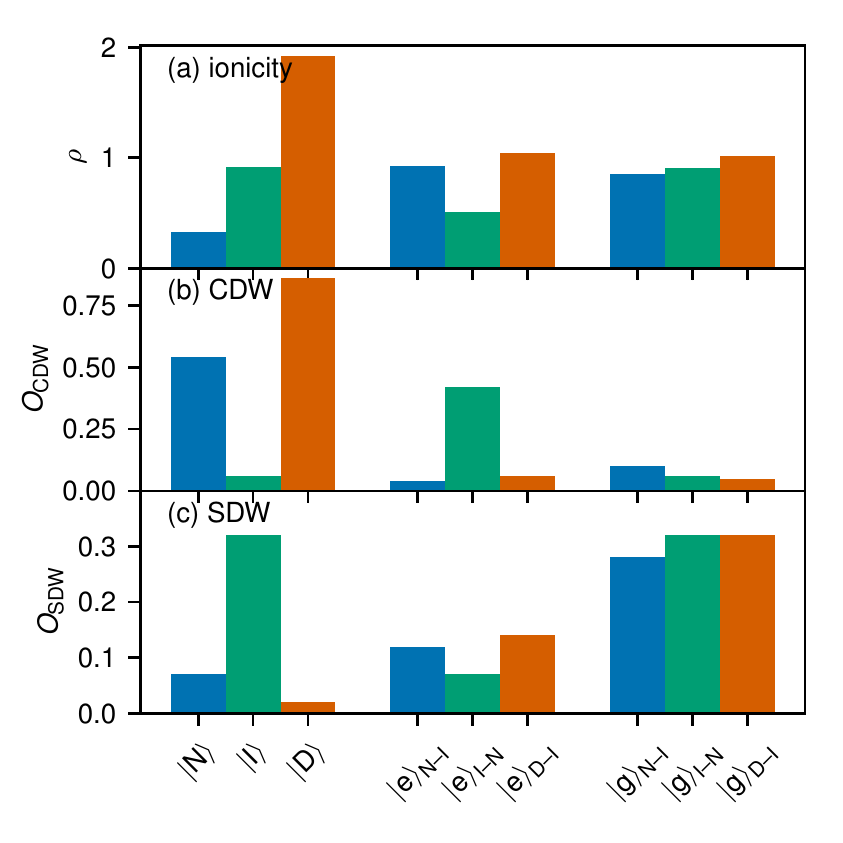}  
    \caption{ Electronic order parameters of the initial ground states (left), the most dominant excited states, $\ket{\rm e}$'s, at $t = 0.8$ ps (middle), and the lowest energy eigenstates, $\ket{\rm g}$'s, at $t = 0.8$ ps (right).}
    \label{fig:bar}
    \end{center}
    \end{figure}

To elucidate the different attributes of the photoinduced phase transitions, we plot the expectation values of the electronic order parameters in terms of the most dominant excited states ($\ket{\rm e}$'s), the new ground states ($\ket{\rm g}$'s), and the initial ground states in Fig.~\ref{fig:bar}. It is clear that the dominant excited states are distinct from their initial ground states; they possess the electronic order parameters that are rather close to those of the target phases. For example, $\ket{\rm{e}}_{\text{N--I}}$ and $\ket{\rm{e}}_{\text{D--I}}$ have the almost ideal ionicity for the ionic phase $\rho \approx 1$. While the ionicity of $\ket{\rm{e}}_{\text{I--N}}$, ($\approx 0.5$) is still larger than the value of the neutral phase ($\approx 0.33$), the other order parameters are close to those of the neutral phase. These observations mean that the intermediate Franck-Condon states already have the properties of the final target phases even though the former is not adiabatically connected to the latter.

The subtle difference among the three transitions can be understood by further considering the new ground states at the new lattice positions, i.e., $\ket{\rm g}$'s in Eq.~\eqref{eq:g_e_approx}. The electronic order parameters of these states shown in Fig~\ref{fig:bar} suggest that all three states show ionic nature. This is because the final dimerization $q_0$ for the three cases is always around $0.03$, where the corresponding ground state is the ionic phase as shown in Fig.~\ref{fig:GS_ED}(a). Therefore, for the N--I and D--I transitions, both the metastable ground state and the dominant excited state are ionic, amounting to $60\% - 70\%$ of the total weight. Thus, the phase transitions are nearly complete in these cases. In contrast, for the I--N transition, the metastable ground state remains ionic, while the dominant excited state is neutral. Therefore, the photoinduced transition to the neutral phase is incomplete. The underlying difficulty of the transition to the neutral phase is that displacement cannot achieve $q_0 = 0$, which is an energy maximum.

Finally, we note that $\ket{\rm{e}}_\text{D--I}$ is not a single-photon excited state corresponding to the peaks in the linear optical conductivity in Fig.~\ref{fig:sigma_GS}(c); $\ket{\rm{e}}_\text{D--I}$ has the excitation energy $\omega \approx 0.1$ eV, which is well below the gap energy in the optical conductivity. On the other hand, $\ket{\rm{e}}_\text{I--N}$ corresponds to the peak at $\omega \approx 0.28$ eV of the linear optical conductivity in Fig.~\ref{fig:sigma_GS}(b). We confirm that the overlap of the two states is over 99\%. Similarly, $\ket{\rm{e}}_\text{N--I}$ has a large overlap with the small peak at $\omega \approx 0.5$ eV in Fig.~\ref{fig:sigma_GS}(a). However, the final weights on $\ket{\rm{e}}_\text{N--I}$ and $\ket{\rm{e}}_\text{I--N}$ in $\ket{\psi(t)}$ are much larger than the ones from the single-photon excitation captured by the linear optical conductivity. These indicate that the photoinduced transitions are triggered by multiphoton absorption \cite{mizuno2000, okamoto2019}, and the optical fields need to have relatively high intensity.

\subsection{Pump-probe optical conductivity}\label{sec:sigma_pp}

\begin{figure}[!tb]
\begin{center}
\includegraphics[width= \columnwidth]{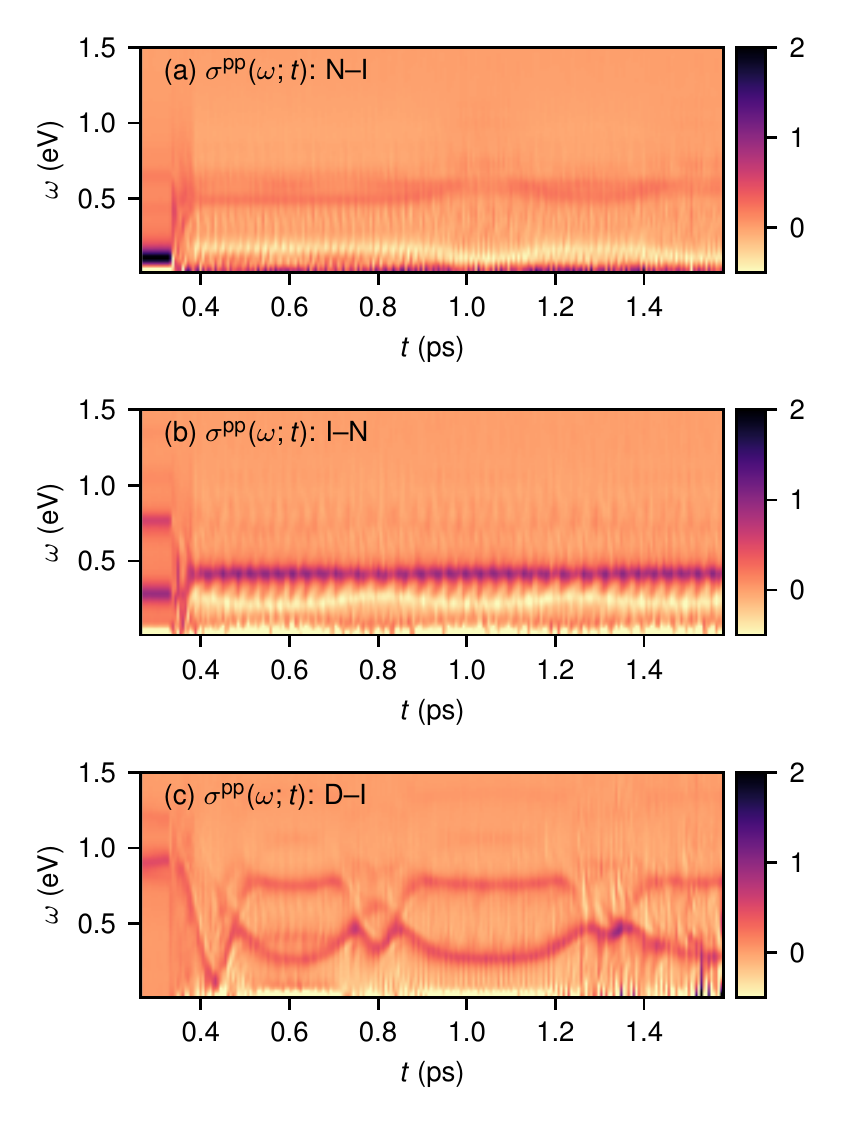}  
\caption{Pump-probe optical conductivity for (a) N--I transition, (b) I--N transition, and (c) D--I transition.}
\label{fig:sigma_pp}
\end{center}
\end{figure}

In Sec.~\ref{sec:Sw}, we have identified the dominant excited states after the pump pulse. The natural question then is on the optical properties of these excited states, which is relevant to the experimentally obtained spectroscopic information. Here, we scrutinize the time-dependent optical conductivity to characterize the photoinduced metastable states. We employ the two-dimensional spectroscopy approach as Refs.~\cite{kindt1999, nemec2002, okamoto2019}, which gives the time-dependent optical conductivity $\sigma^{\rm pp}(\omega; t)$. To obtain this quantity, we repeat the pump-probe simulations with impulsive probe pulses at different $t_{\rm probe}$'s in Eq.~\eqref{eq:probe}. The extra current induced by the probe pulse is given by
\begin{equation}
\begin{split}
\delta j(t, t_{\rm probe}) &= \int_{-\infty}^{t} \Sigma(t, t') E_{\rm probe} (t', t_{\rm probe}) dt' \\
&= \Sigma(t, t_{\rm probe}) E_0,
\end{split}
\end{equation}
where $\Sigma(t, t')$ is the general conductivity response function \cite{vengurlekar1988}. Introducing a new variable $s \equiv t - t_{\rm probe}$, we can write
\begin{equation}
\delta j (t, s) = \Sigma (t, s) E_0.
\end{equation}
Finally, Fourier transforming over $s$, we define the pump-probe conductivity
\begin{equation}
\sigma^{\rm pp}(\omega; t) \equiv \int ds \Sigma(t, s) e^{-i \omega s} = \delta j (t, \omega)/E_{0} .
\label{eq:sigma_pp}
\end{equation}

Let us first consider the N--I transition. The pump-probe conductivity is plotted in Fig.~\ref{fig:sigma_pp}(a). Before the pump pulse, there is an absorption peak around $\omega \approx 0.1$ eV. After the pump pulse, a new absorption peak appears around $\omega \approx 0.5$ eV. At the same time, an emission peak also appears around $\omega \approx 0.15$ eV, which corresponds to the second term of $\sigma^{\rm reg} (\omega)$ in Eq.~\eqref{eq:sigma}. The spectroscopic change is a clear sign of a photoinduced phase transition and occurs in a few subpicoseconds. The Drude weight, which is initially negative due to the finite-size effect [see Fig.~\ref{fig:sigma_GS}(a)], becomes mostly positive, while rapid fluctuations also exist. The change in the Drude weight may indicate metallic behavior. However, the numerical resolution at the low frequency is not good enough to decisively conclude this statement. 

Next, the pump-probe conductivity of the I--N transition is shown in Fig.~\ref{fig:sigma_pp}(b). After the pump pulse, the two absorption peaks become an emission peak and an absorption peak at lower energy. Since $\ket{\rm e}_\text{I--N}$ corresponds to the dominant single-photon absorption peak, we can associate the emission peak with the transition to the ground state and the new absorption peak with the transition to a state near the second largest peak ($\Delta \omega 
\approx 0.5$ eV) in the linear optical conductivity [see Fig.~\ref{fig:sigma_GS}(b)]. The Drude weight retains a finite negative value.

For the D--I transition [Fig.~\ref{fig:sigma_pp}(c)], the high-energy peak at $\omega \approx 1.0$ eV is transformed into low-energy peaks around $\omega \approx 0.3$ and $0.9$ eV. The two-peak structure is the clear signature of the ionic ground state [see Fig.~\ref{fig:sigma_GS}(b)], which arises from  $\ket{\rm{g}}_\text{D--I}$. We do not observe an emission peak since the wave function $\ket{\psi(t)}$ is dominated by $\ket{\rm{g}}_\text{D--I}$ and $\ket{\rm{e}}_\text{D--I}$; the former has no lower energy state and the latter is a multiphoton excited state that cannot transit to $\ket{\rm{g}}_\text{D--I}$ via single-photon emission. The two peaks intersect each other at $t \approx 0.8$ and $1.3$ ps, where dimerization reaches the largest amplitude [Fig.~\ref{fig:obs_dynamics}(c1)]. While the Drude weight is initially ignorable, the strong pump induces a finite negative Drude weight. Again, it is unclear if this indicates metallic behavior due to the severe finite-size effect.

\begin{figure}[!tb]
\begin{center}
\includegraphics[width= \columnwidth]{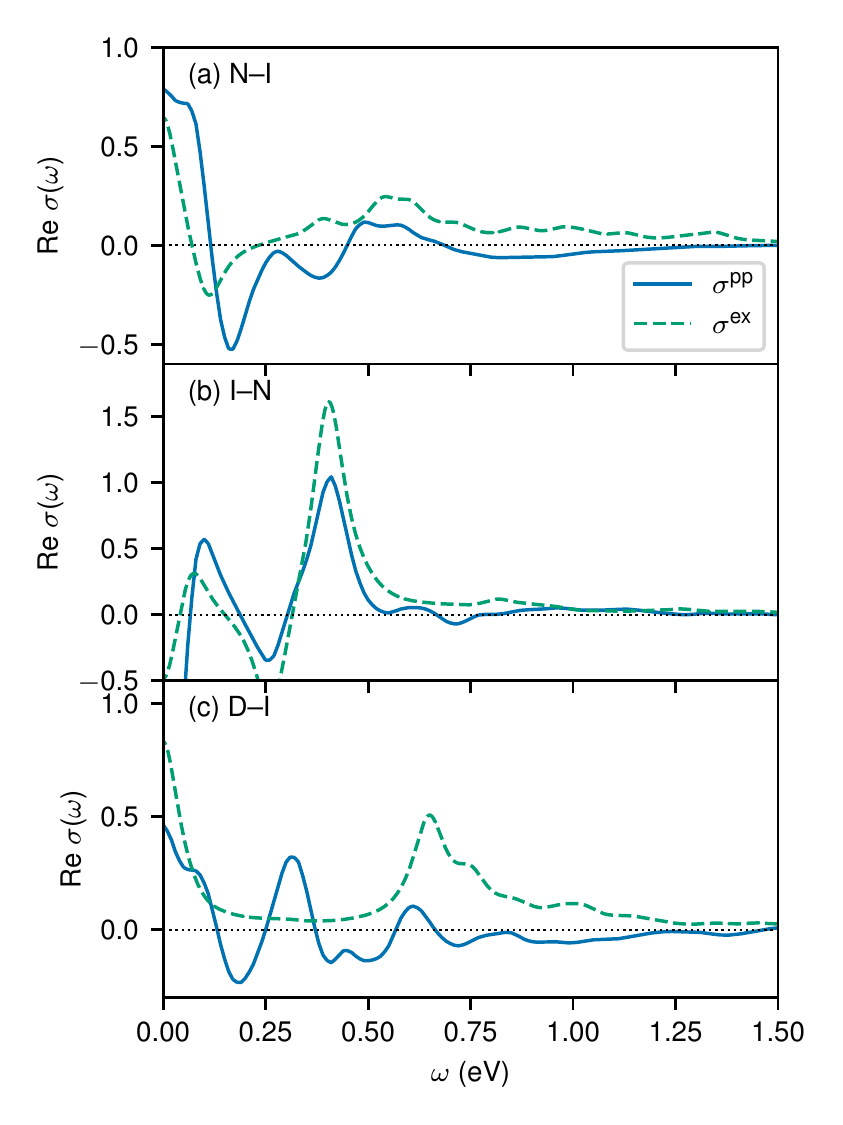}  
\caption{Comparison of the time-dependent conductivity for (a) N--I transition, (b) I--N transition, and (c) D--I transition. Solid lines are the pump-probe conductivity $\sigma^{\rm pp}(\omega)$ in Eq.~\eqref{eq:sigma_pp}. Dashed lines are the static conductivity $\sigma^{\rm ex}(\omega)$ in Eq.~\eqref{eq:sigma} for the most dominant excited states, $\ket{\rm e}$'s. The data are taken at $t = 0.8$ ps.}
\label{fig:sigma_pp_ex}
\end{center}
\end{figure}

In order to better understand these changes, we calculate the optical conductivity of the dominant excited eigenstates $\ket{\rm{e}}_\text{N--I}$, $\ket{\rm{e}}_\text{I--N}$, and $\ket{\rm{e}}_\text{D--I}$ by Eq.~\eqref{eq:sigma}. Since the initial state is an excited state, the second contributions in $\sigma^{\rm reg} (\omega)$ do not vanish even for $\omega >0$. Figure~\ref{fig:sigma_pp_ex} compares the pump-probe conductivity $\sigma^{\rm pp}(\omega; t)$ at $t = 0.8$ ps and the corresponding conductivity of the excited states, $\sigma^{\rm ex}(\omega)$. The two quantities agree for the N--I and I--N transitions, since the quasi-steady states are dominated by $\ket{\rm{e}}$'s. In contrast, for the D--I transition, the pump-probe conductivity $\sigma^{\rm pp} (\omega)$ shows emission/absorption peaks at low frequencies, while the static conductivity of $\ket{\rm{e}}_\text{D--I}$, $\sigma^{\rm ex}(\omega)$, has a large gap $\approx 0.7$ eV. The discrepancy comes from the fact that the quasi-steady state contains a large portion of ionic ground state $\ket{\rm{g}}_\text{D--I}$.

\section{Discussion}\label{sec:discussion}

\begin{figure}[!tb]
\begin{center}
\includegraphics[width= \columnwidth]{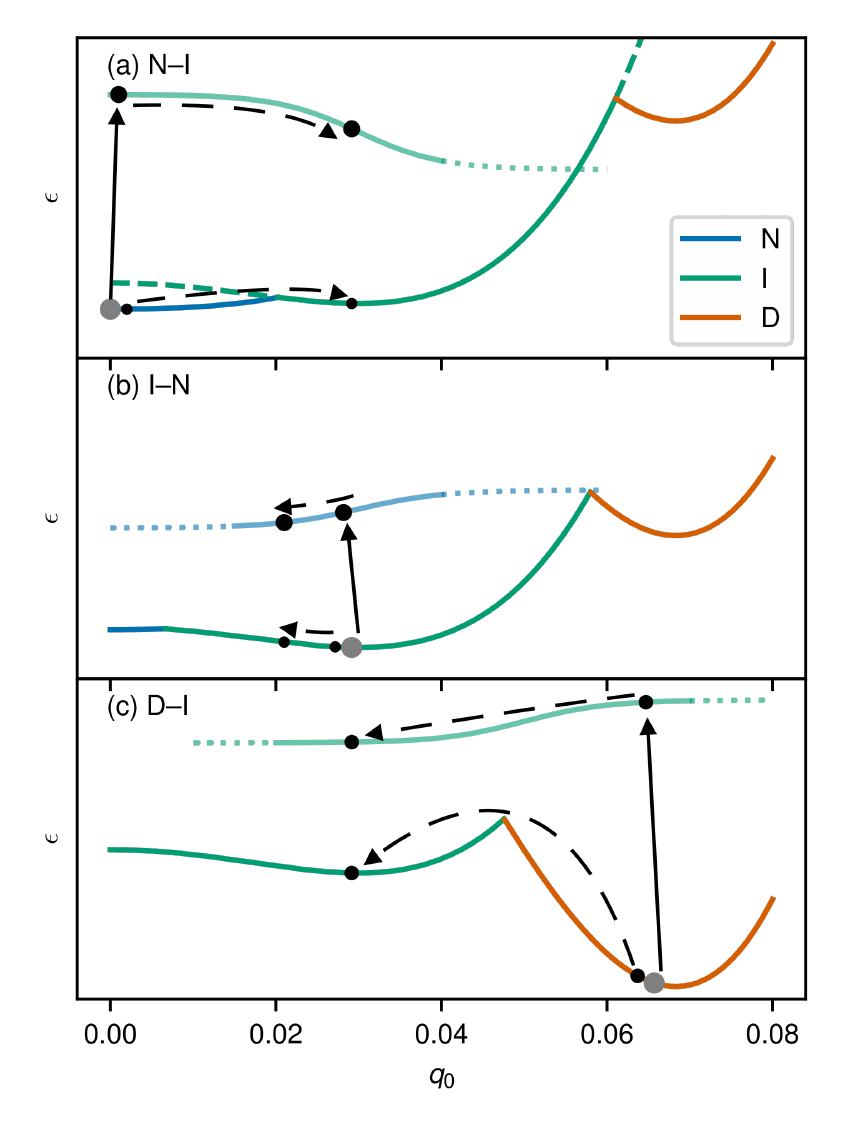}  
\caption{Schematics of the photoinduced dynamics in the potential energy surfaces for (a) N--I transition, (b) I--N transition, and (c) D--I transition. The dashed line in panel (a) shows the diabatic PES of the ionic phase. The size of the bulbs represents the weight of each energy eigenstate. The initial states (the gray dots) are split into the excited states by the optical excitation (solid arrows). The Franck-Condon states then adiabatically move to metastable states by lattice modulation (dashed arrows). }
\label{fig:excitation}
\end{center}
\end{figure}

In this section, we summarize the obtained picture of the photoinduced dynamics in our model and compare it with the previous studies. Since our model considers only the fixed system size and homogeneous excitation, we do not discuss the size-dependence of the CT string~\cite{nagaosa1989, koshihara1999, iwano2006} or the subsequent domain-wall motion~\cite{huai2000, luty2002, yonemitsu2004, yonemitsu2004a, soos2007, peterseim2015, kishida2009}.

Figure~\ref{fig:excitation} shows the schematic excitation paths that our simulations describe for the three transitions that we have discussed in Sec.~\ref{sec:dynamics}. The first step is the transition from the initial ground states to the Franck-Condon states by multiphoton absorption (solid arrows). We find that the excited states already possess similar properties as the final target phases in terms of the electronic order parameters. While this apparently indicates that the Franck-Condon states are on the branch of the diabatic PES for the final phase [the dashed line in Fig.~\ref{fig:excitation}(a)] as considered in Refs.~\cite{koshihara1999, luty2002, collet2003}, we believe that they are on a different PES. This is because the energy of the Franck-Condon state is away from the diabatic PES. For example, for the N--I transition, the Franck-Condon state is roughly $0.4$ eV above the ground state, while the diabatic PES of the ionic phase is only $0.1$ eV above that of the neutral phase at $q_0 = 0$. For the I--N and D--I transitions, the former has smaller energy than the latter. Furthermore, the energy of the excited state is much bigger than the gap due to the avoided crossing $\approx 0.01$ eV (estimated from the PESs for $J_0 = 0$ and $0.17$ eV). 

As the absorbed energy is transferred to the lattice, the Franck-Condon states and the residual ground states at the initial lattice position adiabatically transform into the ones at the new lattice position (see the dashed arrows in Fig.~\ref{fig:excitation}). If the dominant excited states and the ground states at the new lattice position have similar properties, the resultant states have a nearly complete transformation to the final phases as the N--I and D--I transitions in this work. On the other hand, the excited state and the ground state have different attributes in the I--N transition, which results in only partial transformation. In any case, the states after the lattice relaxation are still coherent superpositions of a few dominant energy eigenstates.

Thermal relaxation, which is not included in our model, may lead to two more processes. The first is the relaxation from the dominant excited states towards the metastable ground states at the new lattice position. A similar process is discussed in Refs.~\cite{koshihara1999, collet2003}, while the relaxation is supposed to occur at the avoided crossing point, indicating relatively quick decay to the metastable ground state. On the other hand, the intermediate excited state in our model is away from the avoided crossing, and we expect a longer decay time. After this process, the absorption peaks in the pump-probe optical conductivity will disappear since the system is in the lowest energy eigenstate at the new lattice position. The second process is the relaxation from the metastable ground state to the initial ground state; the system needs to overcome the energy barrier of the adiabatic PES.

Our picture suggests a nonthermal path to the desired long-lived excited states achieved by multiphoton absorption processes. A similar mechanism was discussed for optical switching from a CDW state to a nonequilibrium superconducting state in the extended Hubbard model~\cite{bittner2019}. While the tailored optical pulse targeting the nonequilibrium superconducting state was used there, the photoinduced transitions in our model occur in broad parameter regimes as shown in Fig.~\ref{fig:2d_main}. This is because the different optical pulses change only the weights of a few excited states but not the set of the states. This may be connected to the spin-charge separation, which allows only the discrete number of excited states~ \cite{takahashi2008}, and possible localization in the energy eigenstates~\cite{okamoto2021}.

\section{Conclusion}\label{sec:conclusion}
In this work, we have theoretically investigated the photoinduced microscopic dynamics of the one-dimensional extended Peierls-Hubbard model by exact diagonalization. Aside from the neutral and ionic ground states relevant for TTF-CA, we identified a stable phase (called the dipole phase in this work) with a doubly ionized charge configuration (TTF$^{+2}$CA$^{-2}$) and strong lattice dimerization ($\approx 6$\%). The Zak phases calculated for the three ground states indicate that they are topologically distinct states. We have also shown that the three phases have distinctive spectroscopic characteristics in linear optical conductivity.

The ultrafast dynamics of these phases after monocycle optical pulses has been further investigated. We have detected the neutral--ionic, ionic--neutral, and dipole--ionic transitions and carefully analyzed their real-time dynamics using time-dependent spectral density and pump-probe conductivity. The spectral weights indicate that the Franck-Condon states have the same properties as the final target phases, even though they are away from the diabatic PESs of the latter. Multiphoton absorption achieves such excitation, and the resultant states are long-lived coherent states. This process suggests nonthermal switching of electronic and lattice orders and the generation of coherent CT strings.

Our work focuses on the nucleation of a fixed-size CT string and intra-domain dynamics. A microscopic picture of subsequent slow dynamics involving domain walls and solitons is desirable for precise control of their dynamics, while it requires larger system sizes, thermal relaxation, and inter-chain coupling. The effects of finite temperatures, including dissipation and thermal/radiative relaxation of excited states, are left for future investigation.

\acknowledgments
This work is supported by Georg H. Endress Foundation. The authors acknowledge support by the state of Baden-W\"urttemberg through bwHPC and the German Research Foundation (DFG) through grant no INST 40/575-1 FUGG (JUSTUS 2 cluster). We thank S. Stumper for carefully reading the manuscript.

%


\begin{thebibliography}{96}%
\makeatletter
\providecommand \@ifxundefined [1]{%
 \@ifx{#1\undefined}
}%
\providecommand \@ifnum [1]{%
 \ifnum #1\expandafter \@firstoftwo
 \else \expandafter \@secondoftwo
 \fi
}%
\providecommand \@ifx [1]{%
 \ifx #1\expandafter \@firstoftwo
 \else \expandafter \@secondoftwo
 \fi
}%
\providecommand \natexlab [1]{#1}%
\providecommand \enquote  [1]{``#1''}%
\providecommand \bibnamefont  [1]{#1}%
\providecommand \bibfnamefont [1]{#1}%
\providecommand \citenamefont [1]{#1}%
\providecommand \href@noop [0]{\@secondoftwo}%
\providecommand \href [0]{\begingroup \@sanitize@url \@href}%
\providecommand \@href[1]{\@@startlink{#1}\@@href}%
\providecommand \@@href[1]{\endgroup#1\@@endlink}%
\providecommand \@sanitize@url [0]{\catcode `\\12\catcode `\$12\catcode
  `\&12\catcode `\#12\catcode `\^12\catcode `\_12\catcode `\%12\relax}%
\providecommand \@@startlink[1]{}%
\providecommand \@@endlink[0]{}%
\providecommand \url  [0]{\begingroup\@sanitize@url \@url }%
\providecommand \@url [1]{\endgroup\@href {#1}{\urlprefix }}%
\providecommand \urlprefix  [0]{URL }%
\providecommand \Eprint [0]{\href }%
\providecommand \doibase [0]{https://doi.org/}%
\providecommand \selectlanguage [0]{\@gobble}%
\providecommand \bibinfo  [0]{\@secondoftwo}%
\providecommand \bibfield  [0]{\@secondoftwo}%
\providecommand \translation [1]{[#1]}%
\providecommand \BibitemOpen [0]{}%
\providecommand \bibitemStop [0]{}%
\providecommand \bibitemNoStop [0]{.\EOS\space}%
\providecommand \EOS [0]{\spacefactor3000\relax}%
\providecommand \BibitemShut  [1]{\csname bibitem#1\endcsname}%
\let\auto@bib@innerbib\@empty
\bibitem [{\citenamefont {Tokura}(2006)}]{tokura2006}%
  \BibitemOpen
  \bibfield  {author} {\bibinfo {author} {\bibfnamefont {Y.}~\bibnamefont
  {Tokura}},\ }\bibfield  {title} {\bibinfo {title} {Photoinduced {{Phase
  Transition}}: {{A Tool}} for {{Generating}} a {{Hidden State}} of
  {{Matter}}},\ }\href {https://doi.org/10.1143/JPSJ.75.011001} {\bibfield
  {journal} {\bibinfo  {journal} {J. Phys. Soc. Jpn.}\ }\textbf {\bibinfo
  {volume} {75}},\ \bibinfo {pages} {011001} (\bibinfo {year}
  {2006})}\BibitemShut {NoStop}%
\bibitem [{\citenamefont {Basov}\ \emph {et~al.}(2017)\citenamefont {Basov},
  \citenamefont {Averitt},\ and\ \citenamefont {Hsieh}}]{basov2017}%
  \BibitemOpen
  \bibfield  {author} {\bibinfo {author} {\bibfnamefont {D.~N.}\ \bibnamefont
  {Basov}}, \bibinfo {author} {\bibfnamefont {R.~D.}\ \bibnamefont {Averitt}},\
  and\ \bibinfo {author} {\bibfnamefont {D.}~\bibnamefont {Hsieh}},\ }\bibfield
   {title} {\bibinfo {title} {Towards properties on demand in quantum
  materials},\ }\href {https://doi.org/10.1038/nmat5017} {\bibfield  {journal}
  {\bibinfo  {journal} {Nat. Mater.}\ }\textbf {\bibinfo {volume} {16}},\
  \bibinfo {pages} {1077} (\bibinfo {year} {2017})}\BibitemShut {NoStop}%
\bibitem [{\citenamefont {McIver}\ \emph {et~al.}(2020)\citenamefont {McIver},
  \citenamefont {Schulte}, \citenamefont {Stein}, \citenamefont {Matsuyama},
  \citenamefont {Jotzu}, \citenamefont {Meier},\ and\ \citenamefont
  {Cavalleri}}]{mciver2020}%
  \BibitemOpen
  \bibfield  {author} {\bibinfo {author} {\bibfnamefont {J.~W.}\ \bibnamefont
  {McIver}}, \bibinfo {author} {\bibfnamefont {B.}~\bibnamefont {Schulte}},
  \bibinfo {author} {\bibfnamefont {F.-U.}\ \bibnamefont {Stein}}, \bibinfo
  {author} {\bibfnamefont {T.}~\bibnamefont {Matsuyama}}, \bibinfo {author}
  {\bibfnamefont {G.}~\bibnamefont {Jotzu}}, \bibinfo {author} {\bibfnamefont
  {G.}~\bibnamefont {Meier}},\ and\ \bibinfo {author} {\bibfnamefont
  {A.}~\bibnamefont {Cavalleri}},\ }\bibfield  {title} {\bibinfo {title}
  {Light-induced anomalous {{Hall}} effect in graphene},\ }\href
  {https://doi.org/10.1038/s41567-019-0698-y} {\bibfield  {journal} {\bibinfo
  {journal} {Nat. Phys.}\ }\textbf {\bibinfo {volume} {16}},\ \bibinfo {pages}
  {38} (\bibinfo {year} {2020})}\BibitemShut {NoStop}%
\bibitem [{\citenamefont {Fausti}\ \emph {et~al.}(2011)\citenamefont {Fausti},
  \citenamefont {Tobey}, \citenamefont {Dean}, \citenamefont {Kaiser},
  \citenamefont {Dienst}, \citenamefont {Hoffmann}, \citenamefont {Pyon},
  \citenamefont {Takayama}, \citenamefont {Takagi},\ and\ \citenamefont
  {Cavalleri}}]{fausti2011}%
  \BibitemOpen
  \bibfield  {author} {\bibinfo {author} {\bibfnamefont {D.}~\bibnamefont
  {Fausti}}, \bibinfo {author} {\bibfnamefont {R.~I.}\ \bibnamefont {Tobey}},
  \bibinfo {author} {\bibfnamefont {N.}~\bibnamefont {Dean}}, \bibinfo {author}
  {\bibfnamefont {S.}~\bibnamefont {Kaiser}}, \bibinfo {author} {\bibfnamefont
  {A.}~\bibnamefont {Dienst}}, \bibinfo {author} {\bibfnamefont {M.~C.}\
  \bibnamefont {Hoffmann}}, \bibinfo {author} {\bibfnamefont {S.}~\bibnamefont
  {Pyon}}, \bibinfo {author} {\bibfnamefont {T.}~\bibnamefont {Takayama}},
  \bibinfo {author} {\bibfnamefont {H.}~\bibnamefont {Takagi}},\ and\ \bibinfo
  {author} {\bibfnamefont {A.}~\bibnamefont {Cavalleri}},\ }\bibfield  {title}
  {\bibinfo {title} {Light-{{Induced Superconductivity}} in a {{Stripe-Ordered
  Cuprate}}},\ }\href {https://doi.org/10.1126/science.1197294} {\bibfield
  {journal} {\bibinfo  {journal} {Science}\ }\textbf {\bibinfo {volume}
  {331}},\ \bibinfo {pages} {189} (\bibinfo {year} {2011})}\BibitemShut
  {NoStop}%
\bibitem [{\citenamefont {Hu}\ \emph {et~al.}(2014)\citenamefont {Hu},
  \citenamefont {Kaiser}, \citenamefont {Nicoletti}, \citenamefont {Hunt},
  \citenamefont {Gierz}, \citenamefont {Hoffmann}, \citenamefont {Le~Tacon},
  \citenamefont {Loew}, \citenamefont {Keimer},\ and\ \citenamefont
  {Cavalleri}}]{hu2014}%
  \BibitemOpen
  \bibfield  {author} {\bibinfo {author} {\bibfnamefont {W.}~\bibnamefont
  {Hu}}, \bibinfo {author} {\bibfnamefont {S.}~\bibnamefont {Kaiser}}, \bibinfo
  {author} {\bibfnamefont {D.}~\bibnamefont {Nicoletti}}, \bibinfo {author}
  {\bibfnamefont {C.~R.}\ \bibnamefont {Hunt}}, \bibinfo {author}
  {\bibfnamefont {I.}~\bibnamefont {Gierz}}, \bibinfo {author} {\bibfnamefont
  {M.~C.}\ \bibnamefont {Hoffmann}}, \bibinfo {author} {\bibfnamefont
  {M.}~\bibnamefont {Le~Tacon}}, \bibinfo {author} {\bibfnamefont
  {T.}~\bibnamefont {Loew}}, \bibinfo {author} {\bibfnamefont {B.}~\bibnamefont
  {Keimer}},\ and\ \bibinfo {author} {\bibfnamefont {A.}~\bibnamefont
  {Cavalleri}},\ }\bibfield  {title} {\bibinfo {title} {Optically enhanced
  coherent transport in
  {{YBa}}{\textsubscript{2}}{{Cu}}{\textsubscript{3}}{{O}}{\textsubscript{6.5}}
  by ultrafast redistribution of interlayer coupling},\ }\href
  {https://doi.org/10.1038/nmat3963} {\bibfield  {journal} {\bibinfo  {journal}
  {Nat. Mater.}\ }\textbf {\bibinfo {volume} {13}},\ \bibinfo {pages} {705}
  (\bibinfo {year} {2014})}\BibitemShut {NoStop}%
\bibitem [{\citenamefont {Mitrano}\ \emph {et~al.}(2016)\citenamefont
  {Mitrano}, \citenamefont {Cantaluppi}, \citenamefont {Nicoletti},
  \citenamefont {Kaiser}, \citenamefont {Perucchi}, \citenamefont {Lupi},
  \citenamefont {Di~Pietro}, \citenamefont {Pontiroli}, \citenamefont
  {Ricc{\`o}}, \citenamefont {Clark}, \citenamefont {Jaksch},\ and\
  \citenamefont {Cavalleri}}]{mitrano2016}%
  \BibitemOpen
  \bibfield  {author} {\bibinfo {author} {\bibfnamefont {M.}~\bibnamefont
  {Mitrano}}, \bibinfo {author} {\bibfnamefont {A.}~\bibnamefont {Cantaluppi}},
  \bibinfo {author} {\bibfnamefont {D.}~\bibnamefont {Nicoletti}}, \bibinfo
  {author} {\bibfnamefont {S.}~\bibnamefont {Kaiser}}, \bibinfo {author}
  {\bibfnamefont {A.}~\bibnamefont {Perucchi}}, \bibinfo {author}
  {\bibfnamefont {S.}~\bibnamefont {Lupi}}, \bibinfo {author} {\bibfnamefont
  {P.}~\bibnamefont {Di~Pietro}}, \bibinfo {author} {\bibfnamefont
  {D.}~\bibnamefont {Pontiroli}}, \bibinfo {author} {\bibfnamefont
  {M.}~\bibnamefont {Ricc{\`o}}}, \bibinfo {author} {\bibfnamefont {S.~R.}\
  \bibnamefont {Clark}}, \bibinfo {author} {\bibfnamefont {D.}~\bibnamefont
  {Jaksch}},\ and\ \bibinfo {author} {\bibfnamefont {A.}~\bibnamefont
  {Cavalleri}},\ }\bibfield  {title} {\bibinfo {title} {Possible light-induced
  superconductivity in {{K}}{\textsubscript{3}}{{C}}{\textsubscript{60}} at
  high temperature},\ }\href {https://doi.org/10.1038/nature16522} {\bibfield
  {journal} {\bibinfo  {journal} {Nature}\ }\textbf {\bibinfo {volume} {530}},\
  \bibinfo {pages} {461} (\bibinfo {year} {2016})}\BibitemShut {NoStop}%
\bibitem [{\citenamefont {Stojchevska}\ \emph {et~al.}(2014)\citenamefont
  {Stojchevska}, \citenamefont {Vaskivskyi}, \citenamefont {Mertelj},
  \citenamefont {Kusar}, \citenamefont {Svetin}, \citenamefont {Brazovskii},\
  and\ \citenamefont {Mihailovic}}]{stojchevska2014}%
  \BibitemOpen
  \bibfield  {author} {\bibinfo {author} {\bibfnamefont {L.}~\bibnamefont
  {Stojchevska}}, \bibinfo {author} {\bibfnamefont {I.}~\bibnamefont
  {Vaskivskyi}}, \bibinfo {author} {\bibfnamefont {T.}~\bibnamefont {Mertelj}},
  \bibinfo {author} {\bibfnamefont {P.}~\bibnamefont {Kusar}}, \bibinfo
  {author} {\bibfnamefont {D.}~\bibnamefont {Svetin}}, \bibinfo {author}
  {\bibfnamefont {S.}~\bibnamefont {Brazovskii}},\ and\ \bibinfo {author}
  {\bibfnamefont {D.}~\bibnamefont {Mihailovic}},\ }\bibfield  {title}
  {\bibinfo {title} {Ultrafast {{Switching}} to a {{Stable Hidden Quantum
  State}} in an {{Electronic Crystal}}},\ }\href
  {https://doi.org/10.1126/science.1241591} {\bibfield  {journal} {\bibinfo
  {journal} {Science}\ }\textbf {\bibinfo {volume} {344}},\ \bibinfo {pages}
  {177} (\bibinfo {year} {2014})}\BibitemShut {NoStop}%
\bibitem [{\citenamefont {Basov}\ \emph {et~al.}(2016)\citenamefont {Basov},
  \citenamefont {Fogler},\ and\ \citenamefont {{Garcia de Abajo}}}]{basov2016}%
  \BibitemOpen
  \bibfield  {author} {\bibinfo {author} {\bibfnamefont {D.~N.}\ \bibnamefont
  {Basov}}, \bibinfo {author} {\bibfnamefont {M.~M.}\ \bibnamefont {Fogler}},\
  and\ \bibinfo {author} {\bibfnamefont {F.~J.}\ \bibnamefont {{Garcia de
  Abajo}}},\ }\bibfield  {title} {\bibinfo {title} {Polaritons in van der
  {{Waals}} materials},\ }\href {https://doi.org/10.1126/science.aag1992}
  {\bibfield  {journal} {\bibinfo  {journal} {Science}\ }\textbf {\bibinfo
  {volume} {354}},\ \bibinfo {pages} {aag1992} (\bibinfo {year}
  {2016})}\BibitemShut {NoStop}%
\bibitem [{\citenamefont {Orgiu}\ \emph {et~al.}(2015)\citenamefont {Orgiu},
  \citenamefont {George}, \citenamefont {Hutchison}, \citenamefont {Devaux},
  \citenamefont {Dayen}, \citenamefont {Doudin}, \citenamefont {Stellacci},
  \citenamefont {Genet}, \citenamefont {Schachenmayer}, \citenamefont {Genes},
  \citenamefont {Pupillo}, \citenamefont {Samor{\`i}},\ and\ \citenamefont
  {Ebbesen}}]{orgiu2015}%
  \BibitemOpen
  \bibfield  {author} {\bibinfo {author} {\bibfnamefont {E.}~\bibnamefont
  {Orgiu}}, \bibinfo {author} {\bibfnamefont {J.}~\bibnamefont {George}},
  \bibinfo {author} {\bibfnamefont {J.~A.}\ \bibnamefont {Hutchison}}, \bibinfo
  {author} {\bibfnamefont {E.}~\bibnamefont {Devaux}}, \bibinfo {author}
  {\bibfnamefont {J.~F.}\ \bibnamefont {Dayen}}, \bibinfo {author}
  {\bibfnamefont {B.}~\bibnamefont {Doudin}}, \bibinfo {author} {\bibfnamefont
  {F.}~\bibnamefont {Stellacci}}, \bibinfo {author} {\bibfnamefont
  {C.}~\bibnamefont {Genet}}, \bibinfo {author} {\bibfnamefont
  {J.}~\bibnamefont {Schachenmayer}}, \bibinfo {author} {\bibfnamefont
  {C.}~\bibnamefont {Genes}}, \bibinfo {author} {\bibfnamefont
  {G.}~\bibnamefont {Pupillo}}, \bibinfo {author} {\bibfnamefont
  {P.}~\bibnamefont {Samor{\`i}}},\ and\ \bibinfo {author} {\bibfnamefont
  {T.~W.}\ \bibnamefont {Ebbesen}},\ }\bibfield  {title} {\bibinfo {title}
  {Conductivity in organic semiconductors hybridized with the vacuum field},\
  }\href {https://doi.org/10.1038/nmat4392} {\bibfield  {journal} {\bibinfo
  {journal} {Nat. Mater.}\ }\textbf {\bibinfo {volume} {14}},\ \bibinfo {pages}
  {1123} (\bibinfo {year} {2015})}\BibitemShut {NoStop}%
\bibitem [{\citenamefont {Nagarajan}\ \emph {et~al.}(2020)\citenamefont
  {Nagarajan}, \citenamefont {George}, \citenamefont {Thomas}, \citenamefont
  {Devaux}, \citenamefont {Chervy}, \citenamefont {Azzini}, \citenamefont
  {Joseph}, \citenamefont {Jouaiti}, \citenamefont {Hosseini}, \citenamefont
  {Kumar}, \citenamefont {Genet}, \citenamefont {Bartolo}, \citenamefont
  {Ciuti},\ and\ \citenamefont {Ebbesen}}]{nagarajan2020}%
  \BibitemOpen
  \bibfield  {author} {\bibinfo {author} {\bibfnamefont {K.}~\bibnamefont
  {Nagarajan}}, \bibinfo {author} {\bibfnamefont {J.}~\bibnamefont {George}},
  \bibinfo {author} {\bibfnamefont {A.}~\bibnamefont {Thomas}}, \bibinfo
  {author} {\bibfnamefont {E.}~\bibnamefont {Devaux}}, \bibinfo {author}
  {\bibfnamefont {T.}~\bibnamefont {Chervy}}, \bibinfo {author} {\bibfnamefont
  {S.}~\bibnamefont {Azzini}}, \bibinfo {author} {\bibfnamefont
  {K.}~\bibnamefont {Joseph}}, \bibinfo {author} {\bibfnamefont
  {A.}~\bibnamefont {Jouaiti}}, \bibinfo {author} {\bibfnamefont {M.~W.}\
  \bibnamefont {Hosseini}}, \bibinfo {author} {\bibfnamefont {A.}~\bibnamefont
  {Kumar}}, \bibinfo {author} {\bibfnamefont {C.}~\bibnamefont {Genet}},
  \bibinfo {author} {\bibfnamefont {N.}~\bibnamefont {Bartolo}}, \bibinfo
  {author} {\bibfnamefont {C.}~\bibnamefont {Ciuti}},\ and\ \bibinfo {author}
  {\bibfnamefont {T.~W.}\ \bibnamefont {Ebbesen}},\ }\bibfield  {title}
  {\bibinfo {title} {Conductivity and {{Photoconductivity}} of a p-{{Type
  Organic Semiconductor}} under {{Ultrastrong Coupling}}},\ }\href
  {https://doi.org/10.1021/acsnano.0c03496} {\bibfield  {journal} {\bibinfo
  {journal} {ACS Nano}\ }\textbf {\bibinfo {volume} {14}},\ \bibinfo {pages}
  {10219} (\bibinfo {year} {2020})}\BibitemShut {NoStop}%
\bibitem [{\citenamefont {Nasu}(2004)}]{nasu2004}%
  \BibitemOpen
  \bibfield  {author} {\bibinfo {author} {\bibfnamefont {K.}~\bibnamefont
  {Nasu}},\ }\bibfield  {title} {\bibinfo {title} {Itinerant type many-body
  theories for photo-induced structural phase transitions},\ }\href
  {https://doi.org/10.1088/0034-4885/67/9/R02} {\bibfield  {journal} {\bibinfo
  {journal} {Rep. Prog. Phys.}\ }\textbf {\bibinfo {volume} {67}},\ \bibinfo
  {pages} {1607} (\bibinfo {year} {2004})}\BibitemShut {NoStop}%
\bibitem [{\citenamefont {Yonemitsu}\ and\ \citenamefont
  {Nasu}(2008)}]{yonemitsu2008}%
  \BibitemOpen
  \bibfield  {author} {\bibinfo {author} {\bibfnamefont {K.}~\bibnamefont
  {Yonemitsu}}\ and\ \bibinfo {author} {\bibfnamefont {K.}~\bibnamefont
  {Nasu}},\ }\bibfield  {title} {\bibinfo {title} {Theory of photoinduced phase
  transitions in itinerant electron systems},\ }\href
  {https://doi.org/10.1016/j.physrep.2008.04.008} {\bibfield  {journal}
  {\bibinfo  {journal} {Phys. Rep.}\ }\textbf {\bibinfo {volume} {465}},\
  \bibinfo {pages} {1} (\bibinfo {year} {2008})}\BibitemShut {NoStop}%
\bibitem [{\citenamefont {Dressel}\ and\ \citenamefont
  {Peterseim}(2017)}]{dressel2017}%
  \BibitemOpen
  \bibfield  {author} {\bibinfo {author} {\bibfnamefont {M.}~\bibnamefont
  {Dressel}}\ and\ \bibinfo {author} {\bibfnamefont {T.}~\bibnamefont
  {Peterseim}},\ }\bibfield  {title} {\bibinfo {title} {Infrared
  {{Investigations}} of the {{Neutral-Ionic Phase Transition}} in {{TTF-CA}}
  and {{Its Dynamics}}},\ }\href {https://doi.org/10.3390/cryst7010017}
  {\bibfield  {journal} {\bibinfo  {journal} {Crystals}\ }\textbf {\bibinfo
  {volume} {7}},\ \bibinfo {pages} {17} (\bibinfo {year} {2017})}\BibitemShut
  {NoStop}%
\bibitem [{\citenamefont {Koshihara}\ \emph {et~al.}(1990)\citenamefont
  {Koshihara}, \citenamefont {Tokura}, \citenamefont {Mitani}, \citenamefont
  {Saito},\ and\ \citenamefont {Koda}}]{koshihara1990}%
  \BibitemOpen
  \bibfield  {author} {\bibinfo {author} {\bibfnamefont {S.}~\bibnamefont
  {Koshihara}}, \bibinfo {author} {\bibfnamefont {Y.}~\bibnamefont {Tokura}},
  \bibinfo {author} {\bibfnamefont {T.}~\bibnamefont {Mitani}}, \bibinfo
  {author} {\bibfnamefont {G.}~\bibnamefont {Saito}},\ and\ \bibinfo {author}
  {\bibfnamefont {T.}~\bibnamefont {Koda}},\ }\bibfield  {title} {\bibinfo
  {title} {Photoinduced valence instability in the organic molecular compound
  tetrathiafulvalene-{\emph{p}}-chloranil ({{TTF-CA}})},\ }\href
  {https://doi.org/10.1103/PhysRevB.42.6853} {\bibfield  {journal} {\bibinfo
  {journal} {Phys. Rev. B}\ }\textbf {\bibinfo {volume} {42}},\ \bibinfo
  {pages} {6853} (\bibinfo {year} {1990})}\BibitemShut {NoStop}%
\bibitem [{\citenamefont {Koshihara}\ \emph {et~al.}(1999)\citenamefont
  {Koshihara}, \citenamefont {Takahashi}, \citenamefont {Sakai}, \citenamefont
  {Tokura},\ and\ \citenamefont {Luty}}]{koshihara1999}%
  \BibitemOpen
  \bibfield  {author} {\bibinfo {author} {\bibfnamefont {S.-y.}\ \bibnamefont
  {Koshihara}}, \bibinfo {author} {\bibfnamefont {Y.}~\bibnamefont
  {Takahashi}}, \bibinfo {author} {\bibfnamefont {H.}~\bibnamefont {Sakai}},
  \bibinfo {author} {\bibfnamefont {Y.}~\bibnamefont {Tokura}},\ and\ \bibinfo
  {author} {\bibfnamefont {T.}~\bibnamefont {Luty}},\ }\bibfield  {title}
  {\bibinfo {title} {Photoinduced {{Cooperative Charge Transfer}} in
  {{Low-Dimensional Organic Crystals}}},\ }\href
  {https://doi.org/10.1021/jp984172i} {\bibfield  {journal} {\bibinfo
  {journal} {J. Phys. Chem. B}\ }\textbf {\bibinfo {volume} {103}},\ \bibinfo
  {pages} {2592} (\bibinfo {year} {1999})}\BibitemShut {NoStop}%
\bibitem [{\citenamefont {Collet}(2003)}]{collet2003}%
  \BibitemOpen
  \bibfield  {author} {\bibinfo {author} {\bibfnamefont {E.}~\bibnamefont
  {Collet}},\ }\bibfield  {title} {\bibinfo {title} {Laser-{{Induced
  Ferroelectric Structural Order}} in an {{Organic Charge-Transfer Crystal}}},\
  }\href {https://doi.org/10.1126/science.1082001} {\bibfield  {journal}
  {\bibinfo  {journal} {Science}\ }\textbf {\bibinfo {volume} {300}},\ \bibinfo
  {pages} {612} (\bibinfo {year} {2003})}\BibitemShut {NoStop}%
\bibitem [{\citenamefont {Giovannetti}\ \emph {et~al.}(2009)\citenamefont
  {Giovannetti}, \citenamefont {Kumar}, \citenamefont {Stroppa}, \citenamefont
  {{van den Brink}},\ and\ \citenamefont {Picozzi}}]{giovannetti2009}%
  \BibitemOpen
  \bibfield  {author} {\bibinfo {author} {\bibfnamefont {G.}~\bibnamefont
  {Giovannetti}}, \bibinfo {author} {\bibfnamefont {S.}~\bibnamefont {Kumar}},
  \bibinfo {author} {\bibfnamefont {A.}~\bibnamefont {Stroppa}}, \bibinfo
  {author} {\bibfnamefont {J.}~\bibnamefont {{van den Brink}}},\ and\ \bibinfo
  {author} {\bibfnamefont {S.}~\bibnamefont {Picozzi}},\ }\bibfield  {title}
  {\bibinfo {title} {Multiferroicity in {{TTF-CA Organic Molecular Crystals
  Predicted}} through {{{\emph{Ab Initio}}}} {{Calculations}}},\ }\href
  {https://doi.org/10.1103/PhysRevLett.103.266401} {\bibfield  {journal}
  {\bibinfo  {journal} {Phys. Rev. Lett.}\ }\textbf {\bibinfo {volume} {103}},\
  \bibinfo {pages} {266401} (\bibinfo {year} {2009})}\BibitemShut {NoStop}%
\bibitem [{\citenamefont {Kobayashi}\ \emph {et~al.}(2012)\citenamefont
  {Kobayashi}, \citenamefont {Horiuchi}, \citenamefont {Kumai}, \citenamefont
  {Kagawa}, \citenamefont {Murakami},\ and\ \citenamefont
  {Tokura}}]{kobayashi2012}%
  \BibitemOpen
  \bibfield  {author} {\bibinfo {author} {\bibfnamefont {K.}~\bibnamefont
  {Kobayashi}}, \bibinfo {author} {\bibfnamefont {S.}~\bibnamefont {Horiuchi}},
  \bibinfo {author} {\bibfnamefont {R.}~\bibnamefont {Kumai}}, \bibinfo
  {author} {\bibfnamefont {F.}~\bibnamefont {Kagawa}}, \bibinfo {author}
  {\bibfnamefont {Y.}~\bibnamefont {Murakami}},\ and\ \bibinfo {author}
  {\bibfnamefont {Y.}~\bibnamefont {Tokura}},\ }\bibfield  {title} {\bibinfo
  {title} {Electronic {{Ferroelectricity}} in a {{Molecular Crystal}} with
  {{Large Polarization Directing Antiparallel}} to {{Ionic Displacement}}},\
  }\href {https://doi.org/10.1103/PhysRevLett.108.237601} {\bibfield  {journal}
  {\bibinfo  {journal} {Phys. Rev. Lett.}\ }\textbf {\bibinfo {volume} {108}},\
  \bibinfo {pages} {237601} (\bibinfo {year} {2012})}\BibitemShut {NoStop}%
\bibitem [{\citenamefont {Nakamura}\ \emph {et~al.}(2017)\citenamefont
  {Nakamura}, \citenamefont {Horiuchi}, \citenamefont {Kagawa}, \citenamefont
  {Ogawa}, \citenamefont {Kurumaji}, \citenamefont {Tokura},\ and\
  \citenamefont {Kawasaki}}]{nakamura2017}%
  \BibitemOpen
  \bibfield  {author} {\bibinfo {author} {\bibfnamefont {M.}~\bibnamefont
  {Nakamura}}, \bibinfo {author} {\bibfnamefont {S.}~\bibnamefont {Horiuchi}},
  \bibinfo {author} {\bibfnamefont {F.}~\bibnamefont {Kagawa}}, \bibinfo
  {author} {\bibfnamefont {N.}~\bibnamefont {Ogawa}}, \bibinfo {author}
  {\bibfnamefont {T.}~\bibnamefont {Kurumaji}}, \bibinfo {author}
  {\bibfnamefont {Y.}~\bibnamefont {Tokura}},\ and\ \bibinfo {author}
  {\bibfnamefont {M.}~\bibnamefont {Kawasaki}},\ }\bibfield  {title} {\bibinfo
  {title} {Shift current photovoltaic effect in a ferroelectric charge-transfer
  complex},\ }\href {https://doi.org/10.1038/s41467-017-00250-y} {\bibfield
  {journal} {\bibinfo  {journal} {Nat. Commun.}\ }\textbf {\bibinfo {volume}
  {8}},\ \bibinfo {pages} {281} (\bibinfo {year} {2017})}\BibitemShut {NoStop}%
\bibitem [{\citenamefont {Suzuki}\ \emph {et~al.}(1999)\citenamefont {Suzuki},
  \citenamefont {Sakamaki}, \citenamefont {Tanimura}, \citenamefont
  {Koshihara},\ and\ \citenamefont {Tokura}}]{suzuki1999}%
  \BibitemOpen
  \bibfield  {author} {\bibinfo {author} {\bibfnamefont {T.}~\bibnamefont
  {Suzuki}}, \bibinfo {author} {\bibfnamefont {T.}~\bibnamefont {Sakamaki}},
  \bibinfo {author} {\bibfnamefont {K.}~\bibnamefont {Tanimura}}, \bibinfo
  {author} {\bibfnamefont {S.}~\bibnamefont {Koshihara}},\ and\ \bibinfo
  {author} {\bibfnamefont {Y.}~\bibnamefont {Tokura}},\ }\bibfield  {title}
  {\bibinfo {title} {Ionic-to-neutral phase transformation induced by
  photoexcitation of the charge-transfer band in
  tetrathiafulvalene-{\emph{p}}-chloranil crystals},\ }\href
  {https://doi.org/10.1103/PhysRevB.60.6191} {\bibfield  {journal} {\bibinfo
  {journal} {Phys. Rev. B}\ }\textbf {\bibinfo {volume} {60}},\ \bibinfo
  {pages} {6191} (\bibinfo {year} {1999})}\BibitemShut {NoStop}%
\bibitem [{\citenamefont {Tanimura}\ and\ \citenamefont
  {Akimoto}(2001)}]{tanimura2001}%
  \BibitemOpen
  \bibfield  {author} {\bibinfo {author} {\bibfnamefont {K.}~\bibnamefont
  {Tanimura}}\ and\ \bibinfo {author} {\bibfnamefont {I.}~\bibnamefont
  {Akimoto}},\ }\bibfield  {title} {\bibinfo {title} {Femtosecond time-resolved
  spectroscopy of photoinduced ionic-to-neutral phase transition in
  tetrathiafulvalen-{\emph{p}}-chloranil crystals},\ }\href
  {https://doi.org/10.1016/S0022-2313(01)00314-3} {\bibfield  {journal}
  {\bibinfo  {journal} {J. Lumin.}\ }\textbf {\bibinfo {volume} {94--95}},\
  \bibinfo {pages} {483} (\bibinfo {year} {2001})}\BibitemShut {NoStop}%
\bibitem [{\citenamefont {Iwai}\ \emph {et~al.}(2002)\citenamefont {Iwai},
  \citenamefont {Tanaka}, \citenamefont {Fujinuma}, \citenamefont {Kishida},
  \citenamefont {Okamoto},\ and\ \citenamefont {Tokura}}]{iwai2002}%
  \BibitemOpen
  \bibfield  {author} {\bibinfo {author} {\bibfnamefont {S.}~\bibnamefont
  {Iwai}}, \bibinfo {author} {\bibfnamefont {S.}~\bibnamefont {Tanaka}},
  \bibinfo {author} {\bibfnamefont {K.}~\bibnamefont {Fujinuma}}, \bibinfo
  {author} {\bibfnamefont {H.}~\bibnamefont {Kishida}}, \bibinfo {author}
  {\bibfnamefont {H.}~\bibnamefont {Okamoto}},\ and\ \bibinfo {author}
  {\bibfnamefont {Y.}~\bibnamefont {Tokura}},\ }\bibfield  {title} {\bibinfo
  {title} {Ultrafast {{Optical Switching}} from an {{Ionic}} to a {{Neutral
  State}} in {{Tetrathiafulvalene-}}{\emph{p}}-{{Chloranil}} ({{TTF-CA}})
  {{Observed}} in {{Femtosecond Reflection Spectroscopy}}},\ }\href
  {https://doi.org/10.1103/PhysRevLett.88.057402} {\bibfield  {journal}
  {\bibinfo  {journal} {Phys. Rev. Lett.}\ }\textbf {\bibinfo {volume} {88}},\
  \bibinfo {pages} {057402} (\bibinfo {year} {2002})}\BibitemShut {NoStop}%
\bibitem [{\citenamefont {Morimoto}\ \emph
  {et~al.}(2017{\natexlab{a}})\citenamefont {Morimoto}, \citenamefont
  {Miyamoto}, \citenamefont {Yamakawa}, \citenamefont {Terashige},
  \citenamefont {Ono}, \citenamefont {Kida},\ and\ \citenamefont
  {Okamoto}}]{morimoto2017a}%
  \BibitemOpen
  \bibfield  {author} {\bibinfo {author} {\bibfnamefont {T.}~\bibnamefont
  {Morimoto}}, \bibinfo {author} {\bibfnamefont {T.}~\bibnamefont {Miyamoto}},
  \bibinfo {author} {\bibfnamefont {H.}~\bibnamefont {Yamakawa}}, \bibinfo
  {author} {\bibfnamefont {T.}~\bibnamefont {Terashige}}, \bibinfo {author}
  {\bibfnamefont {T.}~\bibnamefont {Ono}}, \bibinfo {author} {\bibfnamefont
  {N.}~\bibnamefont {Kida}},\ and\ \bibinfo {author} {\bibfnamefont
  {H.}~\bibnamefont {Okamoto}},\ }\bibfield  {title} {\bibinfo {title}
  {Terahertz-{{Field-Induced Large Macroscopic Polarization}} and {{Domain-Wall
  Dynamics}} in an {{Organic Molecular Dielectric}}},\ }\href
  {https://doi.org/10.1103/PhysRevLett.118.107602} {\bibfield  {journal}
  {\bibinfo  {journal} {Phys. Rev. Lett.}\ }\textbf {\bibinfo {volume} {118}},\
  \bibinfo {pages} {107602} (\bibinfo {year} {2017}{\natexlab{a}})}\BibitemShut
  {NoStop}%
\bibitem [{\citenamefont {Kinoshita}\ \emph {et~al.}(2020)\citenamefont
  {Kinoshita}, \citenamefont {Kida}, \citenamefont {Magasaki}, \citenamefont
  {Morimoto}, \citenamefont {Terashige}, \citenamefont {Miyamoto},\ and\
  \citenamefont {Okamoto}}]{kinoshita2020}%
  \BibitemOpen
  \bibfield  {author} {\bibinfo {author} {\bibfnamefont {Y.}~\bibnamefont
  {Kinoshita}}, \bibinfo {author} {\bibfnamefont {N.}~\bibnamefont {Kida}},
  \bibinfo {author} {\bibfnamefont {Y.}~\bibnamefont {Magasaki}}, \bibinfo
  {author} {\bibfnamefont {T.}~\bibnamefont {Morimoto}}, \bibinfo {author}
  {\bibfnamefont {T.}~\bibnamefont {Terashige}}, \bibinfo {author}
  {\bibfnamefont {T.}~\bibnamefont {Miyamoto}},\ and\ \bibinfo {author}
  {\bibfnamefont {H.}~\bibnamefont {Okamoto}},\ }\bibfield  {title} {\bibinfo
  {title} {Strong {{Terahertz Radiation}} via {{Rapid Polarization Reduction}}
  in {{Photoinduced Ionic-To-Neutral Transition}} in
  {{Tetrathiafulvalene-}}{\emph{p}}-{{Chloranil}}},\ }\href
  {https://doi.org/10.1103/PhysRevLett.124.057402} {\bibfield  {journal}
  {\bibinfo  {journal} {Phys. Rev. Lett.}\ }\textbf {\bibinfo {volume} {124}},\
  \bibinfo {pages} {057402} (\bibinfo {year} {2020})}\BibitemShut {NoStop}%
\bibitem [{\citenamefont {Yonemitsu}(2004{\natexlab{a}})}]{yonemitsu2004b}%
  \BibitemOpen
  \bibfield  {author} {\bibinfo {author} {\bibfnamefont {K.}~\bibnamefont
  {Yonemitsu}},\ }\bibfield  {title} {\bibinfo {title} {Phase {{Transition}} in
  a {{One-dimensional Extended Peierls}}\textendash{{Hubbard Model}} with a
  {{Pulse}} of {{Oscillating Electric Field}}: {{III}}. {{Interference Caused}}
  by a {{Double Pulse}}},\ }\href {https://doi.org/10.1143/JPSJ.73.2887}
  {\bibfield  {journal} {\bibinfo  {journal} {J. Phys. Soc. Jpn.}\ }\textbf
  {\bibinfo {volume} {73}},\ \bibinfo {pages} {2887} (\bibinfo {year}
  {2004}{\natexlab{a}})}\BibitemShut {NoStop}%
\bibitem [{\citenamefont {Iwai}\ \emph {et~al.}(2006)\citenamefont {Iwai},
  \citenamefont {Ishige}, \citenamefont {Tanaka}, \citenamefont {Okimoto},
  \citenamefont {Tokura},\ and\ \citenamefont {Okamoto}}]{iwai2006a}%
  \BibitemOpen
  \bibfield  {author} {\bibinfo {author} {\bibfnamefont {S.}~\bibnamefont
  {Iwai}}, \bibinfo {author} {\bibfnamefont {Y.}~\bibnamefont {Ishige}},
  \bibinfo {author} {\bibfnamefont {S.}~\bibnamefont {Tanaka}}, \bibinfo
  {author} {\bibfnamefont {Y.}~\bibnamefont {Okimoto}}, \bibinfo {author}
  {\bibfnamefont {Y.}~\bibnamefont {Tokura}},\ and\ \bibinfo {author}
  {\bibfnamefont {H.}~\bibnamefont {Okamoto}},\ }\bibfield  {title} {\bibinfo
  {title} {Coherent {{Control}} of {{Charge}} and {{Lattice Dynamics}} in a
  {{Photoinduced Neutral-to-Ionic Transition}} of a {{Charge-Transfer
  Compound}}},\ }\href {https://doi.org/10.1103/PhysRevLett.96.057403}
  {\bibfield  {journal} {\bibinfo  {journal} {Phys. Rev. Lett.}\ }\textbf
  {\bibinfo {volume} {96}},\ \bibinfo {pages} {057403} (\bibinfo {year}
  {2006})}\BibitemShut {NoStop}%
\bibitem [{\citenamefont {Miyamoto}\ \emph {et~al.}(2013)\citenamefont
  {Miyamoto}, \citenamefont {Yada}, \citenamefont {Yamakawa},\ and\
  \citenamefont {Okamoto}}]{miyamoto2013a}%
  \BibitemOpen
  \bibfield  {author} {\bibinfo {author} {\bibfnamefont {T.}~\bibnamefont
  {Miyamoto}}, \bibinfo {author} {\bibfnamefont {H.}~\bibnamefont {Yada}},
  \bibinfo {author} {\bibfnamefont {H.}~\bibnamefont {Yamakawa}},\ and\
  \bibinfo {author} {\bibfnamefont {H.}~\bibnamefont {Okamoto}},\ }\bibfield
  {title} {\bibinfo {title} {Ultrafast modulation of polarization amplitude by
  terahertz fields in electronic-type organic ferroelectrics},\ }\href
  {https://doi.org/10.1038/ncomms3586} {\bibfield  {journal} {\bibinfo
  {journal} {Nat. Commun.}\ }\textbf {\bibinfo {volume} {4}},\ \bibinfo {pages}
  {2586} (\bibinfo {year} {2013})}\BibitemShut {NoStop}%
\bibitem [{\citenamefont {Okamoto}\ \emph {et~al.}(2004)\citenamefont
  {Okamoto}, \citenamefont {Ishige}, \citenamefont {Tanaka}, \citenamefont
  {Kishida}, \citenamefont {Iwai},\ and\ \citenamefont {Tokura}}]{okamoto2004}%
  \BibitemOpen
  \bibfield  {author} {\bibinfo {author} {\bibfnamefont {H.}~\bibnamefont
  {Okamoto}}, \bibinfo {author} {\bibfnamefont {Y.}~\bibnamefont {Ishige}},
  \bibinfo {author} {\bibfnamefont {S.}~\bibnamefont {Tanaka}}, \bibinfo
  {author} {\bibfnamefont {H.}~\bibnamefont {Kishida}}, \bibinfo {author}
  {\bibfnamefont {S.}~\bibnamefont {Iwai}},\ and\ \bibinfo {author}
  {\bibfnamefont {Y.}~\bibnamefont {Tokura}},\ }\bibfield  {title} {\bibinfo
  {title} {Photoinduced phase transition in
  tetrathiafulvalene-{\emph{p}}-chloranil observed in femtosecond reflection
  spectroscopy},\ }\href {https://doi.org/10.1103/PhysRevB.70.165202}
  {\bibfield  {journal} {\bibinfo  {journal} {Phys. Rev. B}\ }\textbf {\bibinfo
  {volume} {70}},\ \bibinfo {pages} {165202} (\bibinfo {year}
  {2004})}\BibitemShut {NoStop}%
\bibitem [{\citenamefont {Tanimura}(2004)}]{tanimura2004}%
  \BibitemOpen
  \bibfield  {author} {\bibinfo {author} {\bibfnamefont {K.}~\bibnamefont
  {Tanimura}},\ }\bibfield  {title} {\bibinfo {title} {Femtosecond
  time-resolved reflection spectroscopy of photoinduced ionic-neutral phase
  transition in {{TTF-CA}} crystals},\ }\href
  {https://doi.org/10.1103/PhysRevB.70.144112} {\bibfield  {journal} {\bibinfo
  {journal} {Phys. Rev. B}\ }\textbf {\bibinfo {volume} {70}},\ \bibinfo
  {pages} {144112} (\bibinfo {year} {2004})}\BibitemShut {NoStop}%
\bibitem [{\citenamefont {Uemura}\ and\ \citenamefont
  {Okamoto}(2010)}]{uemura2010}%
  \BibitemOpen
  \bibfield  {author} {\bibinfo {author} {\bibfnamefont {H.}~\bibnamefont
  {Uemura}}\ and\ \bibinfo {author} {\bibfnamefont {H.}~\bibnamefont
  {Okamoto}},\ }\bibfield  {title} {\bibinfo {title} {Direct {{Detection}} of
  the {{Ultrafast Response}} of {{Charges}} and {{Molecules}} in the
  {{Photoinduced Neutral-to-Ionic Transition}} of the {{Organic
  Tetrathiafulvalene-}}{\emph{p}}-{{Chloranil Solid}}},\ }\href
  {https://doi.org/10.1103/PhysRevLett.105.258302} {\bibfield  {journal}
  {\bibinfo  {journal} {Phys. Rev. Lett.}\ }\textbf {\bibinfo {volume} {105}},\
  \bibinfo {pages} {258302} (\bibinfo {year} {2010})}\BibitemShut {NoStop}%
\bibitem [{\citenamefont {Matsubara}\ \emph {et~al.}(2011)\citenamefont
  {Matsubara}, \citenamefont {Okimoto}, \citenamefont {Yoshida}, \citenamefont
  {Ishikawa}, \citenamefont {Koshihara},\ and\ \citenamefont
  {Onda}}]{matsubara2011}%
  \BibitemOpen
  \bibfield  {author} {\bibinfo {author} {\bibfnamefont {Y.}~\bibnamefont
  {Matsubara}}, \bibinfo {author} {\bibfnamefont {Y.}~\bibnamefont {Okimoto}},
  \bibinfo {author} {\bibfnamefont {T.}~\bibnamefont {Yoshida}}, \bibinfo
  {author} {\bibfnamefont {T.}~\bibnamefont {Ishikawa}}, \bibinfo {author}
  {\bibfnamefont {S.-y.}\ \bibnamefont {Koshihara}},\ and\ \bibinfo {author}
  {\bibfnamefont {K.}~\bibnamefont {Onda}},\ }\bibfield  {title} {\bibinfo
  {title} {Photoinduced {{Neutral-to-Ionic Phase Transition}} in
  {{Tetrathiafulvalene-}} {\emph{p}} -chloranil {{Studied}} by {{Time-Resolved
  Vibrational Spectroscopy}}},\ }\href {https://doi.org/10.1143/JPSJ.80.124711}
  {\bibfield  {journal} {\bibinfo  {journal} {J. Phys. Soc. Jpn.}\ }\textbf
  {\bibinfo {volume} {80}},\ \bibinfo {pages} {124711} (\bibinfo {year}
  {2011})}\BibitemShut {NoStop}%
\bibitem [{\citenamefont {Huai}\ \emph {et~al.}(2000)\citenamefont {Huai},
  \citenamefont {Zheng},\ and\ \citenamefont {Nasu}}]{huai2000}%
  \BibitemOpen
  \bibfield  {author} {\bibinfo {author} {\bibfnamefont {P.}~\bibnamefont
  {Huai}}, \bibinfo {author} {\bibfnamefont {H.}~\bibnamefont {Zheng}},\ and\
  \bibinfo {author} {\bibfnamefont {K.}~\bibnamefont {Nasu}},\ }\bibfield
  {title} {\bibinfo {title} {Theory for {{Photoinduced Ionic-Neutral Structural
  Phase Transition}} in {{Quasi One-Dimensional Organic Molecular Crystal
  TTF-CA}}},\ }\href {https://doi.org/10.1143/JPSJ.69.1788} {\bibfield
  {journal} {\bibinfo  {journal} {J. Phys. Soc. Jpn.}\ }\textbf {\bibinfo
  {volume} {69}},\ \bibinfo {pages} {1788} (\bibinfo {year}
  {2000})}\BibitemShut {NoStop}%
\bibitem [{\citenamefont {Luty}\ \emph {et~al.}(2002)\citenamefont {Luty},
  \citenamefont {Cailleau}, \citenamefont {Koshihara}, \citenamefont {Collet},
  \citenamefont {Takesada}, \citenamefont {{Lem{\'e}e-Cailleau}}, \citenamefont
  {Cointe}, \citenamefont {Nagaosa}, \citenamefont {Tokura}, \citenamefont
  {Zienkiewicz},\ and\ \citenamefont {Ouladdiaf}}]{luty2002}%
  \BibitemOpen
  \bibfield  {author} {\bibinfo {author} {\bibfnamefont {T.}~\bibnamefont
  {Luty}}, \bibinfo {author} {\bibfnamefont {H.}~\bibnamefont {Cailleau}},
  \bibinfo {author} {\bibfnamefont {S.}~\bibnamefont {Koshihara}}, \bibinfo
  {author} {\bibfnamefont {E.}~\bibnamefont {Collet}}, \bibinfo {author}
  {\bibfnamefont {M.}~\bibnamefont {Takesada}}, \bibinfo {author}
  {\bibfnamefont {M.~H.}\ \bibnamefont {{Lem{\'e}e-Cailleau}}}, \bibinfo
  {author} {\bibfnamefont {M.~B.-L.}\ \bibnamefont {Cointe}}, \bibinfo {author}
  {\bibfnamefont {N.}~\bibnamefont {Nagaosa}}, \bibinfo {author} {\bibfnamefont
  {Y.}~\bibnamefont {Tokura}}, \bibinfo {author} {\bibfnamefont
  {E.}~\bibnamefont {Zienkiewicz}},\ and\ \bibinfo {author} {\bibfnamefont
  {B.}~\bibnamefont {Ouladdiaf}},\ }\bibfield  {title} {\bibinfo {title}
  {Static and dynamic order of cooperative multi-electron transfer},\ }\href
  {https://doi.org/10.1209/epl/i2002-00149-4} {\bibfield  {journal} {\bibinfo
  {journal} {EPL}\ }\textbf {\bibinfo {volume} {59}},\ \bibinfo {pages} {619}
  (\bibinfo {year} {2002})}\BibitemShut {NoStop}%
\bibitem [{\citenamefont {Yonemitsu}(2004{\natexlab{b}})}]{yonemitsu2004}%
  \BibitemOpen
  \bibfield  {author} {\bibinfo {author} {\bibfnamefont {K.}~\bibnamefont
  {Yonemitsu}},\ }\bibfield  {title} {\bibinfo {title} {Phase {{Transition}} in
  a {{One-dimensional Extended Peierls}}\textendash{{Hubbard Model}} with a
  {{Pulse}} of {{Oscillating Electric Field}}: {{I}}. {{Threshold Behavior}} in
  {{Ionic-to-Neutral Transition}}},\ }\href
  {https://doi.org/10.1143/JPSJ.73.2868} {\bibfield  {journal} {\bibinfo
  {journal} {J. Phys. Soc. Jpn.}\ }\textbf {\bibinfo {volume} {73}},\ \bibinfo
  {pages} {2868} (\bibinfo {year} {2004}{\natexlab{b}})}\BibitemShut {NoStop}%
\bibitem [{\citenamefont {Yonemitsu}(2004{\natexlab{c}})}]{yonemitsu2004a}%
  \BibitemOpen
  \bibfield  {author} {\bibinfo {author} {\bibfnamefont {K.}~\bibnamefont
  {Yonemitsu}},\ }\bibfield  {title} {\bibinfo {title} {Phase {{Transition}} in
  a {{One-dimensional Extended Peierls}}\textendash{{Hubbard Model}} with a
  {{Pulse}} of {{Oscillating Electric Field}}: {{II}}. {{Linear Behavior}} in
  {{Neutral-to-Ionic Transition}}},\ }\href
  {https://doi.org/10.1143/JPSJ.73.2879} {\bibfield  {journal} {\bibinfo
  {journal} {J. Phys. Soc. Jpn.}\ }\textbf {\bibinfo {volume} {73}},\ \bibinfo
  {pages} {2879} (\bibinfo {year} {2004}{\natexlab{c}})}\BibitemShut {NoStop}%
\bibitem [{\citenamefont {Soos}\ and\ \citenamefont
  {Painelli}(2007)}]{soos2007}%
  \BibitemOpen
  \bibfield  {author} {\bibinfo {author} {\bibfnamefont {Z.~G.}\ \bibnamefont
  {Soos}}\ and\ \bibinfo {author} {\bibfnamefont {A.}~\bibnamefont
  {Painelli}},\ }\bibfield  {title} {\bibinfo {title} {Metastable domains and
  potential energy surfaces in organic charge-transfer salts with neutral-ionic
  phase transitions},\ }\href {https://doi.org/10.1103/PhysRevB.75.155119}
  {\bibfield  {journal} {\bibinfo  {journal} {Phys. Rev. B}\ }\textbf {\bibinfo
  {volume} {75}},\ \bibinfo {pages} {155119} (\bibinfo {year}
  {2007})}\BibitemShut {NoStop}%
\bibitem [{\citenamefont {Peterseim}\ \emph {et~al.}(2015)\citenamefont
  {Peterseim}, \citenamefont {Haremski},\ and\ \citenamefont
  {Dressel}}]{peterseim2015}%
  \BibitemOpen
  \bibfield  {author} {\bibinfo {author} {\bibfnamefont {T.}~\bibnamefont
  {Peterseim}}, \bibinfo {author} {\bibfnamefont {P.}~\bibnamefont
  {Haremski}},\ and\ \bibinfo {author} {\bibfnamefont {M.}~\bibnamefont
  {Dressel}},\ }\bibfield  {title} {\bibinfo {title} {Random-walk annihilation
  process of photo-induced neutral-ionic domain walls in {{TTF-CA}}},\ }\href
  {https://doi.org/10.1209/0295-5075/109/67003} {\bibfield  {journal} {\bibinfo
   {journal} {EPL}\ }\textbf {\bibinfo {volume} {109}},\ \bibinfo {pages}
  {67003} (\bibinfo {year} {2015})}\BibitemShut {NoStop}%
\bibitem [{\citenamefont {Kishida}\ \emph {et~al.}(2009)\citenamefont
  {Kishida}, \citenamefont {Takamatsu}, \citenamefont {Fujinuma},\ and\
  \citenamefont {Okamoto}}]{kishida2009}%
  \BibitemOpen
  \bibfield  {author} {\bibinfo {author} {\bibfnamefont {H.}~\bibnamefont
  {Kishida}}, \bibinfo {author} {\bibfnamefont {H.}~\bibnamefont {Takamatsu}},
  \bibinfo {author} {\bibfnamefont {K.}~\bibnamefont {Fujinuma}},\ and\
  \bibinfo {author} {\bibfnamefont {H.}~\bibnamefont {Okamoto}},\ }\bibfield
  {title} {\bibinfo {title} {Ferroelectric nature and real-space observations
  of domain motions in the organic charge-transfer compound
  tetrathiafulvalene-{\emph{p}}-chloranil},\ }\href
  {https://doi.org/10.1103/PhysRevB.80.205201} {\bibfield  {journal} {\bibinfo
  {journal} {Phys. Rev. B}\ }\textbf {\bibinfo {volume} {80}},\ \bibinfo
  {pages} {205201} (\bibinfo {year} {2009})}\BibitemShut {NoStop}%
\bibitem [{\citenamefont {{Lem{\'e}e-Cailleau}}\ \emph
  {et~al.}(1997)\citenamefont {{Lem{\'e}e-Cailleau}}, \citenamefont
  {Le~Cointe}, \citenamefont {Cailleau}, \citenamefont {Luty}, \citenamefont
  {Moussa}, \citenamefont {Roos}, \citenamefont {Brinkmann}, \citenamefont
  {Toudic}, \citenamefont {Ayache},\ and\ \citenamefont
  {Karl}}]{lemee-cailleau1997}%
  \BibitemOpen
  \bibfield  {author} {\bibinfo {author} {\bibfnamefont {M.~H.}\ \bibnamefont
  {{Lem{\'e}e-Cailleau}}}, \bibinfo {author} {\bibfnamefont {M.}~\bibnamefont
  {Le~Cointe}}, \bibinfo {author} {\bibfnamefont {H.}~\bibnamefont {Cailleau}},
  \bibinfo {author} {\bibfnamefont {T.}~\bibnamefont {Luty}}, \bibinfo {author}
  {\bibfnamefont {F.}~\bibnamefont {Moussa}}, \bibinfo {author} {\bibfnamefont
  {J.}~\bibnamefont {Roos}}, \bibinfo {author} {\bibfnamefont {D.}~\bibnamefont
  {Brinkmann}}, \bibinfo {author} {\bibfnamefont {B.}~\bibnamefont {Toudic}},
  \bibinfo {author} {\bibfnamefont {C.}~\bibnamefont {Ayache}},\ and\ \bibinfo
  {author} {\bibfnamefont {N.}~\bibnamefont {Karl}},\ }\bibfield  {title}
  {\bibinfo {title} {Thermodynamics of the {{Neutral-to-Ionic Transition}} as
  {{Condensation}} and {{Crystallization}} of {{Charge-Transfer
  Excitations}}},\ }\href {https://doi.org/10.1103/PhysRevLett.79.1690}
  {\bibfield  {journal} {\bibinfo  {journal} {Phys. Rev. Lett.}\ }\textbf
  {\bibinfo {volume} {79}},\ \bibinfo {pages} {1690} (\bibinfo {year}
  {1997})}\BibitemShut {NoStop}%
\bibitem [{\citenamefont {Gu{\'e}rin}\ \emph {et~al.}(2010)\citenamefont
  {Gu{\'e}rin}, \citenamefont {H{\'e}bert}, \citenamefont {{Buron-Le Cointe}},
  \citenamefont {Adachi}, \citenamefont {Koshihara}, \citenamefont {Cailleau},\
  and\ \citenamefont {Collet}}]{guerin2010}%
  \BibitemOpen
  \bibfield  {author} {\bibinfo {author} {\bibfnamefont {L.}~\bibnamefont
  {Gu{\'e}rin}}, \bibinfo {author} {\bibfnamefont {J.}~\bibnamefont
  {H{\'e}bert}}, \bibinfo {author} {\bibfnamefont {M.}~\bibnamefont {{Buron-Le
  Cointe}}}, \bibinfo {author} {\bibfnamefont {S.-i.}\ \bibnamefont {Adachi}},
  \bibinfo {author} {\bibfnamefont {S.-y.}\ \bibnamefont {Koshihara}}, \bibinfo
  {author} {\bibfnamefont {H.}~\bibnamefont {Cailleau}},\ and\ \bibinfo
  {author} {\bibfnamefont {E.}~\bibnamefont {Collet}},\ }\bibfield  {title}
  {\bibinfo {title} {Capturing {{One-Dimensional Precursors}} of a
  {{Photoinduced Transformation}} in a {{Material}}},\ }\href
  {https://doi.org/10.1103/PhysRevLett.105.246101} {\bibfield  {journal}
  {\bibinfo  {journal} {Phys. Rev. Lett.}\ }\textbf {\bibinfo {volume} {105}},\
  \bibinfo {pages} {246101} (\bibinfo {year} {2010})}\BibitemShut {NoStop}%
\bibitem [{\citenamefont {Cavatorta}\ \emph {et~al.}(2015)\citenamefont
  {Cavatorta}, \citenamefont {Painelli},\ and\ \citenamefont
  {Soos}}]{cavatorta2015}%
  \BibitemOpen
  \bibfield  {author} {\bibinfo {author} {\bibfnamefont {L.}~\bibnamefont
  {Cavatorta}}, \bibinfo {author} {\bibfnamefont {A.}~\bibnamefont
  {Painelli}},\ and\ \bibinfo {author} {\bibfnamefont {Z.~G.}\ \bibnamefont
  {Soos}},\ }\bibfield  {title} {\bibinfo {title} {Coherent excitations at the
  neutral-ionic transition: {{Femtosecond}} dynamics on diabatic potential
  energy surfaces},\ }\href {https://doi.org/10.1103/PhysRevB.91.174301}
  {\bibfield  {journal} {\bibinfo  {journal} {Phys. Rev. B}\ }\textbf {\bibinfo
  {volume} {91}},\ \bibinfo {pages} {174301} (\bibinfo {year}
  {2015})}\BibitemShut {NoStop}%
\bibitem [{\citenamefont {{M{\"o}hr-Vorobeva}}\ \emph
  {et~al.}(2011)\citenamefont {{M{\"o}hr-Vorobeva}}, \citenamefont {Johnson},
  \citenamefont {Beaud}, \citenamefont {Staub}, \citenamefont {De~Souza},
  \citenamefont {Milne}, \citenamefont {Ingold}, \citenamefont {Demsar},
  \citenamefont {Schaefer},\ and\ \citenamefont {Titov}}]{mohr-vorobeva2011}%
  \BibitemOpen
  \bibfield  {author} {\bibinfo {author} {\bibfnamefont {E.}~\bibnamefont
  {{M{\"o}hr-Vorobeva}}}, \bibinfo {author} {\bibfnamefont {S.~L.}\
  \bibnamefont {Johnson}}, \bibinfo {author} {\bibfnamefont {P.}~\bibnamefont
  {Beaud}}, \bibinfo {author} {\bibfnamefont {U.}~\bibnamefont {Staub}},
  \bibinfo {author} {\bibfnamefont {R.}~\bibnamefont {De~Souza}}, \bibinfo
  {author} {\bibfnamefont {C.}~\bibnamefont {Milne}}, \bibinfo {author}
  {\bibfnamefont {G.}~\bibnamefont {Ingold}}, \bibinfo {author} {\bibfnamefont
  {J.}~\bibnamefont {Demsar}}, \bibinfo {author} {\bibfnamefont
  {H.}~\bibnamefont {Schaefer}},\ and\ \bibinfo {author} {\bibfnamefont
  {A.}~\bibnamefont {Titov}},\ }\bibfield  {title} {\bibinfo {title}
  {Nonthermal {{Melting}} of a {{Charge Density Wave}} in
  {{TiSe}}{\textsubscript{2}}},\ }\href
  {https://doi.org/10.1103/PhysRevLett.107.036403} {\bibfield  {journal}
  {\bibinfo  {journal} {Phys. Rev. Lett.}\ }\textbf {\bibinfo {volume} {107}},\
  \bibinfo {pages} {036403} (\bibinfo {year} {2011})}\BibitemShut {NoStop}%
\bibitem [{\citenamefont {Porer}\ \emph {et~al.}(2014)\citenamefont {Porer},
  \citenamefont {Leierseder}, \citenamefont {M{\'e}nard}, \citenamefont
  {Dachraoui}, \citenamefont {Mouchliadis}, \citenamefont {Perakis},
  \citenamefont {Heinzmann}, \citenamefont {Demsar}, \citenamefont
  {Rossnagel},\ and\ \citenamefont {Huber}}]{porer2014}%
  \BibitemOpen
  \bibfield  {author} {\bibinfo {author} {\bibfnamefont {M.}~\bibnamefont
  {Porer}}, \bibinfo {author} {\bibfnamefont {U.}~\bibnamefont {Leierseder}},
  \bibinfo {author} {\bibfnamefont {J.-M.}\ \bibnamefont {M{\'e}nard}},
  \bibinfo {author} {\bibfnamefont {H.}~\bibnamefont {Dachraoui}}, \bibinfo
  {author} {\bibfnamefont {L.}~\bibnamefont {Mouchliadis}}, \bibinfo {author}
  {\bibfnamefont {I.~E.}\ \bibnamefont {Perakis}}, \bibinfo {author}
  {\bibfnamefont {U.}~\bibnamefont {Heinzmann}}, \bibinfo {author}
  {\bibfnamefont {J.}~\bibnamefont {Demsar}}, \bibinfo {author} {\bibfnamefont
  {K.}~\bibnamefont {Rossnagel}},\ and\ \bibinfo {author} {\bibfnamefont
  {R.}~\bibnamefont {Huber}},\ }\bibfield  {title} {\bibinfo {title}
  {Non-thermal separation of electronic and structural orders in a persisting
  charge density wave},\ }\href {https://doi.org/10.1038/nmat4042} {\bibfield
  {journal} {\bibinfo  {journal} {Nat. Mater.}\ }\textbf {\bibinfo {volume}
  {13}},\ \bibinfo {pages} {857} (\bibinfo {year} {2014})}\BibitemShut
  {NoStop}%
\bibitem [{\citenamefont {Sch{\"u}ler}\ \emph {et~al.}(2018)\citenamefont
  {Sch{\"u}ler}, \citenamefont {Murakami},\ and\ \citenamefont
  {Werner}}]{schuler2018}%
  \BibitemOpen
  \bibfield  {author} {\bibinfo {author} {\bibfnamefont {M.}~\bibnamefont
  {Sch{\"u}ler}}, \bibinfo {author} {\bibfnamefont {Y.}~\bibnamefont
  {Murakami}},\ and\ \bibinfo {author} {\bibfnamefont {P.}~\bibnamefont
  {Werner}},\ }\bibfield  {title} {\bibinfo {title} {Nonthermal switching of
  charge order: {{Dynamical}} slowing down and optimal control},\ }\href
  {https://doi.org/10.1103/PhysRevB.97.155136} {\bibfield  {journal} {\bibinfo
  {journal} {Phys. Rev. B}\ }\textbf {\bibinfo {volume} {97}},\ \bibinfo
  {pages} {155136} (\bibinfo {year} {2018})}\BibitemShut {NoStop}%
\bibitem [{\citenamefont {Zak}(1989)}]{zak1989}%
  \BibitemOpen
  \bibfield  {author} {\bibinfo {author} {\bibfnamefont {J.}~\bibnamefont
  {Zak}},\ }\bibfield  {title} {\bibinfo {title} {Berry's phase for energy
  bands in solids},\ }\href {https://doi.org/10.1103/PhysRevLett.62.2747}
  {\bibfield  {journal} {\bibinfo  {journal} {Phys. Rev. Lett.}\ }\textbf
  {\bibinfo {volume} {62}},\ \bibinfo {pages} {2747} (\bibinfo {year}
  {1989})}\BibitemShut {NoStop}%
\bibitem [{\citenamefont {Resta}(2002)}]{resta2002}%
  \BibitemOpen
  \bibfield  {author} {\bibinfo {author} {\bibfnamefont {R.}~\bibnamefont
  {Resta}},\ }\bibfield  {title} {\bibinfo {title} {Why are insulators
  insulating and metals conducting?},\ }\href
  {https://doi.org/10.1088/0953-8984/14/20/201} {\bibfield  {journal} {\bibinfo
   {journal} {J. Phys. Condens. Matter}\ }\textbf {\bibinfo {volume} {14}},\
  \bibinfo {pages} {33} (\bibinfo {year} {2002})}\BibitemShut {NoStop}%
\bibitem [{\citenamefont {Torrance}\ \emph
  {et~al.}(1981{\natexlab{a}})\citenamefont {Torrance}, \citenamefont
  {Girlando}, \citenamefont {Mayerle}, \citenamefont {Crowley}, \citenamefont
  {Lee}, \citenamefont {Batail},\ and\ \citenamefont {LaPlaca}}]{torrance1981}%
  \BibitemOpen
  \bibfield  {author} {\bibinfo {author} {\bibfnamefont {J.~B.}\ \bibnamefont
  {Torrance}}, \bibinfo {author} {\bibfnamefont {A.}~\bibnamefont {Girlando}},
  \bibinfo {author} {\bibfnamefont {J.~J.}\ \bibnamefont {Mayerle}}, \bibinfo
  {author} {\bibfnamefont {J.~I.}\ \bibnamefont {Crowley}}, \bibinfo {author}
  {\bibfnamefont {V.~Y.}\ \bibnamefont {Lee}}, \bibinfo {author} {\bibfnamefont
  {P.}~\bibnamefont {Batail}},\ and\ \bibinfo {author} {\bibfnamefont {S.~J.}\
  \bibnamefont {LaPlaca}},\ }\bibfield  {title} {\bibinfo {title} {Anomalous
  {{Nature}} of {{Neutral-to-Ionic Phase Transition}} in
  {{Tetrathiafulvalene-Chloranil}}},\ }\href
  {https://doi.org/10.1103/PhysRevLett.47.1747} {\bibfield  {journal} {\bibinfo
   {journal} {Phys. Rev. Lett.}\ }\textbf {\bibinfo {volume} {47}},\ \bibinfo
  {pages} {1747} (\bibinfo {year} {1981}{\natexlab{a}})}\BibitemShut {NoStop}%
\bibitem [{\citenamefont {Torrance}\ \emph
  {et~al.}(1981{\natexlab{b}})\citenamefont {Torrance}, \citenamefont
  {Vazquez}, \citenamefont {Mayerle},\ and\ \citenamefont
  {Lee}}]{torrance1981a}%
  \BibitemOpen
  \bibfield  {author} {\bibinfo {author} {\bibfnamefont {J.~B.}\ \bibnamefont
  {Torrance}}, \bibinfo {author} {\bibfnamefont {J.~E.}\ \bibnamefont
  {Vazquez}}, \bibinfo {author} {\bibfnamefont {J.~J.}\ \bibnamefont
  {Mayerle}},\ and\ \bibinfo {author} {\bibfnamefont {V.~Y.}\ \bibnamefont
  {Lee}},\ }\bibfield  {title} {\bibinfo {title} {Discovery of a
  {{Neutral-to-Ionic Phase Transition}} in {{Organic Materials}}},\ }\href
  {https://doi.org/10.1103/PhysRevLett.46.253} {\bibfield  {journal} {\bibinfo
  {journal} {Phys. Rev. Lett.}\ }\textbf {\bibinfo {volume} {46}},\ \bibinfo
  {pages} {253} (\bibinfo {year} {1981}{\natexlab{b}})}\BibitemShut {NoStop}%
\bibitem [{\citenamefont {Soos}(1974)}]{soos1974}%
  \BibitemOpen
  \bibfield  {author} {\bibinfo {author} {\bibfnamefont {Z.~G.}\ \bibnamefont
  {Soos}},\ }\bibfield  {title} {\bibinfo {title} {Theory of
  {$\pi$}-{{Molecular Charge-Transfer Crystals}}},\ }\href
  {https://doi.org/10.1146/annurev.pc.25.100174.001005} {\bibfield  {journal}
  {\bibinfo  {journal} {Annu. Rev. Phys. Chem.}\ }\textbf {\bibinfo {volume}
  {25}},\ \bibinfo {pages} {121} (\bibinfo {year} {1974})}\BibitemShut
  {NoStop}%
\bibitem [{\citenamefont {Bruinsma}\ \emph {et~al.}(1983)\citenamefont
  {Bruinsma}, \citenamefont {Bak},\ and\ \citenamefont
  {Torrance}}]{bruinsma1983}%
  \BibitemOpen
  \bibfield  {author} {\bibinfo {author} {\bibfnamefont {R.}~\bibnamefont
  {Bruinsma}}, \bibinfo {author} {\bibfnamefont {P.}~\bibnamefont {Bak}},\ and\
  \bibinfo {author} {\bibfnamefont {J.~B.}\ \bibnamefont {Torrance}},\
  }\bibfield  {title} {\bibinfo {title} {Neutral-ionic transitions in organic
  mixed-stack compounds},\ }\href {https://doi.org/10.1103/PhysRevB.27.456}
  {\bibfield  {journal} {\bibinfo  {journal} {Phys. Rev. B}\ }\textbf {\bibinfo
  {volume} {27}},\ \bibinfo {pages} {456} (\bibinfo {year} {1983})}\BibitemShut
  {NoStop}%
\bibitem [{\citenamefont {Avignon}\ \emph {et~al.}(1986)\citenamefont
  {Avignon}, \citenamefont {Balseiro}, \citenamefont {Proetto},\ and\
  \citenamefont {Alascio}}]{avignon1986}%
  \BibitemOpen
  \bibfield  {author} {\bibinfo {author} {\bibfnamefont {M.}~\bibnamefont
  {Avignon}}, \bibinfo {author} {\bibfnamefont {C.~A.}\ \bibnamefont
  {Balseiro}}, \bibinfo {author} {\bibfnamefont {C.~R.}\ \bibnamefont
  {Proetto}},\ and\ \bibinfo {author} {\bibfnamefont {B.}~\bibnamefont
  {Alascio}},\ }\bibfield  {title} {\bibinfo {title} {Neutral-ionic transition
  and dimerization in organic mixed-stack compounds},\ }\href
  {https://doi.org/10.1103/PhysRevB.33.205} {\bibfield  {journal} {\bibinfo
  {journal} {Phys. Rev. B}\ }\textbf {\bibinfo {volume} {33}},\ \bibinfo
  {pages} {205} (\bibinfo {year} {1986})}\BibitemShut {NoStop}%
\bibitem [{\citenamefont {{Iizuka-Sakano}}\ and\ \citenamefont
  {Toyozawa}(1996)}]{iizuka-sakano1996}%
  \BibitemOpen
  \bibfield  {author} {\bibinfo {author} {\bibfnamefont {T.}~\bibnamefont
  {{Iizuka-Sakano}}}\ and\ \bibinfo {author} {\bibfnamefont {Y.}~\bibnamefont
  {Toyozawa}},\ }\bibfield  {title} {\bibinfo {title} {The {{Role}} of
  {{Long-Range Coulomb Interaction}} in the {{Neutral-to-Ionic Transition}} of
  {{Quasi-One-Dimensional Charge Transfer Compounds}}},\ }\href
  {https://doi.org/10.1143/JPSJ.65.671} {\bibfield  {journal} {\bibinfo
  {journal} {J. Phys. Soc. Jpn.}\ }\textbf {\bibinfo {volume} {65}},\ \bibinfo
  {pages} {671} (\bibinfo {year} {1996})}\BibitemShut {NoStop}%
\bibitem [{\citenamefont {Miyashita}\ \emph {et~al.}(2003)\citenamefont
  {Miyashita}, \citenamefont {Kuwabara},\ and\ \citenamefont
  {Yonemitsu}}]{miyashita2003}%
  \BibitemOpen
  \bibfield  {author} {\bibinfo {author} {\bibfnamefont {N.}~\bibnamefont
  {Miyashita}}, \bibinfo {author} {\bibfnamefont {M.}~\bibnamefont
  {Kuwabara}},\ and\ \bibinfo {author} {\bibfnamefont {K.}~\bibnamefont
  {Yonemitsu}},\ }\bibfield  {title} {\bibinfo {title} {Electronic and
  {{Lattice Dynamics}} in the {{Photoinduced Ionic-to-Neutral Phase
  Transition}} in a {{One-Dimensional Extended Peierls}}\textendash{{Hubbard
  Model}}},\ }\href {https://doi.org/10.1143/JPSJ.72.2282} {\bibfield
  {journal} {\bibinfo  {journal} {J. Phys. Soc. Jpn.}\ }\textbf {\bibinfo
  {volume} {72}},\ \bibinfo {pages} {2282} (\bibinfo {year}
  {2003})}\BibitemShut {NoStop}%
\bibitem [{\citenamefont {Su}\ \emph {et~al.}(1980)\citenamefont {Su},
  \citenamefont {Schrieffer},\ and\ \citenamefont {Heeger}}]{su1980}%
  \BibitemOpen
  \bibfield  {author} {\bibinfo {author} {\bibfnamefont {W.~P.}\ \bibnamefont
  {Su}}, \bibinfo {author} {\bibfnamefont {J.~R.}\ \bibnamefont {Schrieffer}},\
  and\ \bibinfo {author} {\bibfnamefont {A.~J.}\ \bibnamefont {Heeger}},\
  }\bibfield  {title} {\bibinfo {title} {Soliton excitations in
  polyacetylene},\ }\href {https://doi.org/10.1103/PhysRevB.22.2099} {\bibfield
   {journal} {\bibinfo  {journal} {Phys. Rev. B}\ }\textbf {\bibinfo {volume}
  {22}},\ \bibinfo {pages} {2099} (\bibinfo {year} {1980})}\BibitemShut
  {NoStop}%
\bibitem [{\citenamefont {Horovitz}\ and\ \citenamefont
  {S{\'o}lyom}(1987)}]{horovitz1987}%
  \BibitemOpen
  \bibfield  {author} {\bibinfo {author} {\bibfnamefont {B.}~\bibnamefont
  {Horovitz}}\ and\ \bibinfo {author} {\bibfnamefont {J.}~\bibnamefont
  {S{\'o}lyom}},\ }\bibfield  {title} {\bibinfo {title} {Dimerization
  transition versus neutral-ionic transition in organic mixed-stack
  compounds},\ }\href {https://doi.org/10.1103/PhysRevB.35.7081} {\bibfield
  {journal} {\bibinfo  {journal} {Phys. Rev. B}\ }\textbf {\bibinfo {volume}
  {35}},\ \bibinfo {pages} {7081} (\bibinfo {year} {1987})}\BibitemShut
  {NoStop}%
\bibitem [{\citenamefont {Luty}\ and\ \citenamefont {Kuchta}(1987)}]{luty1987}%
  \BibitemOpen
  \bibfield  {author} {\bibinfo {author} {\bibfnamefont {T.}~\bibnamefont
  {Luty}}\ and\ \bibinfo {author} {\bibfnamefont {B.}~\bibnamefont {Kuchta}},\
  }\bibfield  {title} {\bibinfo {title} {Ground-state properties and
  neutral-to-ionic transformation of organic mixed-stack compounds:
  {{Mean-field}} approximation},\ }\href
  {https://doi.org/10.1103/PhysRevB.35.8542} {\bibfield  {journal} {\bibinfo
  {journal} {Phys. Rev. B}\ }\textbf {\bibinfo {volume} {35}},\ \bibinfo
  {pages} {8542} (\bibinfo {year} {1987})}\BibitemShut {NoStop}%
\bibitem [{\citenamefont {Gomi}\ \emph {et~al.}(2017)\citenamefont {Gomi},
  \citenamefont {Yamagishi}, \citenamefont {Mase}, \citenamefont {Inagaki},\
  and\ \citenamefont {Takahashi}}]{gomi2017}%
  \BibitemOpen
  \bibfield  {author} {\bibinfo {author} {\bibfnamefont {H.}~\bibnamefont
  {Gomi}}, \bibinfo {author} {\bibfnamefont {N.}~\bibnamefont {Yamagishi}},
  \bibinfo {author} {\bibfnamefont {T.}~\bibnamefont {Mase}}, \bibinfo {author}
  {\bibfnamefont {T.~J.}\ \bibnamefont {Inagaki}},\ and\ \bibinfo {author}
  {\bibfnamefont {A.}~\bibnamefont {Takahashi}},\ }\bibfield  {title} {\bibinfo
  {title} {Instantaneous charge and dielectric response to terahertz pulse
  excitation in {{TTF-CA}}},\ }\href
  {https://doi.org/10.1103/PhysRevB.95.094116} {\bibfield  {journal} {\bibinfo
  {journal} {Phys. Rev. B}\ }\textbf {\bibinfo {volume} {95}},\ \bibinfo
  {pages} {094116} (\bibinfo {year} {2017})}\BibitemShut {NoStop}%
\bibitem [{\citenamefont {Ohmura}\ \emph {et~al.}(2019)\citenamefont {Ohmura},
  \citenamefont {Mase},\ and\ \citenamefont {Takahashi}}]{ohmura2019}%
  \BibitemOpen
  \bibfield  {author} {\bibinfo {author} {\bibfnamefont {S.}~\bibnamefont
  {Ohmura}}, \bibinfo {author} {\bibfnamefont {T.}~\bibnamefont {Mase}},\ and\
  \bibinfo {author} {\bibfnamefont {A.}~\bibnamefont {Takahashi}},\ }\bibfield
  {title} {\bibinfo {title} {Terahertz pulse induced transitions between ionic
  and neutral phases and electronic polarization reversal in {{TTF-CA}}},\
  }\href {https://doi.org/10.1103/PhysRevB.100.035116} {\bibfield  {journal}
  {\bibinfo  {journal} {Phys. Rev. B}\ }\textbf {\bibinfo {volume} {100}},\
  \bibinfo {pages} {035116} (\bibinfo {year} {2019})}\BibitemShut {NoStop}%
\bibitem [{\citenamefont {Resta}\ and\ \citenamefont
  {Sorella}(1995)}]{resta1995}%
  \BibitemOpen
  \bibfield  {author} {\bibinfo {author} {\bibfnamefont {R.}~\bibnamefont
  {Resta}}\ and\ \bibinfo {author} {\bibfnamefont {S.}~\bibnamefont
  {Sorella}},\ }\bibfield  {title} {\bibinfo {title} {Many-{{Body Effects}} on
  {{Polarization}} and {{Dynamical Charges}} in a {{Partly Covalent Polar
  Insulator}}},\ }\href {https://doi.org/10.1103/PhysRevLett.74.4738}
  {\bibfield  {journal} {\bibinfo  {journal} {Phys. Rev. Lett.}\ }\textbf
  {\bibinfo {volume} {74}},\ \bibinfo {pages} {4738} (\bibinfo {year}
  {1995})}\BibitemShut {NoStop}%
\bibitem [{\citenamefont {Resta}\ and\ \citenamefont
  {Sorella}(1999)}]{resta1999}%
  \BibitemOpen
  \bibfield  {author} {\bibinfo {author} {\bibfnamefont {R.}~\bibnamefont
  {Resta}}\ and\ \bibinfo {author} {\bibfnamefont {S.}~\bibnamefont
  {Sorella}},\ }\bibfield  {title} {\bibinfo {title} {Electron {{Localization}}
  in the {{Insulating State}}},\ }\href
  {https://doi.org/10.1103/PhysRevLett.82.370} {\bibfield  {journal} {\bibinfo
  {journal} {Phys. Rev. Lett.}\ }\textbf {\bibinfo {volume} {82}},\ \bibinfo
  {pages} {370} (\bibinfo {year} {1999})}\BibitemShut {NoStop}%
\bibitem [{\citenamefont {Kumar}\ and\ \citenamefont {Soos}(2010)}]{kumar2010}%
  \BibitemOpen
  \bibfield  {author} {\bibinfo {author} {\bibfnamefont {M.}~\bibnamefont
  {Kumar}}\ and\ \bibinfo {author} {\bibfnamefont {Z.~G.}\ \bibnamefont
  {Soos}},\ }\bibfield  {title} {\bibinfo {title} {Bond-order wave phase of the
  extended {{Hubbard}} model: {{Electronic}} solitons, paramagnetism, and
  coupling to {{Peierls}} and {{Holstein}} phonons},\ }\href
  {https://doi.org/10.1103/PhysRevB.82.155144} {\bibfield  {journal} {\bibinfo
  {journal} {Phys. Rev. B}\ }\textbf {\bibinfo {volume} {82}},\ \bibinfo
  {pages} {155144} (\bibinfo {year} {2010})}\BibitemShut {NoStop}%
\bibitem [{\citenamefont {Painelli}\ and\ \citenamefont
  {Girlando}(1988)}]{painelli1988}%
  \BibitemOpen
  \bibfield  {author} {\bibinfo {author} {\bibfnamefont {A.}~\bibnamefont
  {Painelli}}\ and\ \bibinfo {author} {\bibfnamefont {A.}~\bibnamefont
  {Girlando}},\ }\bibfield  {title} {\bibinfo {title} {Zero-temperature phase
  diagram of mixed-stack charge-transfer crystals},\ }\href
  {https://doi.org/10.1103/PhysRevB.37.5748} {\bibfield  {journal} {\bibinfo
  {journal} {Phys. Rev. B}\ }\textbf {\bibinfo {volume} {37}},\ \bibinfo
  {pages} {5748} (\bibinfo {year} {1988})}\BibitemShut {NoStop}%
\bibitem [{\citenamefont {Caprara}\ \emph {et~al.}(2000)\citenamefont
  {Caprara}, \citenamefont {Avignon},\ and\ \citenamefont
  {Navarro}}]{caprara2000}%
  \BibitemOpen
  \bibfield  {author} {\bibinfo {author} {\bibfnamefont {S.}~\bibnamefont
  {Caprara}}, \bibinfo {author} {\bibfnamefont {M.}~\bibnamefont {Avignon}},\
  and\ \bibinfo {author} {\bibfnamefont {O.}~\bibnamefont {Navarro}},\
  }\bibfield  {title} {\bibinfo {title} {Spin and charge ordering in the
  dimerized {{Hubbard}} model},\ }\href
  {https://doi.org/10.1103/PhysRevB.61.15667} {\bibfield  {journal} {\bibinfo
  {journal} {Phys. Rev. B}\ }\textbf {\bibinfo {volume} {61}},\ \bibinfo
  {pages} {15667} (\bibinfo {year} {2000})}\BibitemShut {NoStop}%
\bibitem [{\citenamefont {Yonemitsu}(2011)}]{yonemitsu2011}%
  \BibitemOpen
  \bibfield  {author} {\bibinfo {author} {\bibfnamefont {K.}~\bibnamefont
  {Yonemitsu}},\ }\bibfield  {title} {\bibinfo {title} {Effects of {{Lattice}}
  and {{Molecular Phonons}} on {{Photoinduced Neutral-to-Ionic Transition
  Dynamics}} in {{Tetrathiafulvalene-}}{\emph{p}}-chloranil},\ }\href
  {https://doi.org/10.1143/JPSJ.80.084707} {\bibfield  {journal} {\bibinfo
  {journal} {J. Phys. Soc. Jpn.}\ }\textbf {\bibinfo {volume} {80}},\ \bibinfo
  {pages} {084707} (\bibinfo {year} {2011})}\BibitemShut {NoStop}%
\bibitem [{\citenamefont {D'Avino}\ \emph {et~al.}(2017)\citenamefont
  {D'Avino}, \citenamefont {Painelli},\ and\ \citenamefont
  {Soos}}]{davino2017}%
  \BibitemOpen
  \bibfield  {author} {\bibinfo {author} {\bibfnamefont {G.}~\bibnamefont
  {D'Avino}}, \bibinfo {author} {\bibfnamefont {A.}~\bibnamefont {Painelli}},\
  and\ \bibinfo {author} {\bibfnamefont {Z.}~\bibnamefont {Soos}},\ }\bibfield
  {title} {\bibinfo {title} {Modeling the {{Neutral-Ionic Transition}} with
  {{Correlated Electrons Coupled}} to {{Soft Lattices}} and {{Molecules}}},\
  }\href {https://doi.org/10.3390/cryst7050144} {\bibfield  {journal} {\bibinfo
   {journal} {Crystals}\ }\textbf {\bibinfo {volume} {7}},\ \bibinfo {pages}
  {144} (\bibinfo {year} {2017})}\BibitemShut {NoStop}%
\bibitem [{\citenamefont {Nagaosa}\ and\ \citenamefont
  {Takimoto}(1986)}]{nagaosa1986}%
  \BibitemOpen
  \bibfield  {author} {\bibinfo {author} {\bibfnamefont {N.}~\bibnamefont
  {Nagaosa}}\ and\ \bibinfo {author} {\bibfnamefont {J.-i.}\ \bibnamefont
  {Takimoto}},\ }\bibfield  {title} {\bibinfo {title} {Theory of
  {{Neutral-Ionic Transition}} in {{Organic Crystals}}. {{I}}. {{Monte Carlo
  Simulation}} of {{Modified Hubbard Model}}},\ }\href
  {https://doi.org/10.1143/JPSJ.55.2735} {\bibfield  {journal} {\bibinfo
  {journal} {J. Phys. Soc. Jpn.}\ }\textbf {\bibinfo {volume} {55}},\ \bibinfo
  {pages} {2735} (\bibinfo {year} {1986})}\BibitemShut {NoStop}%
\bibitem [{\citenamefont {Egami}\ \emph {et~al.}(1993)\citenamefont {Egami},
  \citenamefont {Ishihara},\ and\ \citenamefont {Tachiki}}]{egami1993}%
  \BibitemOpen
  \bibfield  {author} {\bibinfo {author} {\bibfnamefont {T.}~\bibnamefont
  {Egami}}, \bibinfo {author} {\bibfnamefont {S.}~\bibnamefont {Ishihara}},\
  and\ \bibinfo {author} {\bibfnamefont {M.}~\bibnamefont {Tachiki}},\
  }\bibfield  {title} {\bibinfo {title} {Lattice {{Effect}} of {{Strong
  Electron Correlation}}: {{Implication}} for {{Ferroelectricity}} and
  {{Superconductivity}}},\ }\href
  {https://doi.org/10.1126/science.261.5126.1307} {\bibfield  {journal}
  {\bibinfo  {journal} {Science}\ }\textbf {\bibinfo {volume} {261}},\ \bibinfo
  {pages} {1307} (\bibinfo {year} {1993})}\BibitemShut {NoStop}%
\bibitem [{\citenamefont {Ishihara}\ \emph {et~al.}(1994)\citenamefont
  {Ishihara}, \citenamefont {Egami},\ and\ \citenamefont
  {Tachiki}}]{ishihara1994}%
  \BibitemOpen
  \bibfield  {author} {\bibinfo {author} {\bibfnamefont {S.}~\bibnamefont
  {Ishihara}}, \bibinfo {author} {\bibfnamefont {T.}~\bibnamefont {Egami}},\
  and\ \bibinfo {author} {\bibfnamefont {M.}~\bibnamefont {Tachiki}},\
  }\bibfield  {title} {\bibinfo {title} {Enhancement of the electron-lattice
  interaction due to strong electron correlation},\ }\href
  {https://doi.org/10.1103/PhysRevB.49.8944} {\bibfield  {journal} {\bibinfo
  {journal} {Phys. Rev. B}\ }\textbf {\bibinfo {volume} {49}},\ \bibinfo
  {pages} {8944} (\bibinfo {year} {1994})}\BibitemShut {NoStop}%
\bibitem [{\citenamefont {Maeshima}\ and\ \citenamefont
  {Yonemitsu}(2005)}]{maeshima2005a}%
  \BibitemOpen
  \bibfield  {author} {\bibinfo {author} {\bibfnamefont {N.}~\bibnamefont
  {Maeshima}}\ and\ \bibinfo {author} {\bibfnamefont {K.}~\bibnamefont
  {Yonemitsu}},\ }\bibfield  {title} {\bibinfo {title} {Photoinduced {{Metallic
  Properties}} of {{One-Dimensional Strongly Correlated Electron Systems}}},\
  }\href {https://doi.org/10.1143/JPSJ.74.2671} {\bibfield  {journal} {\bibinfo
   {journal} {J. Phys. Soc. Jpn.}\ }\textbf {\bibinfo {volume} {74}},\ \bibinfo
  {pages} {2671} (\bibinfo {year} {2005})}\BibitemShut {NoStop}%
\bibitem [{\citenamefont {Morimoto}\ \emph
  {et~al.}(2017{\natexlab{b}})\citenamefont {Morimoto}, \citenamefont
  {Miyamoto},\ and\ \citenamefont {Okamoto}}]{morimoto2017}%
  \BibitemOpen
  \bibfield  {author} {\bibinfo {author} {\bibfnamefont {T.}~\bibnamefont
  {Morimoto}}, \bibinfo {author} {\bibfnamefont {T.}~\bibnamefont {Miyamoto}},\
  and\ \bibinfo {author} {\bibfnamefont {H.}~\bibnamefont {Okamoto}},\
  }\bibfield  {title} {\bibinfo {title} {Ultrafast {{Electron}} and {{Molecular
  Dynamics}} in {{Photoinduced}} and {{Electric-Field-Induced
  Neutral}}\textendash{{Ionic Transitions}}},\ }\href
  {https://doi.org/10.3390/cryst7050132} {\bibfield  {journal} {\bibinfo
  {journal} {Crystals}\ }\textbf {\bibinfo {volume} {7}},\ \bibinfo {pages}
  {132} (\bibinfo {year} {2017}{\natexlab{b}})}\BibitemShut {NoStop}%
\bibitem [{\citenamefont {Shao}\ \emph {et~al.}(2016)\citenamefont {Shao},
  \citenamefont {Tohyama}, \citenamefont {Luo},\ and\ \citenamefont
  {Lu}}]{shao2016}%
  \BibitemOpen
  \bibfield  {author} {\bibinfo {author} {\bibfnamefont {C.}~\bibnamefont
  {Shao}}, \bibinfo {author} {\bibfnamefont {T.}~\bibnamefont {Tohyama}},
  \bibinfo {author} {\bibfnamefont {H.-G.}\ \bibnamefont {Luo}},\ and\ \bibinfo
  {author} {\bibfnamefont {H.}~\bibnamefont {Lu}},\ }\bibfield  {title}
  {\bibinfo {title} {Numerical method to compute optical conductivity based on
  pump-probe simulations},\ }\href {https://doi.org/10.1103/PhysRevB.93.195144}
  {\bibfield  {journal} {\bibinfo  {journal} {Phys. Rev. B}\ }\textbf {\bibinfo
  {volume} {93}},\ \bibinfo {pages} {195144} (\bibinfo {year}
  {2016})}\BibitemShut {NoStop}%
\bibitem [{\citenamefont {Lanczos}(1950)}]{lanczos1950}%
  \BibitemOpen
  \bibfield  {author} {\bibinfo {author} {\bibfnamefont {C.}~\bibnamefont
  {Lanczos}},\ }\bibfield  {title} {\bibinfo {title} {An iteration method for
  the solution of the eigenvalue problem of linear differential and integral
  operators},\ }\href {https://doi.org/10.6028/jres.045.026} {\bibfield
  {journal} {\bibinfo  {journal} {J. Res. Natl. Bur. Stand.}\ }\textbf
  {\bibinfo {volume} {45}},\ \bibinfo {pages} {255} (\bibinfo {year}
  {1950})}\BibitemShut {NoStop}%
\bibitem [{\citenamefont {Mitani}\ \emph {et~al.}(1984)\citenamefont {Mitani},
  \citenamefont {Saito}, \citenamefont {Tokura},\ and\ \citenamefont
  {Koda}}]{mitani1984}%
  \BibitemOpen
  \bibfield  {author} {\bibinfo {author} {\bibfnamefont {T.}~\bibnamefont
  {Mitani}}, \bibinfo {author} {\bibfnamefont {G.}~\bibnamefont {Saito}},
  \bibinfo {author} {\bibfnamefont {Y.}~\bibnamefont {Tokura}},\ and\ \bibinfo
  {author} {\bibfnamefont {T.}~\bibnamefont {Koda}},\ }\bibfield  {title}
  {\bibinfo {title} {Soliton {{Formation}} at the {{Neutral-to-Ionic Phase
  Transition}} in the {{Mixed-Stack Charge-Transfer Crystal
  Tetrathiafulvalene-}}{\emph{p}}-{{Chloranil}}},\ }\href
  {https://doi.org/10.1103/PhysRevLett.53.842} {\bibfield  {journal} {\bibinfo
  {journal} {Phys. Rev. Lett.}\ }\textbf {\bibinfo {volume} {53}},\ \bibinfo
  {pages} {842} (\bibinfo {year} {1984})}\BibitemShut {NoStop}%
\bibitem [{\citenamefont {Manmana}\ \emph {et~al.}(2004)\citenamefont
  {Manmana}, \citenamefont {Meden}, \citenamefont {Noack},\ and\ \citenamefont
  {Sch{\"o}nhammer}}]{manmana2004}%
  \BibitemOpen
  \bibfield  {author} {\bibinfo {author} {\bibfnamefont {S.~R.}\ \bibnamefont
  {Manmana}}, \bibinfo {author} {\bibfnamefont {V.}~\bibnamefont {Meden}},
  \bibinfo {author} {\bibfnamefont {R.~M.}\ \bibnamefont {Noack}},\ and\
  \bibinfo {author} {\bibfnamefont {K.}~\bibnamefont {Sch{\"o}nhammer}},\
  }\bibfield  {title} {\bibinfo {title} {Quantum critical behavior of the
  one-dimensional ionic {{Hubbard}} model},\ }\href
  {https://doi.org/10.1103/PhysRevB.70.155115} {\bibfield  {journal} {\bibinfo
  {journal} {Phys. Rev. B}\ }\textbf {\bibinfo {volume} {70}},\ \bibinfo
  {pages} {155115} (\bibinfo {year} {2004})}\BibitemShut {NoStop}%
\bibitem [{\citenamefont {{King-Smith}}\ and\ \citenamefont
  {Vanderbilt}(1993)}]{king-smith1993}%
  \BibitemOpen
  \bibfield  {author} {\bibinfo {author} {\bibfnamefont {R.~D.}\ \bibnamefont
  {{King-Smith}}}\ and\ \bibinfo {author} {\bibfnamefont {D.}~\bibnamefont
  {Vanderbilt}},\ }\bibfield  {title} {\bibinfo {title} {Theory of polarization
  of crystalline solids},\ }\href {https://doi.org/10.1103/PhysRevB.47.1651}
  {\bibfield  {journal} {\bibinfo  {journal} {Phys. Rev. B}\ }\textbf {\bibinfo
  {volume} {47}},\ \bibinfo {pages} {1651} (\bibinfo {year}
  {1993})}\BibitemShut {NoStop}%
\bibitem [{\citenamefont {Resta}(1992)}]{resta1992}%
  \BibitemOpen
  \bibfield  {author} {\bibinfo {author} {\bibfnamefont {R.}~\bibnamefont
  {Resta}},\ }\bibfield  {title} {\bibinfo {title} {Theory of the electric
  polarization in crystals},\ }\href
  {https://doi.org/10.1080/00150199208016065} {\bibfield  {journal} {\bibinfo
  {journal} {Ferroelectrics}\ }\textbf {\bibinfo {volume} {136}},\ \bibinfo
  {pages} {51} (\bibinfo {year} {1992})}\BibitemShut {NoStop}%
\bibitem [{\citenamefont {Soos}\ \emph {et~al.}(2004)\citenamefont {Soos},
  \citenamefont {Bewick}, \citenamefont {Peri},\ and\ \citenamefont
  {Painelli}}]{soos2004}%
  \BibitemOpen
  \bibfield  {author} {\bibinfo {author} {\bibfnamefont {Z.~G.}\ \bibnamefont
  {Soos}}, \bibinfo {author} {\bibfnamefont {S.~A.}\ \bibnamefont {Bewick}},
  \bibinfo {author} {\bibfnamefont {A.}~\bibnamefont {Peri}},\ and\ \bibinfo
  {author} {\bibfnamefont {A.}~\bibnamefont {Painelli}},\ }\bibfield  {title}
  {\bibinfo {title} {Dielectric response of modified {{Hubbard}} models with
  neutral-ionic and {{Peierls}} transitions},\ }\href
  {https://doi.org/10.1063/1.1665824} {\bibfield  {journal} {\bibinfo
  {journal} {J. Chem. Phys.}\ }\textbf {\bibinfo {volume} {120}},\ \bibinfo
  {pages} {6712} (\bibinfo {year} {2004})}\BibitemShut {NoStop}%
\bibitem [{\citenamefont {Ortiz}\ and\ \citenamefont
  {Martin}(1994)}]{ortiz1994}%
  \BibitemOpen
  \bibfield  {author} {\bibinfo {author} {\bibfnamefont {G.}~\bibnamefont
  {Ortiz}}\ and\ \bibinfo {author} {\bibfnamefont {R.~M.}\ \bibnamefont
  {Martin}},\ }\bibfield  {title} {\bibinfo {title} {Macroscopic polarization
  as a geometric quantum phase: {{Many-body}} formulation},\ }\href
  {https://doi.org/10.1103/PhysRevB.49.14202} {\bibfield  {journal} {\bibinfo
  {journal} {Phys. Rev. B}\ }\textbf {\bibinfo {volume} {49}},\ \bibinfo
  {pages} {14202} (\bibinfo {year} {1994})}\BibitemShut {NoStop}%
\bibitem [{\citenamefont {Resta}(1998)}]{resta1998}%
  \BibitemOpen
  \bibfield  {author} {\bibinfo {author} {\bibfnamefont {R.}~\bibnamefont
  {Resta}},\ }\bibfield  {title} {\bibinfo {title} {Quantum-{{Mechanical
  Position Operator}} in {{Extended Systems}}},\ }\href
  {https://doi.org/10.1103/PhysRevLett.80.1800} {\bibfield  {journal} {\bibinfo
   {journal} {Phys. Rev. Lett.}\ }\textbf {\bibinfo {volume} {80}},\ \bibinfo
  {pages} {1800} (\bibinfo {year} {1998})}\BibitemShut {NoStop}%
\bibitem [{\citenamefont {Souza}\ \emph {et~al.}(2000)\citenamefont {Souza},
  \citenamefont {Wilkens},\ and\ \citenamefont {Martin}}]{souza2000}%
  \BibitemOpen
  \bibfield  {author} {\bibinfo {author} {\bibfnamefont {I.}~\bibnamefont
  {Souza}}, \bibinfo {author} {\bibfnamefont {T.}~\bibnamefont {Wilkens}},\
  and\ \bibinfo {author} {\bibfnamefont {R.~M.}\ \bibnamefont {Martin}},\
  }\bibfield  {title} {\bibinfo {title} {Polarization and localization in
  insulators: {{Generating}} function approach},\ }\href
  {https://doi.org/10.1103/PhysRevB.62.1666} {\bibfield  {journal} {\bibinfo
  {journal} {Phys. Rev. B}\ }\textbf {\bibinfo {volume} {62}},\ \bibinfo
  {pages} {1666} (\bibinfo {year} {2000})}\BibitemShut {NoStop}%
\bibitem [{\citenamefont {Atala}\ \emph {et~al.}(2013)\citenamefont {Atala},
  \citenamefont {Aidelsburger}, \citenamefont {Barreiro}, \citenamefont
  {Abanin}, \citenamefont {Kitagawa}, \citenamefont {Demler},\ and\
  \citenamefont {Bloch}}]{atala2013}%
  \BibitemOpen
  \bibfield  {author} {\bibinfo {author} {\bibfnamefont {M.}~\bibnamefont
  {Atala}}, \bibinfo {author} {\bibfnamefont {M.}~\bibnamefont {Aidelsburger}},
  \bibinfo {author} {\bibfnamefont {J.~T.}\ \bibnamefont {Barreiro}}, \bibinfo
  {author} {\bibfnamefont {D.}~\bibnamefont {Abanin}}, \bibinfo {author}
  {\bibfnamefont {T.}~\bibnamefont {Kitagawa}}, \bibinfo {author}
  {\bibfnamefont {E.}~\bibnamefont {Demler}},\ and\ \bibinfo {author}
  {\bibfnamefont {I.}~\bibnamefont {Bloch}},\ }\bibfield  {title} {\bibinfo
  {title} {Direct measurement of the {{Zak}} phase in topological {{Bloch}}
  bands},\ }\href {https://doi.org/10.1038/nphys2790} {\bibfield  {journal}
  {\bibinfo  {journal} {Nat. Phys.}\ }\textbf {\bibinfo {volume} {9}},\
  \bibinfo {pages} {795} (\bibinfo {year} {2013})}\BibitemShut {NoStop}%
\bibitem [{\citenamefont {Dagotto}(1994)}]{dagotto1994}%
  \BibitemOpen
  \bibfield  {author} {\bibinfo {author} {\bibfnamefont {E.}~\bibnamefont
  {Dagotto}},\ }\bibfield  {title} {\bibinfo {title} {Correlated electrons in
  high-temperature superconductors},\ }\href
  {https://doi.org/10.1103/RevModPhys.66.763} {\bibfield  {journal} {\bibinfo
  {journal} {Rev. Mod. Phys.}\ }\textbf {\bibinfo {volume} {66}},\ \bibinfo
  {pages} {763} (\bibinfo {year} {1994})}\BibitemShut {NoStop}%
\bibitem [{\citenamefont {Takahashi}\ \emph {et~al.}(2008)\citenamefont
  {Takahashi}, \citenamefont {Itoh},\ and\ \citenamefont
  {Aihara}}]{takahashi2008}%
  \BibitemOpen
  \bibfield  {author} {\bibinfo {author} {\bibfnamefont {A.}~\bibnamefont
  {Takahashi}}, \bibinfo {author} {\bibfnamefont {H.}~\bibnamefont {Itoh}},\
  and\ \bibinfo {author} {\bibfnamefont {M.}~\bibnamefont {Aihara}},\
  }\bibfield  {title} {\bibinfo {title} {Photoinduced insulator-metal
  transition in one-dimensional {{Mott}} insulators},\ }\href
  {https://doi.org/10.1103/PhysRevB.77.205105} {\bibfield  {journal} {\bibinfo
  {journal} {Phys. Rev. B}\ }\textbf {\bibinfo {volume} {77}},\ \bibinfo
  {pages} {205105} (\bibinfo {year} {2008})}\BibitemShut {NoStop}%
\bibitem [{\citenamefont {Jacobsen}\ and\ \citenamefont
  {Torrance}(1983)}]{jacobsen1983}%
  \BibitemOpen
  \bibfield  {author} {\bibinfo {author} {\bibfnamefont {C.~S.}\ \bibnamefont
  {Jacobsen}}\ and\ \bibinfo {author} {\bibfnamefont {J.~B.}\ \bibnamefont
  {Torrance}},\ }\bibfield  {title} {\bibinfo {title} {Behavior of
  charge-transfer absorption upon passing through the neutral-ionic phase
  transition},\ }\href {https://doi.org/10.1063/1.444530} {\bibfield  {journal}
  {\bibinfo  {journal} {J. Chem. Phys.}\ }\textbf {\bibinfo {volume} {78}},\
  \bibinfo {pages} {112} (\bibinfo {year} {1983})}\BibitemShut {NoStop}%
\bibitem [{\citenamefont {Fye}\ \emph {et~al.}(1991)\citenamefont {Fye},
  \citenamefont {Martins}, \citenamefont {Scalapino}, \citenamefont {Wagner},\
  and\ \citenamefont {Hanke}}]{fye1991}%
  \BibitemOpen
  \bibfield  {author} {\bibinfo {author} {\bibfnamefont {R.~M.}\ \bibnamefont
  {Fye}}, \bibinfo {author} {\bibfnamefont {M.~J.}\ \bibnamefont {Martins}},
  \bibinfo {author} {\bibfnamefont {D.~J.}\ \bibnamefont {Scalapino}}, \bibinfo
  {author} {\bibfnamefont {J.}~\bibnamefont {Wagner}},\ and\ \bibinfo {author}
  {\bibfnamefont {W.}~\bibnamefont {Hanke}},\ }\bibfield  {title} {\bibinfo
  {title} {Drude weight, optical conductivity, and flux properties of
  one-dimensional {{Hubbard}} rings},\ }\href
  {https://doi.org/10.1103/PhysRevB.44.6909} {\bibfield  {journal} {\bibinfo
  {journal} {Phys. Rev. B}\ }\textbf {\bibinfo {volume} {44}},\ \bibinfo
  {pages} {6909} (\bibinfo {year} {1991})}\BibitemShut {NoStop}%
\bibitem [{\citenamefont {Tokura}\ \emph {et~al.}(1982)\citenamefont {Tokura},
  \citenamefont {Koda}, \citenamefont {Mitani},\ and\ \citenamefont
  {Saito}}]{tokura1982}%
  \BibitemOpen
  \bibfield  {author} {\bibinfo {author} {\bibfnamefont {Y.}~\bibnamefont
  {Tokura}}, \bibinfo {author} {\bibfnamefont {T.}~\bibnamefont {Koda}},
  \bibinfo {author} {\bibfnamefont {T.}~\bibnamefont {Mitani}},\ and\ \bibinfo
  {author} {\bibfnamefont {G.}~\bibnamefont {Saito}},\ }\bibfield  {title}
  {\bibinfo {title} {Neutral-to-ionic transition in
  tetrathiafulvalene-{\emph{p}}-chloranil as investigated by optical reflection
  spectra},\ }\href {https://doi.org/10.1016/0038-1098(82)90986-3} {\bibfield
  {journal} {\bibinfo  {journal} {Solid State Commun.}\ }\textbf {\bibinfo
  {volume} {43}},\ \bibinfo {pages} {757} (\bibinfo {year} {1982})}\BibitemShut
  {NoStop}%
\bibitem [{\citenamefont {Okamoto}\ and\ \citenamefont
  {Peronaci}(2021)}]{okamoto2021}%
  \BibitemOpen
  \bibfield  {author} {\bibinfo {author} {\bibfnamefont {J.}~\bibnamefont
  {Okamoto}}\ and\ \bibinfo {author} {\bibfnamefont {F.}~\bibnamefont
  {Peronaci}},\ }\bibfield  {title} {\bibinfo {title} {Floquet
  prethermalization and {{Rabi}} oscillations in optically excited {{Hubbard}}
  clusters},\ }\href {https://doi.org/10.1038/s41598-021-97104-x} {\bibfield
  {journal} {\bibinfo  {journal} {Sci. Rep.}\ }\textbf {\bibinfo {volume}
  {11}},\ \bibinfo {pages} {17994} (\bibinfo {year} {2021})}\BibitemShut
  {NoStop}%
\bibitem [{\citenamefont {Okamoto}(2019)}]{okamoto2019}%
  \BibitemOpen
  \bibfield  {author} {\bibinfo {author} {\bibfnamefont {J.}~\bibnamefont
  {Okamoto}},\ }\bibfield  {title} {\bibinfo {title} {Time-dependent spectral
  properties of a photoexcited one-dimensional ionic {{Hubbard}} model: An
  exact diagonalization study},\ }\href
  {https://doi.org/10.1088/1367-2630/ab5c54} {\bibfield  {journal} {\bibinfo
  {journal} {New J. Phys.}\ }\textbf {\bibinfo {volume} {21}},\ \bibinfo
  {pages} {123040} (\bibinfo {year} {2019})}\BibitemShut {NoStop}%
\bibitem [{\citenamefont {Sunami}\ \emph {et~al.}(2018)\citenamefont {Sunami},
  \citenamefont {Nishikawa}, \citenamefont {Miyagawa}, \citenamefont
  {Horiuchi}, \citenamefont {Kato}, \citenamefont {Miyamoto}, \citenamefont
  {Okamoto},\ and\ \citenamefont {Kanoda}}]{sunami2018}%
  \BibitemOpen
  \bibfield  {author} {\bibinfo {author} {\bibfnamefont {K.}~\bibnamefont
  {Sunami}}, \bibinfo {author} {\bibfnamefont {T.}~\bibnamefont {Nishikawa}},
  \bibinfo {author} {\bibfnamefont {K.}~\bibnamefont {Miyagawa}}, \bibinfo
  {author} {\bibfnamefont {S.}~\bibnamefont {Horiuchi}}, \bibinfo {author}
  {\bibfnamefont {R.}~\bibnamefont {Kato}}, \bibinfo {author} {\bibfnamefont
  {T.}~\bibnamefont {Miyamoto}}, \bibinfo {author} {\bibfnamefont
  {H.}~\bibnamefont {Okamoto}},\ and\ \bibinfo {author} {\bibfnamefont
  {K.}~\bibnamefont {Kanoda}},\ }\bibfield  {title} {\bibinfo {title} {Evidence
  for solitonic spin excitations from a charge-lattice\textendash coupled
  ferroelectric order},\ }\href {https://doi.org/10.1126/sciadv.aau7725}
  {\bibfield  {journal} {\bibinfo  {journal} {Sci. Adv.}\ }\textbf {\bibinfo
  {volume} {4}},\ \bibinfo {pages} {eaau7725} (\bibinfo {year}
  {2018})}\BibitemShut {NoStop}%
\bibitem [{\citenamefont {Mizuno}\ \emph {et~al.}(2000)\citenamefont {Mizuno},
  \citenamefont {Tsutsui}, \citenamefont {Tohyama},\ and\ \citenamefont
  {Maekawa}}]{mizuno2000}%
  \BibitemOpen
  \bibfield  {author} {\bibinfo {author} {\bibfnamefont {Y.}~\bibnamefont
  {Mizuno}}, \bibinfo {author} {\bibfnamefont {K.}~\bibnamefont {Tsutsui}},
  \bibinfo {author} {\bibfnamefont {T.}~\bibnamefont {Tohyama}},\ and\ \bibinfo
  {author} {\bibfnamefont {S.}~\bibnamefont {Maekawa}},\ }\bibfield  {title}
  {\bibinfo {title} {Nonlinear optical response and spin-charge separation in
  one-dimensional {{Mott}} insulators},\ }\href
  {https://doi.org/10.1103/PhysRevB.62.R4769} {\bibfield  {journal} {\bibinfo
  {journal} {Phys. Rev. B}\ }\textbf {\bibinfo {volume} {62}},\ \bibinfo
  {pages} {R4769} (\bibinfo {year} {2000})}\BibitemShut {NoStop}%
\bibitem [{\citenamefont {Kindt}\ and\ \citenamefont
  {Schmuttenmaer}(1999)}]{kindt1999}%
  \BibitemOpen
  \bibfield  {author} {\bibinfo {author} {\bibfnamefont {J.~T.}\ \bibnamefont
  {Kindt}}\ and\ \bibinfo {author} {\bibfnamefont {C.~A.}\ \bibnamefont
  {Schmuttenmaer}},\ }\bibfield  {title} {\bibinfo {title} {Theory for
  determination of the low-frequency time-dependent response function in
  liquids using time-resolved terahertz pulse spectroscopy},\ }\href
  {https://doi.org/10.1063/1.478766} {\bibfield  {journal} {\bibinfo  {journal}
  {J. Chem. Phys.}\ }\textbf {\bibinfo {volume} {110}},\ \bibinfo {pages}
  {8589} (\bibinfo {year} {1999})}\BibitemShut {NoStop}%
\bibitem [{\citenamefont {N{\v e}mec}\ \emph {et~al.}(2002)\citenamefont {N{\v
  e}mec}, \citenamefont {Kadlec},\ and\ \citenamefont {Ku{\v
  z}el}}]{nemec2002}%
  \BibitemOpen
  \bibfield  {author} {\bibinfo {author} {\bibfnamefont {H.}~\bibnamefont {N{\v
  e}mec}}, \bibinfo {author} {\bibfnamefont {F.}~\bibnamefont {Kadlec}},\ and\
  \bibinfo {author} {\bibfnamefont {P.}~\bibnamefont {Ku{\v z}el}},\ }\bibfield
   {title} {\bibinfo {title} {Methodology of an optical pump-terahertz probe
  experiment: {{An}} analytical frequency-domain approach},\ }\href
  {https://doi.org/10.1063/1.1512648} {\bibfield  {journal} {\bibinfo
  {journal} {J. Chem. Phys.}\ }\textbf {\bibinfo {volume} {117}},\ \bibinfo
  {pages} {8454} (\bibinfo {year} {2002})}\BibitemShut {NoStop}%
\bibitem [{\citenamefont {Vengurlekar}\ and\ \citenamefont
  {Jha}(1988)}]{vengurlekar1988}%
  \BibitemOpen
  \bibfield  {author} {\bibinfo {author} {\bibfnamefont {A.~S.}\ \bibnamefont
  {Vengurlekar}}\ and\ \bibinfo {author} {\bibfnamefont {S.~S.}\ \bibnamefont
  {Jha}},\ }\bibfield  {title} {\bibinfo {title} {Conductivity response of
  nonthermal hot carriers photoexcited by subpicosecond pulses in {{GaAs}}},\
  }\href {https://doi.org/10.1103/PhysRevB.38.2044} {\bibfield  {journal}
  {\bibinfo  {journal} {Phys. Rev. B}\ }\textbf {\bibinfo {volume} {38}},\
  \bibinfo {pages} {2044} (\bibinfo {year} {1988})}\BibitemShut {NoStop}%
\bibitem [{\citenamefont {Nagaosa}\ and\ \citenamefont
  {Ogawa}(1989)}]{nagaosa1989}%
  \BibitemOpen
  \bibfield  {author} {\bibinfo {author} {\bibfnamefont {N.}~\bibnamefont
  {Nagaosa}}\ and\ \bibinfo {author} {\bibfnamefont {T.}~\bibnamefont
  {Ogawa}},\ }\bibfield  {title} {\bibinfo {title} {Theory of photoinduced
  structure changes},\ }\href {https://doi.org/10.1103/PhysRevB.39.4472}
  {\bibfield  {journal} {\bibinfo  {journal} {Phys. Rev. B}\ }\textbf {\bibinfo
  {volume} {39}},\ \bibinfo {pages} {4472} (\bibinfo {year}
  {1989})}\BibitemShut {NoStop}%
\bibitem [{\citenamefont {Iwano}(2006)}]{iwano2006}%
  \BibitemOpen
  \bibfield  {author} {\bibinfo {author} {\bibfnamefont {K.}~\bibnamefont
  {Iwano}},\ }\bibfield  {title} {\bibinfo {title} {Direct {{Photoexcitation}}
  of {{Appreciable Size}} of {{Domains}} without {{Lattice Motion}} in
  {{Neutral-Ionic Transition Systems}}},\ }\href
  {https://doi.org/10.1103/PhysRevLett.97.226404} {\bibfield  {journal}
  {\bibinfo  {journal} {Phys. Rev. Lett.}\ }\textbf {\bibinfo {volume} {97}},\
  \bibinfo {pages} {226404} (\bibinfo {year} {2006})}\BibitemShut {NoStop}%
\bibitem [{\citenamefont {Bittner}\ \emph {et~al.}(2019)\citenamefont
  {Bittner}, \citenamefont {Tohyama}, \citenamefont {Kaiser},\ and\
  \citenamefont {Manske}}]{bittner2019}%
  \BibitemOpen
  \bibfield  {author} {\bibinfo {author} {\bibfnamefont {N.}~\bibnamefont
  {Bittner}}, \bibinfo {author} {\bibfnamefont {T.}~\bibnamefont {Tohyama}},
  \bibinfo {author} {\bibfnamefont {S.}~\bibnamefont {Kaiser}},\ and\ \bibinfo
  {author} {\bibfnamefont {D.}~\bibnamefont {Manske}},\ }\bibfield  {title}
  {\bibinfo {title} {Possible {{Light-Induced Superconductivity}} in a
  {{Strongly Correlated Electron System}}},\ }\href
  {https://doi.org/10.7566/JPSJ.88.044704} {\bibfield  {journal} {\bibinfo
  {journal} {J. Phys. Soc. Jpn.}\ }\textbf {\bibinfo {volume} {88}},\ \bibinfo
  {pages} {044704} (\bibinfo {year} {2019})}\BibitemShut {NoStop}%
\end{thebibliography}

%

\end{document}